\RequirePackage{fix-cm}
\documentclass{svjour3}   
\smartqed 

\usepackage{graphicx}
\usepackage{color}
\usepackage[breakwords]{truncate}
\usepackage[font={small}]{caption}
\usepackage{graphicx}
\usepackage{cite} 
\usepackage{subcaption}
\usepackage{etoolbox}
\usepackage[section]{placeins}
\usepackage[sectionbib]{chapterbib}
\usepackage{amsfonts}
\usepackage{amsmath}
\usepackage[bookmarks,bookmarksopen,bookmarksdepth=2]{hyperref}
\usepackage{bm}

\newcommand{\bea}{\begin{eqnarray}}
\newcommand{\eea}{\end{eqnarray}}

\begin{document}

\title{Universal order statistics for random walks \& L\'evy flights}

\author{Benjamin De Bruyne \and
        Satya N. Majumdar \and
        Gr\'egory Schehr
}

\institute{Benjamin De Bruyne, Satya N. Majumdar \at
              LPTMS, CNRS, Univ.\ Paris-Sud, Universit\'e Paris-Saclay, 91405 Orsay, France\\
              \email{benjamin.debruyne@centraliens.net} \\
              \email{satya.majumdar@universite-paris-saclay.fr}          
              \and
           Gr\'egory Schehr \at
           Sorbonne Universit\'e, Laboratoire de Physique Th\'eorique et Hautes Energies, CNRS UMR 7589, 4 Place Jussieu, 75252 Paris Cedex 05, France \\
           \email{gregory.schehr@lpthe.jussieu.fr}
}

\date{Received: date / Accepted: date}
\maketitle

\begin{abstract}
We consider one-dimensional discrete-time random walks (RWs) of $n$ steps, starting from $x_0=0$, with
arbitrary symmetric and continuous jump distributions $f(\eta)$, including the important case of L\'evy flights. We study the statistics of the gaps $\Delta_{k,n}$ between the $k^\text{th}$ and $(k+1)^\text{th}$ maximum of the set of positions $\{x_1,\ldots,x_n\}$. We obtain an exact analytical expression for the probability distribution $P_{k,n}(\Delta)$ valid for any $k$ and $n$, and jump distribution $f(\eta)$, which we then analyse in the large $n$ limit. For jump distributions whose Fourier transform behaves, for small $q$, as $\hat f (q) \sim 1 - |q|^\mu$ with a L\'evy index $0< \mu \leq 2$, we find that, the distribution becomes stationary in the limit of $n\to \infty$, i.e. $\lim_{n\to \infty} P_{k,n}(\Delta)=P_k(\Delta)$. We obtain an explicit expression for its first moment $\mathbb{E}[\Delta_{k}]$, valid for any $k$ and jump distribution $f(\eta)$ with $\mu>1$, and show that it exhibits a universal algebraic decay $ \mathbb{E}[\Delta_{k}]\sim k^{1/\mu-1} \Gamma\left(1-1/\mu\right)/\pi$ for large $k$. Furthermore, for $\mu>1$, we show that in the limit of $k\to\infty$ the stationary distribution exhibits a universal scaling form $P_k(\Delta) \sim  k^{1-1/\mu} \mathcal{P}_\mu(k^{1-1/\mu}\Delta)$ which depends only on the L\'evy index $\mu$, but not on the details of the jump distribution. We compute explicitly the limiting scaling function $\mathcal{P}_\mu(x)$ in terms of Mittag-Leffler functions. For $1< \mu <2$, we show that, while this scaling function captures the distribution of the typical gaps on the scale $k^{1/\mu-1}$, the atypical large gaps are not described by this scaling function since they occur at a larger scale of order $k^{1/\mu}$. This atypical part of the distribution is reminiscent of a ``condensation bump'' that one often encounters in several mass transport models. 
\end{abstract}

\section{Introduction}
\begin{figure}[t]
  \begin{center}
    \includegraphics[width=0.33\textwidth]{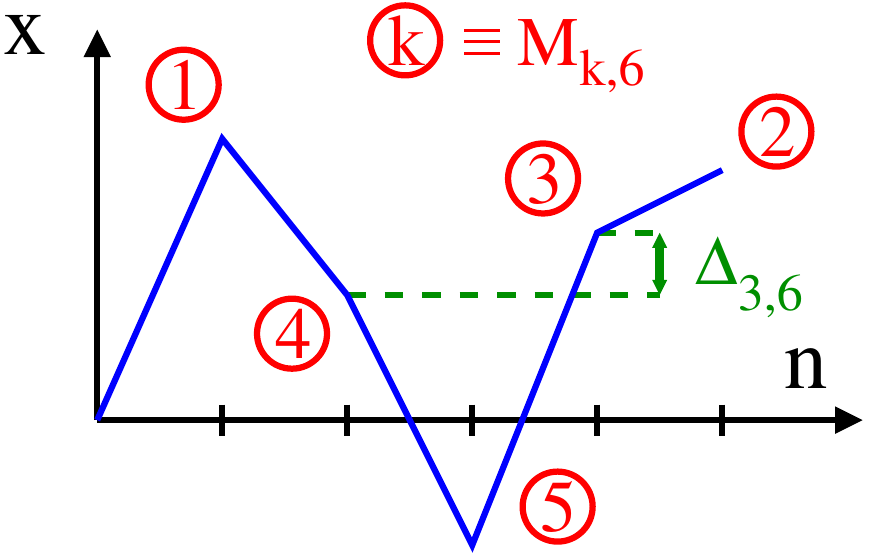}
    \caption{A trajectory of a random walk of $n=5$ steps. The positions are ordered by ascending order $M_{1,n} > \ldots> M_{n,n}$, where $M_{k,n}$ is the $k^\text{th}$ maximum of the set of positions $\{x_1,\ldots,x_n\}$. Note that the initial position $x_0=0$ is not included. The gap $\Delta_{k,n}$ is the difference between two consecutive maxima $\Delta_{k,n} = M_{k,n} - M_{k+1,n}$ . }
    \label{fig:model}
  \end{center}
\end{figure}

Extreme events are rare but can have drastic consequences. The field of
extreme value statistics (EVS) is devoted to their study and has shown to have a wide variety of applications ranging from environmental sciences \cite{gumbel,katz} to finance \cite{embrecht,bouchaud_satya}. EVS also play a key role in physics, especially in the description of disordered systems \cite{bm97,PLDCecile,Dahmen,sg}, fluctuating interfaces \cite{Shapir,GHPZ,Satya_Airy1,Satya_Airy2,SOS_Airy}, and random matrices \cite{TW,SMS14} (for a recent review see \cite{reviewMPS}). One of the simplest example of EVS is the statistics of the maximum of a set of independent and identically distributed (IID) random variables drawn from the same probability distribution function (PDF). This example has been widely studied and has shown to exhibit three universality classes that describe the statistics of the maximum in the limit of a large number of random variables \cite{gumbel} (see \cite{reviewMPS,SMS14r,Vivo15} for pedagogical reviews). While the statistics of the maximum is already of great interest on its own, it is sometimes necessary to know whether this maximum is an isolated event or if there exists other events with a similar magnitude \cite{SSM07,SSM08}. A natural way to shed some light on this matter is to not only study the global maximum, but also the second maximum, the third, etc. These ordered maxima can be seen as a new set of random variables obtained by ordering the original variables by decreasing order of magnitude. The ordered maxima usually become strongly correlated due to the ordering. While the statistics of ordered IID random variables are well understood \cite{Arnold,Nagaraja}, there exists very few results on the statistics of ordered strongly correlated random variables, which often appear in practical contexts. However, several specific models of correlated random variables have been studied and have shown to exhibit very rich behaviors \cite{Feller,dean_majumdar,PLDCecile,pld_carpentier,satya_airy,schehr_airy,gyorgyi,satya_yor,comtet_precise,schehr_rsrg,SM12,MMS13,MMS14,BM17,Lacroix,BSG21a,BSG21b,PT20,PT21,Mori1,Mori2,Mori3} (for a recent review see \cite{reviewMPS}).
In particular, it was shown that one-dimensional discrete-time random walks constitute a very useful playground to investigate EVS of strongly correlated random variables \cite{SM12,MMS13,MMS14,BM17,Lacroix,PT20,PT21,Feller}.  In its simplest form, a one-dimensional discrete-time RW $x_n$ is described by a Markov rule
\begin{align} 
x_{n} = x_{n-1} + \eta_n\,,
\label{def_RW}
\end{align}  
starting from $x_0=0$ where the jumps $\eta_n$'s are IID random variables drawn from a symmetric and continuous PDF $f(\eta)$. The set of positions $\{x_1,\ldots,x_n\}$ is strongly correlated and constitutes a simple, yet non-trivial, example to investigate the order statistics of correlated random variables. The ordered variables are obtained upon
arranging the random variables $x_n$'s in decreasing order of magnitude and define the $k^\text{th}$ maximum $M_{k,n}$ of the set of positions $\{x_1,\ldots,x_n\}$ with $k=1, 2, \ldots, n$ such that (see figure \ref{fig:model})
\begin{align} 
M_{1,n} > M_{2,n} > \ldots> M_{n,n} \;.
\label{def_order}
\end{align}
Note that, at variance with previous studies (see e. g. \cite{SM12, MMS13, MMS14, BM17, PT20, PT21}), we have not included the initial position $x_0$ in the ordered set of positions. As we will see, this choice makes our
analytical approach somewhat easier. Besides, in the limit of large $n$ (which is our main focus here), we expect (and check it explicitly in some cases) that including $x_0$ or not in the list of ordered positions does not affect the statistics of $M_{k,n}$. Therefore $M_{1,n}$ and $M_{n,n}$ are respectively the global maximum and minimum of the walk (excluding the initial position $x_0=0$). While the marginal distribution of the global maximum and minimum are well-known and have be thoroughly studied \cite{Erdos46,Darling56}, for instance by using the Pollaczek-Spitzer formula, the statistics of the $k$-th maximum are much less known.  They can however be characterised through the Pollaczek-Wendel identity \cite{Pollackzek,Wendel}, which interestingly relates, in distribution, the $k^\text{th}$ maximum to the global maximum and minimum of two independent copies of the random walk. This surprising identity has been revisited in several other works and has been employed to study $\alpha$-quantiles, i.e., the statistics of $M_{k,n}$ with $k = \alpha n$ in the limit $n \to \infty$ with $0<\alpha<1$ \cite{Chaumont,Port,Dassios,Embrechts,Dassiosa}.  

A natural set of variables to describe how close the maxima are to each others are the gaps $\Delta_{k,n}$ between them, i.e., the difference between two consecutive maxima
\begin{align}
  \Delta_{k,n} = M_{k,n} - M_{k+1,n}\,,\quad k=1,\ldots,n-1\,.\label{eq:defD}
\end{align}
By definition, $\Delta_{k,n}$ are positive random variables. Contrary to the $k^\text{th}$ maximum, there does not exist, to our knowledge, any general formula for the distribution $P_{k,n}(\Delta)$ of the gap $\Delta_{k,n}$. Note that it does not suffice to know the marginal distributions of $M_{k,n}$ and $M_{k+1,n}$ to devise the distribution of $\Delta_{k,n}$ in (\ref{eq:defD}) as $M_{k,n}$ and $M_{k+1,n}$ are correlated. Recently, it has nevertheless been possible to obtain some analytical results for the statistics of $\Delta_{k,n}$. By using the linearity of the expectation value, one can connect the expected gap to the expected $k^\text{th}$ maximum through $ \mathbb{E}[\Delta_{k,n}]=\mathbb{E}[M_{k,n}]-\mathbb{E}[M_{k+1,n}]$. Using this identity, it was shown that, for symmetric and continuous jump distributions with a finite variance $\sigma^2=\int_{-\infty}^\infty \eta^2 f(\eta)$, the expected gap has a well-defined stationary limit $\mathbb{E}[\Delta_{k}]$ given by \cite{SM12}
\begin{eqnarray}\label{exact_mean_gap}
&& \mathbb{E}[\Delta_{k}]=\lim_{n\to \infty}\mathbb{E}[\Delta_{k,n}] = 
\frac{\sigma}{\sqrt{2 \pi}} \frac{\Gamma(k+\frac{1}{2})}{\Gamma(k+1)}  - \frac{1}{\pi k} \int_0^\infty \frac{dq}{q^2} 
\left[ [\hat f (q) ]^k - \frac{1}{(1 + \frac{\sigma^2}{2}q^2)^k} \right] 
\label{exact_gap} \;,
\end{eqnarray} 
where $\hat f(q)=\int_{-\infty}^\infty d\eta f(\eta) e^{i\eta q}$ is the Fourier transform of the jump distribution.
While the expression (\ref{exact_mean_gap}) depends explicitly on the jump distribution, it was shown that, for large $k$, the expected stationary gap decays~as
\begin{align}
  \mathbb{E}[\Delta_{k}] \sim \frac{\sigma}{\sqrt{2\pi k}}\,,\quad k\to \infty\,,\label{eq:decmu2}
\end{align}
which depends only on $\sigma$ but not on other details of the jump distribution, as long as $\sigma$ is finite. This universal decay is quite remarkable and raises the question whether this universal behavior of the first moment extends to the full PDF $P_{k,n}(\Delta)$. This question turns out to be quite challenging given the absence of analytical tools to go beyond the first moment. In Ref. \cite{SM12}, the full gap distribution $P_{k,n}(\Delta)$ was computed in the special case of the double sided exponential distribution $f(\eta) = 1/(2b) \, e^{-|\eta|/b}$, using a backward Fokker-Planck approach. It was indeed shown that in this case $P_{k,n}(\Delta)$ converges towards a limiting distribution as $n \to \infty$, i.e.,
\bea \label{limiting_pkn}
\lim_{n \to \infty} P_{k,n}(\Delta) = P_k(\Delta) \;,
\eea
where the generating function of $P_k(\Delta)$ (with respect to $k$) was computed explicitly. Furthermore, in the scaling limit $k \to \infty$, $\Delta \to 0$ keeping $\sqrt{k}\, \Delta$ fixed, it was shown that the stationary PDF $P_k(\Delta)$ takes the scaling form \cite{SM12}
\begin{align}
  P_k(\Delta) \sim \frac{\sqrt{k}}{\sigma}\mathcal{P}_2\left(\frac{\sqrt{k}\Delta}{\sigma}\right)\,,\quad k\to \infty\,,\label{eq:Pscal2}
\end{align}
where $\mathcal{P}_2(x)$ is given by  
\begin{equation}\label{exact_F} 
\mathcal{P}_2(x) = 4\left[\sqrt{\frac{2}{\pi}}(1+2x^2) -
e^{2x^2}x(4x^2+3) {\rm erfc}(\sqrt{2}x)\right] 
\,, 
\end{equation} 
with ${\rm erfc}(z) = (2/\sqrt{\pi})\int_z^\infty e^{-t^2} \, dt$ being the complementary error function. In particular, for large $x$, ${\cal P}_2(x)$ has a power law tail
\begin{align}
\mathcal{P}_2(x)\sim \frac{3}{2\sqrt{2\pi}\,x^4}\,,\quad x\to \infty\,.\label{eq:P2as}
\end{align} 
Note also that the average value of the scaling function $\int_0^\infty dx\, x\, \mathcal{P}_2(x) = 1/\sqrt{2\pi}$, together with the scaling form (\ref{eq:Pscal2}) allows to recover the asymptotic behavior of the expected gap in (\ref{eq:decmu2}).

Remarkably, it was conjectured in \cite{SM12}, based on numerical simulations, that the limiting behaviors in Eqs.~(\ref{limiting_pkn}) and (\ref{eq:Pscal2}) actually hold for any continuous and symmetric jump distribution with a finite variance $\sigma^2$, with the {\it same} universal scaling function $\mathcal{P}_2(x)$ as
given in (\ref{exact_F}). This conjecture was later on corroborated by exact analytical results for a rather wide class of jump distributions, namely Erlang distributions of the form $f(\eta)\propto |\eta|^p e^{-|\eta|}$, with $p$ being an integer \cite{BM17}.   

Recently, this question has attracted some attention in the probability literature where the existence of the stationary distribution $P_k(\Delta)$ as in (\ref{limiting_pkn}) holds for any continuous jump distribution \cite{PT20}. In addition, for the case of the double sided exponential distribution, the result in (\ref{eq:Pscal2}) and (\ref{exact_F}) was proved rigorously \cite{PT21} in the framework of fluctuation theory for random walks \cite{Feller}. In particular, the authors of Ref. \cite{PT21} obtained a probabilistic interpretation of this distribution (\ref{eq:P2as}) by showing that  \cite{PT21}
\bea \label{proba_interp}
\frac{\sqrt{k} \Delta}{\sigma} \overset{d}{\longrightarrow} \frac{\varepsilon}{\chi_3} \quad, \quad k \to \infty  \;,
\eea
where $\varepsilon$ is an exponential random variable with mean $1$ and $\chi_3$ is an independent chi-distribution of parameter~$3$ (i.e., $\chi_3^2$ is the sum of three independent Gaussians of zero mean and unit variance). Up to now, the question of the universality of these results 
in (\ref{eq:Pscal2}) and (\ref{exact_F}) for symmetric and continuous jump distributions with finite variance $\sigma^2$, beyond the cases of Erlang distributions, remains open. 

Much less is known for the case of a jump distribution which has heavy tails, i.e., for L\'evy fights such that $f(\eta)\propto |\eta|^{-1-\mu}$ for $|\eta|\to \infty$ with $0<\mu<2$. In this case, as far as we know, only the distribution of the first gap has been computed \cite{MMS13, MMS14}. In this paper, we extend the results   
in (\ref{eq:Pscal2}) and (\ref{exact_F}) to L\'evy flights with index $1 \leq \mu < 2$. In this case, we compute the scaling function ${\cal P}_\mu(x)$, that generalises the function ${\cal P}_2(x)$ in Eq. (\ref{exact_F}), and show that it is universal, i.e., it depends only on $\mu$. In addition, we also show that the result in (\ref{eq:Pscal2}) and (\ref{exact_F}), 
previously obtained for the double sided exponential and Erlang jump distributions, is indeed universal, i.e., hold for any jump distribution with a finite $\sigma$. Our method extends an original idea developed by Spitzer  
in a paper \cite{Spi56} which seems to have attracted little notice. In doing so, we unveil a rich analytical structure with an effective random walk that governs the gap statistics of the underlying random walk.

\subsection{Summary of the main results}
It is useful to summarise our main results. We assume a general expansion of the Fourier transform of the jump distribution $\hat f(q)$ of the form 
\begin{align}
  \hat f(q) = \int_{-\infty}^\infty d\eta\, e^{i\eta q} f(\eta)  \sim 1 - |q|^\mu + O(|q|^{2\mu})\,,\quad q\rightarrow 0\,,\label{Fourier}
\end{align}
where $1\leq \mu\leq 2$ is the L\'evy index and where we have set the typical jump to one by rescaling the distribution. While $\mu=2$ corresponds to finite variance distributions, $\mu<2$ corresponds to infinite variance distributions, with fat tails that decay like $f(\eta)\sim\eta^{-1-\mu}$ for $\eta\rightarrow\infty$. We leave aside jump distributions with $\mu< 1$ as they yield to transient random walks where the gap statistics behave quite differently, as we briefly discuss below. Note also that we assume sufficiently ``regular'' jump distributions where the next-to-leading order term is $O(|k|^{\nu})$ with $\nu=2\mu$ and leave aside the more singular case where $\mu<\nu<2\mu$ which might require a bit more care (although we expect that this should not modify the leading large $n$ behaviour). 

Similarly to the case of finite variance distributions discussed in the introduction, we find that the expected gap, which is only defined for $\mu>1$, has a stationary limit $\mathbb{E}[\Delta_{k}]$ given by
\begin{align}
   \mathbb{E}[\Delta_{k}]=\lim_{n\to \infty}\mathbb{E}[\Delta_{k,n}] =\frac{1}{k} \int_0^\infty \frac{dq}{\pi q^2}\left[1-\hat f(q)^k\right]\,,\label{eq:statg}
\end{align}
where $\hat f (q)$ is the Fourier transform of the jump distribution defined in (\ref{Fourier}). We have checked that the expression (\ref{eq:statg}) coincides with the result in equation (1.14) obtained in \cite{PT20} (see the end of section \ref{sec:seg}).
The expression (\ref{eq:statg}) is strikingly simple and one can recover the expression for $\mu=2$ discussed in the introduction in (\ref{exact_gap}) by using an identity given below in (\ref{eq:idcheck}). For a few notable distributions, the integral in (\ref{eq:statg}) can be computed explicitly, e.g., 
\begin{align}
\begin{array}{lll}
  \mathbb{E}[\Delta_{k}] &= \dfrac{\Gamma \left(k+\frac{1}{2}\right)}{\sqrt{\pi } k
   \Gamma (k)}\,, & \text{for} \quad \hat f(q)=\dfrac{1}{1+q^2}\quad \text{(double sided exponential) \;,}\\[1em]
    \mathbb{E}[\Delta_{k}] &= \dfrac{1}{\sqrt{\pi k}}\,, & \text{for} \quad \hat f(q)=e^{-q^2}\quad \text{(Gaussian)\;,}\\[1em]
    \mathbb{E}[\Delta_{k}] &=\Gamma\left(1-\dfrac{1}{\mu}\right)\dfrac{k^{\frac{1}{\mu}-1}}{\pi}\,, & \text{for} \quad \hat f(q)=e^{-|q|^\mu}\quad \text{(stable with $\mu>1$)\;.}
    \end{array}
\end{align}
Although the expression in (\ref{eq:statg}) depends on the full details of the jump distribution $f(\eta)$, it becomes universal in the limit $k\to\infty$ and behaves as
\begin{align}
   \mathbb{E}[\Delta_{k}]   & \sim \Gamma\left(1-\frac{1}{\mu}\right)\frac{k^{\frac{1}{\mu}-1}}{\pi}\,,\quad k\to \infty\,,\label{eq:avgGass}
\end{align}
which depends only on the L\'evy index $\mu>1$ of the jump distribution.

We find that the full distribution $P_{k,n}(\Delta)$ reaches a stationary limit $P_k(\Delta)$ when $n\to \infty$, as in Eq.~(\ref{limiting_pkn}) -- in agreement with the rigorous result from \cite{PT20}. As it was conjectured for the case of finite variance distributions, we find that the limiting probability distribution $ P_k(\Delta)$ becomes universal, not only for $\mu=2$ but also for $1\leq \mu\leq 2$. In the scaling limit $k\to\infty$ with $\Delta=O(k^{1/\mu-1})$, we find that, for $\mu>1$, the distribution $ P_k(\Delta)$ behaves as
\begin{align}
  P_k(\Delta) &\sim  \frac{1}{k^{\frac{1}{\mu}-1}} \mathcal{P}_\mu\left(\frac{\Delta}{k^{\frac{1}{\mu}-1}}\right)\,,\quad \Delta=O(k^{\frac{1}{\mu}-1})\,,\quad k\to \infty\,,\label{eq:Pkas}
\end{align}
where $\mathcal{P}_\mu(x)$ is a universal scaling function given by
\begin{align}
\mathcal{P}_\mu(x) = \frac{\mu B_\mu}{(\mu-1)^2}\left[\mu\, E_{\frac{\mu-1}{\mu},-\frac{1}{\mu}}(-B_\mu x)+(2\mu-1)E_{\frac{\mu-1}{\mu},\frac{\mu-1}{\mu}}(- B_\mu x)\right]\,,\label{eq:Px}
\end{align}
where 
\begin{align}
B_\mu=[\sin(\frac{\pi}{\mu})]^{-1} \quad \quad {\rm and} \quad \quad E_{\alpha,\beta}(z)=\sum_{k=0}^\infty \frac{z^k}{\Gamma(\alpha k+\beta)}\,,
\label{eq:MittagLeff}
\end{align} 
is the two-parameter Mittag-Leffler function (sometimes called Wiman's function). The expression (\ref{eq:Px}) simplifies for $\mu=2$, upon using the identities of Mittag-Leffler functions in (\ref{eq:Mittag2}), and rescaling $x$ by $\sigma=\sqrt{2}$,  we recover the expression in (\ref{exact_F}). 
The asymptotic behaviors of the universal scaling function ${\cal P}_\mu(x)$, for $1 < \mu < 2$ are given by
\begin{align}
  \mathcal{P}_\mu(x) \sim \left\{\begin{array}{ll}
    \dfrac{2 }{\sin \left(\frac{\pi }{\mu }\right)\Gamma
   \left(2-\frac{1}{\mu }\right)}\,, & \; x\to 0\,,
   \\
   \\
    \dfrac{2 \sin ^2\left(\frac{\pi }{\mu }\right)}{
   \Gamma \left(\frac{2}{\mu }-1\right)} \dfrac{1}{x^3} \,,& \; x\to \infty\,.\\[1em]
  \end{array}\right.\label{eq:Pxa}
\end{align}
Note that the $1/x^3$ tail is universal for all $1 < \mu < 2$. However, exactly at $\mu = 2$, the tail behaves as $1/x^4$, as in Eq.~(\ref{eq:P2as}). Thus there is a discontinuous jump in the exponent from $3$ to $4$ as $\mu$ approaches $2$. This is consistent with the fact that, in the second line of Eq. (\ref{eq:Pxa}) the amplitude vanishes as $\mu \to 2$ and then the leading order decay comes from the subleading term scaling as $1/x^4$. 

In the opposite limit when $\mu \to 1$, the scaling form in Eq. (\ref{eq:Pkas}) is no longer valid and the typical scale of $\Delta$ changes from $k^{1/\mu-1}$ to $1/\ln k$ as $\mu \to 1$. In this case, the scaling form of the distribution $P_k(\Delta)$ reads
\begin{align}
  P_{k}(\Delta) \sim  \ln(k)\, \mathcal{P}_1\left(\ln(k)\Delta\right)\,,\label{eq:pkmu1}
\end{align}
where we find that the scaling function ${\cal P}_1(x)$ is given explicitly by
\begin{align}
  \mathcal{P}_1(x) = \frac{2\pi^2}{(\pi+x)^3}\,.\label{eq:mathcalp1}
\end{align}
One can check that the scaling function ${\cal P}_\mu(x)$ is normalised to unity, i.e., $\int_0^\infty {\cal P}_\mu(x) dx = 1$, which indicates that ${\cal P}_\mu(x)$ indeed describes the {\it typical} behavior of $P_k(\Delta)$, corresponding to $\Delta = O(k^{1/\mu-1})$ for large $k$. However, the first moment of ${\cal P}_\mu(x)$ reads
\begin{align}
  \int_0^\infty dx \,x \mathcal{P}_\mu(x) = \frac{\sin\left(\frac{\pi}{\mu}\right)}{\Gamma\left(\frac{1}{\mu}\right)}\,,\label{eq:avgPmu}
\end{align}
which, together with the scaling form in (\ref{eq:Pkas}) does not yield the correct value for the expected stationary gap given in (\ref{eq:avgGass}) for $1<\mu<2$. This apparent paradox can be resolved by noticing that, for $\mu <2$, there are atypically large gaps of scale $k^{1/\mu}$ that are not captured by the scaling form in (\ref{eq:Pkas}), which only describes the typical gaps, of order $k^{1/\mu-1}$. More precisely, we find that for $1 < \mu <2$, the distribution of the gap for large $k$ has two parts:  a typical one where $\Delta=O(k^{\frac{1}{\mu}-1})$ described as in (\ref{eq:Pkas}) and an additional atypical one where $\Delta=O(k^{\frac{1}{\mu}})$, e.g.,
\begin{align}
  P_k(\Delta) \sim \left\{\begin{array}{lll}
  \frac{1}{k^{\frac{1}{\mu}-1}} \mathcal{P}_\mu\left(\frac{\Delta}{k^{\frac{1}{\mu}-1}}\right)\,,  & \Delta = O(k^{\frac{1}{\mu}-1})\,,&\text{(typical gap/``fluid'')}\\
  \frac{1}{k^{1+\frac{1}{\mu}}} \mathcal{M}_\mu\left(\frac{\Delta}{k^{\frac{1}{\mu}}}\right)\,,   & \Delta = O(k^{\frac{1}{\mu}})\,,&\text{(atypical large gap/``condensate'')}
  \end{array}\right.\quad k\to \infty\,, \quad 1<\mu < 2 \;, \label{eq:Pksum}
\end{align}
where $\mathcal{P}_\mu(x)$ is given in (\ref{eq:Px}) and $\mathcal{M}_\mu(u)$ is a universal scaling function. Note that the contribution of $\mathcal{M}_\mu(u)$ to the normalisation vanishes in the limit $k\to\infty$, while its contribution to the average value is of the same order as the scaling function (\ref{eq:Pkas}). This is actually reminiscent of the ``condensation'' phenomena in the distribution of the mass in a class of mass transport models, such as in zero-range processes \cite{EH05,MEZ05,EMZ06,Satya_Houches} or even in the active run-and-tumble particles and related models \cite{Gradenigo,MGM21,MoriPRE,Smith}. Here, the gaps play the role of ``mass'' at a given site in the transport models or the ``runs'' in run-and-tumble models. In this condensed phase, there is a coexistence of a ``fluid'' regime where the masses (or the runs) are of the typical size and a ``condensation'' part which contains atypically large masses or runs. In our model, for $\mu <2$, we also see the coexistence of the ``fluid'' regime, consisting of typical gaps of order $\Delta = O(k^{1/\mu-1})$ and a ``condensate'' consisting of large gaps of order $\Delta = O(k^{1/\mu})$. 

The scaling function ${\cal M}_\mu(u)$ in Eq. (\ref{eq:Pksum}) that describes the condensate part is hard to compute analytically. Fortunately, we still managed to extract the asymptotic tail of $\mathcal{M}_\mu(u)$ which behaves as
\begin{align}
  \mathcal{M}_\mu(u)\sim \frac{  \mu }{\Gamma \left(1-\frac{\mu }{2}\right)^2}\frac{1}{u^{\mu+1}}\,,\quad u\to \infty\,.\label{eq:expMmu}
 \end{align}
  Interestingly, the amplitude in (\ref{eq:expMmu}) is the same as the one for the distribution of the first gap obtained in \cite{MMS13, MMS14}.

The rest of the paper is organised as follows. In section \ref{sec:exp}, we compute the expected stationary gap $\mathbb{E}[\Delta_{k}]$. We first obtain a closed-form expression for the double generating function of $\mathbb{E}[\Delta_{k,n}]$, valid for all $k$ and $n$, which we then analyse in the limit $n\to \infty$. This section will serve as a ``warm-up'' to illustrate our method based on an idea developed by Spitzer in \cite{Spi56}. In section \ref{sec:pro}, we provide the derivation of the universal scaling function for the PDF of the stationary gap. We first obtain a closed-form expression for the  double generating function of $P_{k,n}(\Delta)$, valid for all $k$ and $n$, which we then analyse in the limit $n\to \infty$. Finally, we show the existence of a condensed phase in the distribution. Some detailed calculations are presented in Appendix \ref{app:SN} to~\ref{app:scalC}.
\section{Expected $k^\text{th}$ gap}
\label{sec:exp}
\subsection{Exact results for the expected gap $ \mathbb{E}[\Delta_{k,n}]$}
\begin{figure}[t]
  \begin{center}
    \includegraphics[width=0.6\textwidth]{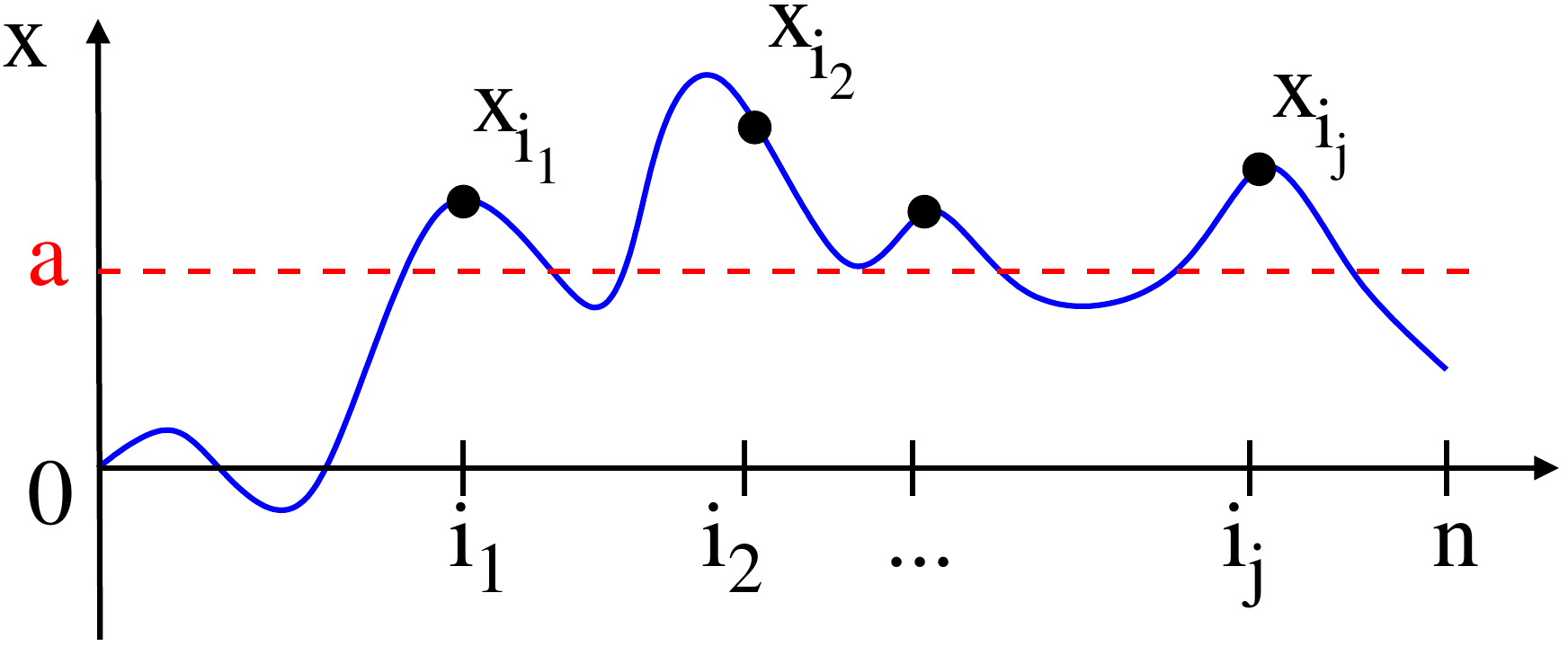}
    \caption{The probability $\text{Prob.}\left(\bigcap_{t=1}^j A_{i_t}(a)\right)$ in (\ref{eq:probpa}) corresponds to all the trajectories of $n$ steps that start at the origin and are above the level $a$ at the intermediate times $i_1,\ldots,i_j$. }
    \label{fig:figAvg}
  \end{center}
\end{figure}

In this section, we derive the expression of the stationary expected gap in (\ref{eq:statg}). We start by introducing $p_{k,n}(a,\Delta)$ as the joint probability 
\begin{align}
p_{k,n}(a,\Delta) = {\rm Prob.}\left(M_{k+1,n} < a-\frac{\Delta}{2}, M_{k,n}> a+\frac{\Delta}{2}\right) \:,  \quad \quad k = 1,\ldots, n-1\;,
 \label{def_S}
\end{align}
together with $p_{0,n}(a,\Delta) = {\rm Prob.}\left(M_{1,n}<a-\frac{\Delta}{2}\right)$ as well as $p_{n,n}(a,\Delta) = {\rm Prob.}\left(M_{n,n}>a+\frac{\Delta}{2}\right)$. 
 If we view the positions of the random walk after $n$ steps as a collection of $n$ points on the real line, excluding the initial position, the probability  
$p_{k,n}(a,\Delta)$ is the probability that there is no point in the interval $\left[a-\frac{\Delta}{2}, a +\frac{\Delta}{2}\right]$, $k$ points above $a +\frac{\Delta}{2}$, and $n-k$ points below $a-\frac{\Delta}{2}$. 
The probability $p_{k,n}(a,\Delta)$ is an interesting observable  as it is related to the PDF of the gap between two consecutive maxima $P_{k,n}(\Delta)=\text{Prob.}\left(M_{k,n}-M_{k+1,n}=\Delta\right)$ discussed in the introduction.
To connect $P_{k,n}(\Delta)$ and $p_{k,n}(a,\Delta)$, we relate $p_{k,n}(a,\Delta)$ to the cumulative joint distribution of the $k$-th and $(k+1)$-th maximum $S_{k,n}(x,y)$, which is denoted by
\begin{align} 
S_{k,n}(x,y) = {\rm Prob.}(M_{k+1,n} < x, M_{k,n}>y) \:, \quad {\rm with} \quad \quad k = 1, 2, \ldots, n-1\;.
\label{def_S2}
\end{align}
One has indeed
\begin{align} 
p_{k,n}\left(a,{\Delta}\right) = S_{k,n} \left(a- \frac{\Delta}{2}, a +\frac{\Delta}{2} \right)  \;.
\label{rel_gen}
\end{align}
The PDF $P_{k,n}(\Delta)$ of the $k$-th gap is obtained from the cumulative joint distribution $S_{k,n}(x,y)$ as
\begin{align} 
P_{k,n}(\Delta) = - \int_{-\infty}^\infty dx \int_{-\infty}^\infty dy \frac{\partial^2 S_{k,n}(x,y)}{\partial x \partial y} \theta(y-x) \delta(y-x-\Delta) \;.
\label{rel_SPDF}
\end{align}
Taking derivatives with respect to $x$ and $y$ in (\ref{rel_gen}), one finds
\begin{align} 
\frac{\partial^2 S_{k,n}(x,y)}{\partial x \partial y} = \frac{1}{4}{\partial_1^2}p_{k,n}\left(\frac{x+y}{2},\frac{y-x}{2} \right) - \frac{1}{4}\partial_2^2 p_{k,n}\left(\frac{x+y}{2},\frac{y-x}{2} \right) \;,
\label{sec_derivative}
\end{align}
where $\partial_1^2$ and $\partial_2^2$ refer to the double partial derivative with respect to the first and second argument respectively.
Upon inserting this expression (\ref{sec_derivative}) in  (\ref{rel_SPDF}) and shifting the integration variable by $\Delta/2$, we find that the first term $\propto {\partial_1^2}p_{k,n}\left(\frac{x+y}{2},\frac{x-y}{2} \right)$ does not contribute, as the boundary terms vanish, and we are left with 
\begin{align} 
P_{k,n}(\Delta) = \partial^2_{\Delta} \int_{-\infty}^\infty  p_{k,n}\left(a ,\Delta\right) \;da \;,\quad k=1,\ldots,n-1\,.
\label{rel_SPDF2}
\end{align}
The expression (\ref{rel_SPDF2}) thus relates the PDF of the gap $P_{k,n}(\Delta)$ and the probability (\ref{def_S}). It will serve as a starting point to study the PDF of the gap in the next section. As we are only interested in the first moment in this section, we obtain the expected gap from (\ref{rel_SPDF2}) as 
\begin{align}
  \mathbb{E}[\Delta_{k,n}] &= \int_0^\infty d\Delta\, \Delta P_{k,n}(\Delta)=\int_{-\infty}^\infty  p_{k,n}\left(a ,0\right) \;da\,,\label{eq:avgf}
\end{align}
where we integrated twice by parts and used that the boundary terms vanish. This expression (\ref{eq:avgf}) is our starting point to compute the expected gap.
The quantity $p_{k,n}\left(a ,0\right)\equiv p_{k,n}(a)$ is simply the probability that exactly $k$ steps out of $n$ of the random walk are above the level $a$. Let us compute it by building upon an idea developed by Spitzer in \cite{Spi56}. To do so, we define the events
\begin{align}
  A_m(a) = \{x_m > a\}\,,\label{eq:Cm}
\end{align}
which corresponds to the event where the $m^\text{th}$ step of the random walk is above the level $a$.
We have then 
\begin{align}
   p_{k,n}(a) = \text{Prob.}\left[\left(\sum_{m=1}^n I[A_m(a)]\right)=k\right]\,,\label{eq:pkn}
\end{align}
where $I(A)$ is the indicator function which takes the value $1$ if the event $A$ happened and $0$ otherwise. The right-hand side (rhs) in (\ref{eq:pkn}) is cumbersome to compute as one needs to sum over all random walk trajectories that spend \emph{exactly} $k$ steps above the level $a$. Spitzer's idea in \cite{Spi56} consists in rewriting this probability in terms of the complementary events which are easier to compute. One can do so by using a formula known as the Schuette-Nesbitt formula, which can be seen as the generalisation of the ``inclusion-exclusion'' formula in probability theory (see Appendix \ref{app:SN} for a sketch of the proof). In the present case, the formula for the double generating function of $p_{k,n}(a)$ with respect to both $n$ and $k$ reads
\begin{align}
\bar p_{z,s}(a) =  \sum_{n=0}^\infty \sum_{k=0}^{n} s^n   z^k p_{k,n}(a) = \sum_{n=0}^\infty \sum_{j=0}^n s^n (z-1)^j \sum_{1\leq i_1<\ldots<i_j\leq n}\text{Prob.}\left(\bigcap_{t=1}^j A_{i_t}(a)\right)  \,.\label{eq:SN}
\end{align}
Note that the $(z-1)$ factor in the rhs in (\ref{eq:SN}) is quite unusual but arises from the Schuette-Nesbitt formula which consists in an alternating sum, as in the ``inclusion-exclusion'' formula. Furthermore, note that for $j=0$, the multiple sum is just $1$ by convention. The key observation is that the probability in the rhs can now be easily computed as it is the probability that the random walk is above the level $a$ at the discrete times $i_1,\ldots,i_j$. Note that the random walk is therefore allowed to propagate freely in-between those times (see figure \ref{fig:figAvg}). 
Hence, this probability reads 
\begin{align}
  \text{Prob.}\left(\bigcap_{t=1}^j A_{i_t}(a)\right) = \int_{a}^\infty dx_{i_1}\ldots  \int_a^\infty dx_{i_j} G_{i_1}(x_{i_1})G_{i_2-i_1}(x_{i_2}-x_{i_1})\ldots G_{i_j-i_{j-1}}(x_{i_j}-x_{i_{j-1}})\,,\label{eq:probpa}
\end{align}
where $G_j(x)=\int_{-\infty}^\infty \frac{dq}{2\pi} e^{-iqx} \hat [f(q)]^j$ is the free propagator of the random walk starting from the origin and $\hat f(q)$ is the Fourier transform of the jump distribution defined in (\ref{Fourier}). Shifting all the integration variables by $a$ gives
\begin{align}
   \text{Prob.}\left(\bigcap_{t=1}^j A_{i_t}(a)\right) = \int_{0}^\infty dx_{i_1}\ldots  \int_0^\infty dx_{i_j} G_{i_1}(x_{i_1}+a)G_{i_2-i_1}(x_{i_2}-x_{i_1})\ldots G_{i_j-i_{j-1}}(x_{i_j}-x_{i_{j-1}})\,.\label{eq:probpa2}
\end{align}
Inserting the expression (\ref{eq:probpa2}) in (\ref{eq:SN}) and recognizing the discrete convolution structure over the $i$'s, we find
\begin{align}
  \bar p_{z,s}(a) = \frac{1}{1-s}\sum_{j=0}^\infty (z-1)^j \int_{0}^\infty dx_{i_1}\ldots  \int_0^\infty dx_{i_j} \bar G_s(x_{i_1}+a)\bar G_s(x_{i_2}-x_{i_1})\ldots \bar G_s(x_{i_j}-x_{i_{j-1}})\label{eq:barpzsa}\,,
\end{align}
where $\bar G_s(x)$ is the generating function of the propagator
\begin{align}
  \bar G_s(x)=\sum_{n=1}^\infty s^n G_n(x)\,.\label{eq:barP}
\end{align}
Note that the $1/(1-s)$ prefactor in (\ref{eq:barpzsa}) comes from the last sum between $i_j$ and $n$ in (\ref{eq:SN}), and that the sum in (\ref{eq:barP}) starts from $n=1$ as the $i$'s in (\ref{eq:SN}) are not allowed to take the same value. 

\begin{figure}[t]
 \centering
     \includegraphics[width=0.4\textwidth]{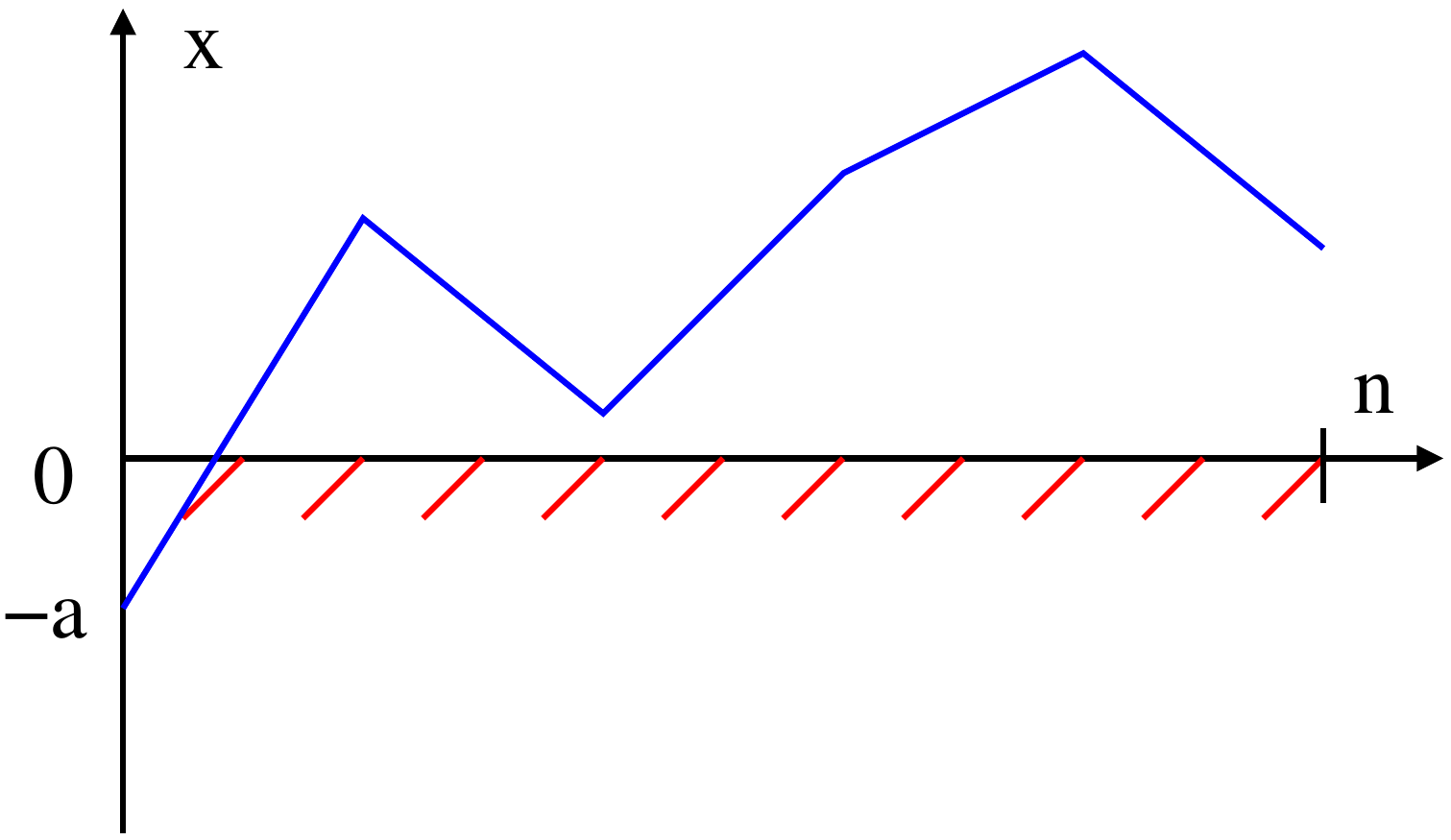} \includegraphics[width=0.4\textwidth]{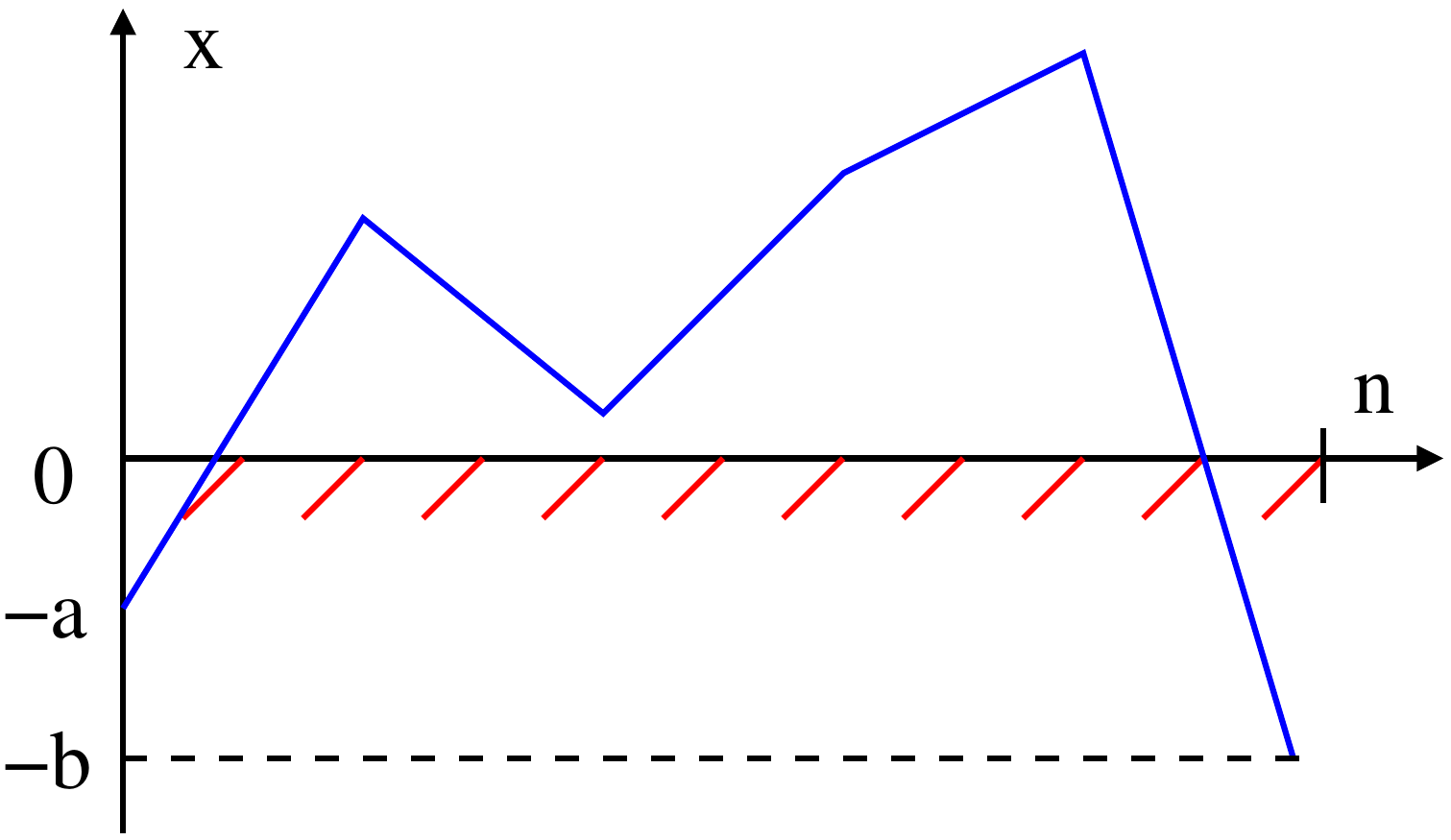}
     \caption{\textbf{Left panel}: The survival probability $S_s(j|-a)$, is the probability that a random walk starting from $-a<0$ with a jump distribution $ \mathrm{f}_s(\eta)$ did not cross the origin except at the first step. \textbf{Right panel}: The excursion probability $E_s(j,-b,|-a)$ is the probability that a random walk starting from $-a<0$ with a jump distribution $ \mathrm{f}_s(\eta)$ reaches $-b<0$ after $j$ steps while remaining above the origin during the intermediate steps.}
     \label{fig:surv}
 \end{figure}

We now show that the expression in (\ref{eq:barpzsa}) can be interpreted in terms of an effective random walk. To this purpose, we define an effective jump distribution, parametrised by $s$, which reads
\begin{align}
  \mathrm{f}_s(\eta)=\frac{1-s}{s}\bar G_s(\eta)\,,\label{eq:Fse}
\end{align}
where the factor $(1-s)/s$ in the definition of $\mathrm{f}_s(\eta)$ has been added for normalization. One can check, by using that $\int_{-\infty}^\infty dx  \; \bar G_s(x) = \sum_{n=1}^\infty s^n \int_{-\infty}^\infty dx  \, G_n(x)= s/(1-s)$, that the distribution $\mathrm{f}_s(\eta)$ is properly normalized. The expression (\ref{eq:barpzsa}) now reads
\begin{align}
\bar p_{z,s}(a) = \frac{1}{1-s} \sum_{j=0}^\infty u^j \int_{0}^\infty dx_{i_1}\ldots  \int_0^\infty dx_{i_j} \mathrm{f}_s(x_{i_1}+a)\mathrm{f}_s(x_{i_2}-x_{i_1})\ldots \mathrm{f}_s(x_{i_j}-x_{i_{j-1}})\,,\label{eq:probpa3}
\end{align}
where to ease notation we defined
\begin{align}
u=  \frac{s(z-1)}{1-s}\,.\label{eq:defU}
\end{align}
 We recognize that the multiple integral in (\ref{eq:probpa3}) is the survival probability $S_s(j|-a)$ after $j$ steps for a random walk with a jump distribution $ \mathrm{f}_s(\eta)$ starting from $-a$:
 \begin{align}
 S_s(j|-a) &= \int_{0}^\infty dy_{1}\ldots dy_{j} \mathrm{f}_{s}(y_{1}+a)\mathrm{f}_{s}(y_{2}-y_{1})\ldots \mathrm{f}_{s}(y_{j}-y_{j-1})\,.\label{eq:Ssse}
 \end{align}
The survival probability $S_s(j|-a)$ is the probability that the random walk, starting from $-a<0$, did not cross the origin during $j$ steps except at the first step (see left panel in figure \ref{fig:surv}). Note that, contrary to the usual definition of the survival probability where the initial position is positive, the current computation requires to extend it to a negative initial position.
 Denoting the generating function of the survival probability
 \begin{align}
 \bar S_s(z|-a)=\sum_{j=0}^\infty z^j S(j|-a)\,,\label{eq:genSszma}
 \end{align}
 the expression (\ref{eq:probpa3}) reads
\begin{align}
  \bar p_{z,s}(a) = \frac{1}{1-s}  \bar S_s\left(u|-a\right)\,.\label{eq:probpa4}
\end{align}
One can check that upon setting $a=0$ and using the Sparre-Andersen result $\bar S_s(z|0)=1/\sqrt{1-z}$, we recover the occupation time distribution
\begin{align}
  \bar p_{z,s}(a=0) = \frac{1}{\sqrt{1-sz}}\,,\label{eq:a0barp}
\end{align}
which is the generating function of the discrete-time version of the arc-sine law.

We now compute the generating function of the expected gap in (\ref{eq:avgf}). We first use the symmetry $p_{k,n}(a)=p_{n-k,n}(-a)$, which is a consequence of the $x\to-x$ symmetry of the jump distribution, to rewrite (\ref{eq:avgf}) as 
\begin{align}
   \mathbb{E}[\Delta_{k,n}] =\int_{-\infty}^\infty  p_{k,n}\left(a ,0\right) \;da = \int_{-\infty}^0 \left[p_{k,n}(a) + p_{n-k,n}(a)\right]\,da\,.\label{eq:sym}
\end{align}
Note that if we had chosen to include the initial point $x_0=0$ in the definition of the $k^\text{th}$ maximum, the symmetry $p_{k,n}(a)=p_{n-k,n}(-a)$ would slightly change depending on the sign of $a$, which is quite cumbersome and is the reason why we did not choose to include $x_0$ here. Taking the double generating function of (\ref{eq:sym}) gives
\begin{align}
  \mathbb{E}[\tilde \Delta_{z,s}] =  \sum_{n=0}^\infty s^n   \sum_{k=1}^{n-1} z^k\mathbb{E}[\Delta_{k,n}] = \int_{-\infty}^0 \left[ \tilde p_{z,s}(a) +  \tilde p_{\frac{1}{z},sz}(a)\right]\,da\,,\label{eq:sym2}
\end{align}
where 
\begin{align}
\tilde p_{z,s}(a)= \sum_{n=0}^\infty s^n   \sum_{k=1}^{n-1} z^k p_{k,n}(a)\,.\label{eq:tildepzsa}
\end{align} Note that the generating function $\tilde p_{z,s}(a)$ starts from $k=1$ and ends at $k=n-1$. We use an upper tilde to distinguish it from $\bar p_{z,s}(a)$ (with an upper bar) defined in (\ref{eq:SN}) which starts from $k=0$ and goes to $k=n$. Hence, to relate $\tilde p_{z,s}(a)$ with $\bar p_{z,s}(a)$, one must remove the terms $k=0$ and $k=n$, which reads 
\begin{align}
   \tilde p_{z,s}(a) &=  \bar p_{z,s}(a)-\sum_{n=0}^\infty s^n p_{0,n}(a)-\sum_{n=0}^\infty (zs)^n p_{n,n}(a)+1\nonumber\\
   &= \bar p_{z,s}(a)  - \bar S_0(s|a) - \bar S_0(sz|-a)+1\,,\label{eq:pknag2}
\end{align}
where we used that $p_{0,n}(a)=S_0(n|a)$ and $p_{n,n}(a)=S_0(n|-a)$. The quantity $S_0(n|a)$ is the survival probability of a random walk with the original jump distribution $f(\eta)$ starting from $a$. We denote its generating function by $\bar S_0(s|a)=\sum_{n=0}^\infty s^n S_0(n|a)$. Note that the $+1$ term in (\ref{eq:pknag2}) arises from the double counting of $p_{0,0}(a)=1$ in the two sums in the first line of (\ref{eq:pknag2}). Computing the first term in the rhs in (\ref{eq:sym2}) by inserting (\ref{eq:pknag2}) gives
\begin{align}
  \int_{-\infty}^0 da\, \tilde p_{z,s}(a)  &= \int_0^\infty da \left[\frac{1}{1-s} \bar S_s\left(u|a\right)- \bar S_0(s|-a) - \bar S_0(sz|a)+1\right]\,.\label{eq:minf}
\end{align}
Of course, it is difficult to compute the survival probability for a random walk with a generic jump distribution. However, there exists a formula, known as the Pollaczek-Spitzer formula \cite{Pollackzek}, which gives the Laplace transform of the generating function of the survival probability and reads
\begin{align}
   \int_0^\infty da\, e^{-\lambda a} \bar S_s(z|a) &= \frac{1}{\lambda\sqrt{1-z}}\exp\left(-\lambda\int_0^\infty \frac{dq}{\pi} \frac{\ln[1-z\,\mathrm{\hat f}_s(q)]}{\lambda^2+q^2}\right)\,,
 \label{eq:HI}
\end{align}
where $\mathrm{\hat f}_s(q)$ is the Fourier transform of the effective jump distribution in (\ref{eq:Fse}), which is related to the original distribution through
\begin{align}
  \mathrm{\hat f}_s(q) = \frac{(1-s)\hat f(q)}{1-s\hat f(q)}\,.\label{eq:effFb}
\end{align}
Using the Pollaczek-Spitzer formula (\ref{eq:HI}), one can compute the integral in (\ref{eq:minf}) and find that it reads (see Appendix \ref{app:pzsai})
\begin{align}
   \int_{-\infty}^0 da\, \tilde p_{z,s}(a)   &= \frac{1}{1-sz}\int_0^\infty \frac{dq}{\pi q^2}\ln\left(\frac{1-s \hat f(q)}{1-s}\right) \nonumber\\
   &- \frac{s}{\sqrt{1-s}}\int_{-\infty}^0 da \int_0^\infty dy f(y-a)\int_{\gamma_B} \frac{d\lambda e^{\lambda a}}{2\pi i \lambda}\exp\left(-\lambda \int_0^\infty \frac{dq}{\pi} \frac{\ln(1-s \hat f(q))}{\lambda^2+q^2}\right)\,,\label{eq:int2av}
\end{align}
where $\gamma_B$ is the usual Bromwich contour in the complex $\lambda$-plane.
Inserting this result in (\ref{eq:sym2}), we get that the double generating function of the expected gap is given by
\begin{align}
   \mathbb{E}[\tilde \Delta_{z,s}]  &=  \frac{1}{1-sz}\int_0^\infty \frac{dq}{\pi q^2}\ln\left(\frac{1-s \hat f(q)}{1-s}\right) +  \frac{1}{1-s}\int_0^\infty \frac{dq}{\pi q^2}\ln\left(\frac{1-sz \hat f(q)}{1-sz}\right)\nonumber\\
   &- \frac{s}{\sqrt{1-s}}\int_{-\infty}^0 da \int_0^\infty dy f(y-a)\int_{\gamma_B} \frac{d\lambda e^{\lambda a}}{2\pi i \lambda}\exp\left(-\lambda \int_0^\infty \frac{dq}{\pi} \frac{\ln(1-s \hat f(q))}{\lambda^2+q^2}\right)\nonumber\\
   &- \frac{sz}{\sqrt{1-sz}}\int_{-\infty}^0 da \int_0^\infty dy f(y-a)\int_{\gamma_B} \frac{d\lambda e^{\lambda a}}{2\pi i \lambda}\exp\left(-\lambda \int_0^\infty \frac{dq}{\pi} \frac{\ln(1-sz \hat f(q))}{\lambda^2+q^2}\right) \,.\label{eq:genavg}
\end{align}

\subsection{Stationary expected gap  $\mathbb{E}[\Delta_{k}]$} \label{sec:seg}
The expression (\ref{eq:genavg}) is exact and in principle, one can get $E[\Delta_{k,n}]$ by expanding formally on powers of $z$ and $s$, for any jump distribution. In addition, one can extract its asymptotic limits. In particular, in the limit $n\to\infty$, which corresponds to $s\to 1$, we get that the leading order term in (\ref{eq:genavg}) is the second term in the first line which reads
\begin{align}
   \mathbb{E}[\tilde \Delta_{z,s}] \sim   \frac{1}{1-s}\int_0^\infty \frac{dq}{\pi q^2}\ln\left(\frac{1-z \hat f(q)}{1-z}\right)\,,\quad s\to 1\,.\label{eq:genstat}
\end{align}
One can indeed show that the second line in (\ref{eq:genavg}) is subleading (see (\ref{eq:I2sfl2}) in Appendix \ref{app:pzsai}).
Using the series expansion $-\ln(1-z)=\sum_{k=1}^\infty z^k/k$, we recover the expected stationary gap given in Eq. (\ref{eq:statg}), namely
\begin{align}
   \mathbb{E}[\Delta_{k}]=\lim_{n\to \infty}\mathbb{E}[\Delta_{k,n}] =\frac{1}{k} \int_0^\infty \frac{dq}{\pi q^2}\left[1-\hat f(q)^k\right]\,.\label{eq:statg_2}
\end{align}
One can recover the expression for $\mu=2$ in (\ref{exact_mean_gap}) obtained in \cite{SM12} by using the following identity
\begin{align}
  \int_0^\infty \frac{dq}{q^2}\left(1-\frac{1}{(1+q^\mu)^k}\right) = \frac{\Gamma\left(1-\frac{1}{\mu}\right)\Gamma\left(\frac{1}{\mu}+k\right)}{\Gamma(k)}\,,\label{eq:idcheck}
\end{align}
which can be checked by using Mathematica.
In Appendix \ref{app:alt}, we show that this result can also be obtained from a more standard approach using the Pollaczeck-Wendel identity \cite{Pollackzek,Wendel}, known in the context of fluctuation theory. Although the expression (\ref{eq:statg}) depends on the full details of the jump distribution, it becomes universal in the limit $k\to\infty$. Indeed in this limit one has
\begin{align}
   \mathbb{E}[\Delta_{k}]   &=\frac{1}{k} \int_0^\infty \frac{dq}{\pi q^2}\left[1-\exp\left(k\log(\hat f(q))\right)\right]\nonumber\\
   &\sim \frac{1}{k} \int_0^\infty \frac{dq}{\pi q^2}\left[1-\exp\left(-k q^\mu\right)\right]\,,
\end{align}
where we have taken the small $q$ limit of $\hat f(q)$ in (\ref{Fourier}) and which upon performing the integral, yields the expression (\ref{eq:avgGass}) given in the introduction. 

Note that the expression for the expected stationary gap in (\ref{eq:statg}) coincides with the rigorous result obtained by a completely different method in \cite{PT20} (see Eq. (1.14) there). Their result reads (in their notation)
\begin{align}
 \lim_{n\to\infty}  \mathbb{E}[D_{k,n}] =   \mathbb{E}[D_{k}]=\frac{\mathbb{E}[S_k^+] }{k} +\mu^{-}\,,\label{eq:PT}
 \end{align}
 where $D_{k,n}$ is the $k^\text{th}$ gap of a random walk of $n$ steps ($\Delta_{k,n}$ in our notation), $S_k$ is the position of the random walk after the $k^\text{th}$ step ($x_k$ in our notation), $\mu$ is the mean step (which is equal to zero in our case since we consider  symmetric jump distributions), and $x^+=\max(x,0)$ and $x^-=-\min(x,0)$. For a continuous and symmetric jump distribution $f(\eta)$, the expected value $\mathbb{E}[S_k^+]$ can be obtained by averaging the propagator $G_k(x)$ over positive $x$, namely
 \begin{align}
   \mathbb{E}[S_k^+] = \int_0^\infty dx\, x\, G_k(x)\,.\label{eq:avgGkx}
 \end{align}
Using the expression of the Fourier transform of the propagator gives
\begin{align}
  \mathbb{E}[S_k^+] = \int_0^\infty dx\, x\, \int_{-\infty}^\infty \frac{dq}{2\pi}\, e^{-i q x} \hat f(q)^k\,.\label{eq:avgGkx2}
\end{align}
Switching the order of integration and performing the integral over $x$ gives
\begin{align}
   \mathbb{E}[S_k^+] = \lim_{\epsilon\to 0^+}  \int_{-\infty}^\infty \frac{dq}{2\pi}\, \hat f(q)^k \int_0^\infty x e^{-i q x - \epsilon x} = \lim_{\epsilon\to 0^+}\int_{-\infty}^\infty \frac{dq}{2\pi} \frac{-\hat f (q)^k}{(q-i \epsilon)^2}\,dx \,,\label{eq:avgx}
\end{align}
where $\epsilon>0$ is a regularisation parameter. Finally, adding and subtracting $1$ in the numerator in (\ref{eq:avgx}) and taking the limit $\epsilon\to 0$ gives
\begin{align}
   \mathbb{E}[S_k^+] &= \int_{-\infty}^\infty \frac{dq}{2\pi} \frac{1-\hat f (q)^k}{q^2} -\lim_{\epsilon\to 0^+}\int_{-\infty}^\infty \frac{dq}{2\pi} \frac{1}{(q-i \epsilon)^2}\nonumber\\
   &=\int_{-\infty}^\infty \frac{dq}{2\pi} \frac{1-\hat f (q)^k}{q^2} \,.\label{eq:ESkf}
\end{align}
where we used that $\int_{-\infty}^\infty \frac{dq}{2\pi} \frac{1}{(q-i \epsilon)^2}=0$ for $\epsilon>0$. Using the symmetry of the integral $q\to -q$ and inserting the result in (\ref{eq:PT}), we recover (\ref{eq:statg}).
\section{Probability distribution of the $k^\text{th}$ gap}
\label{sec:pro}
\subsection{Exact results on the probability distribution $P_{k,n}(\Delta)$} \label{sec:3.1}

\begin{figure}[t]
  \begin{center}
    \includegraphics[width=0.6\textwidth]{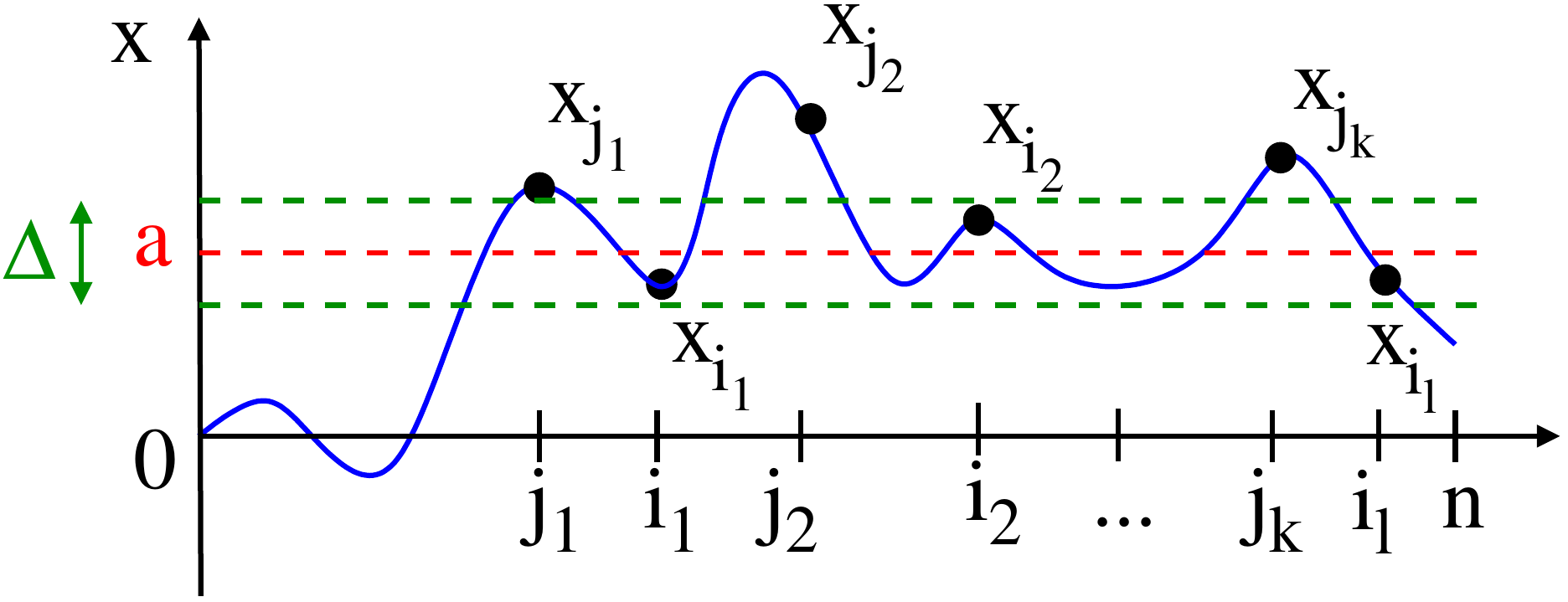}
    \caption{The probability $\text{Prob.}\left(\bigcap_{t=1}^l  B_{i_t}(a,\Delta)\bigcap_{u=1}^k C_{j_u}(a,\Delta)\right)$ in (\ref{eq:SN2a}) corresponds to all the trajectories of $n$ steps that start at the origin and are above the level $a+\Delta/2$ at the intermediate times $j_1,\ldots,j_k$ and in the interval $]a-\Delta/2,a+\Delta/2[$ at the intermediate times $i_1,\ldots,i_l$. }
    \label{fig:figDist}
  \end{center}
\end{figure}
\begin{figure}[t]
  \begin{center}
    \includegraphics[width=0.6\textwidth]{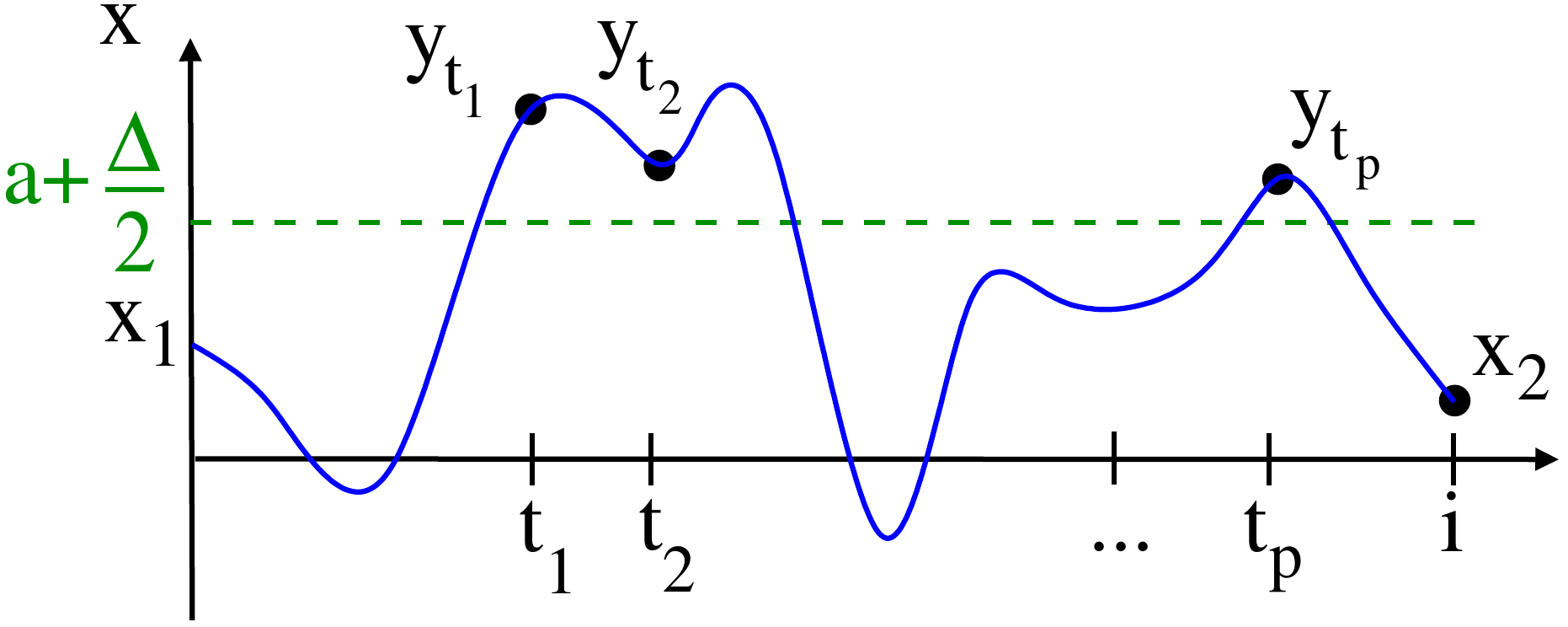}
    \caption{The propagator $K_i(x_2,p|x_1)$ is the probability that a random walk starting from $x_1$ reaches $x_2$ after $i$ steps with $p$ steps at times $t_1,\ldots,t_p$ above $a+\frac{\Delta}{2}$ summed over all possible locations of the $p$ steps. }
    \label{fig:G}
  \end{center}
\end{figure}

We now turn to the computation of the gap distribution (\ref{rel_SPDF2}). To do so, let us go back to the probability $p_{k,n}(a,\Delta)$ defined in (\ref{def_S}). To study this object using similar ideas as for the expected gap, it is convenient to introduce $p_{l,k,n}(a,\Delta)$, which is the probability that there are $k$ points above the level $a+\Delta/2$, $l$ points in the interval $[a-\Delta/2,a+\Delta/2]$ and the remaining ones below $a-\Delta/2$. We will first compute $p_{l,k,n}(a,\Delta)$ for an arbitrary $l$ and then obtain $p_{k,n}(a,\Delta)=p_{l=0,k,n}(a,\Delta)$ by setting $l=0$.
Similarly to the computation for the expected gap, we define the events $B_m(a,\Delta)$ that the $m^\text{th}$ step is located in $[a-\Delta/2,a+\Delta/2]$, i.e.,
\begin{align}
    B_m(a,\Delta) = \left\{|x_m-a|<\frac{\Delta}{2}\right\}\,,\label{eq:Bk}
    \end{align}
    and the events $C_m(a,\Delta)$ that the $m^\text{th}$ step is located above $a+\Delta/2$, i.e.,
\begin{align}
    C_m(a,\Delta) = \left\{x_m>a+\frac{\Delta}{2}\right\}\,.\label{eq:Ck}
    \end{align}
   One can then write $p_{l,k,n}(a,\Delta)$ as
\begin{align}
   p_{l,k,n}(a,\Delta) = \text{Prob.}\left(\sum_{m=1}^n I[B_m(a,\Delta)]=l\,,\sum_{m=1}^n I[C_m(a,\Delta)]=k\right)\,.\label{eq:plkn}
\end{align}
As in the previous section, we rely on the Schuette-Nesbitt formula to rewrite the probability in the rhs in (\ref{eq:plkn}) in terms of the complementary events. Using this formula, the triple generating function of $p_{l,k,n}(a,\Delta)$ reads 
\begin{align}
\bar p_{r,z,s}(a,\Delta) &=  \sum_{n=0}^\infty \sum_{k=0}^n \sum_{l=0}^{n-k} s^n   z^k r^l p_{l,k,n}(a,\Delta)\nonumber\\
&=\sum_{n=0}^\infty \sum_{k=0}^n \sum_{l=0}^{n-k} s^n (z-1)^k (r-1)^l\nonumber \\
&\qquad \times \sum_{1\leq i_1<\ldots<i_l\leq n}\sum_{1\leq j_1<\ldots<j_k\leq n}\text{Prob.}\left(\bigcap_{t=1}^l  B_{i_t}(a,\Delta)\bigcap_{u=1}^k C_{j_u}(a,\Delta)\right) \,.\label{eq:SN2a}
\end{align}
The probability in the rhs in (\ref{eq:SN2a}) is the probability that the random walk is above the level $a+\Delta/2$ at the intermediate times $j_1,\ldots,j_k$ and in the interval $]a-\Delta/2,a+\Delta/2[$ at the intermediate times $i_1,\ldots,i_l$ (see figure \ref{fig:figDist}).
The additional difficulty compared to the computation performed in the previous section is the presence of the double sum over the $i$'s and the $j$'s, which are ordered among themselves, i.e. $i_1  < \ldots < i_l$ and $j_1<\ldots<j_k$, but not among each others. We choose to order the sum over the $j$'s with respect to the sum over the $i$'s. To do so, we denote by $p_q$ the number of $j's$ between $i_{q-1}$ and $i_{q}$ (with $i_0=1$ and $i_{l+1}=n$). By the Markov property of the random walk, one can write the probability in (\ref{eq:SN2a}) as
\begin{align}
& \sum_{1\leq j_1<\ldots<j_k\leq n}\text{Prob.}\left(\bigcap_{t=1}^l  B_{i_t}(a,\Delta)\bigcap_{u=1}^k C_{j_u}(a,\Delta)\right) \nonumber\\
  & = \sum_{p_1+\ldots+p_{l+1}=k}  \int_{a-\frac{\Delta}{2}}^{a+\frac{\Delta}{2}}dx_{i_1}\ldots dx_{i_l}K_{i_1}(x_{i_1},p_1|0) K_{i_2-i_1}(x_{i_2},p_2|x_{i_1})\ldots K_{i_l-i_{l-1}}(x_{i_l},p_l|x_{i_{l-1}})H_{n-i_l}(p_{l+1}|x_{i_{l}}) \,,\label{eq:SN2b}
\end{align}
where $K_i(x_2,p|x_1)$ is the propagator from $x_1$ to $x_2$ during $i$ steps with $p$ steps above $a+\frac{\Delta}{2}$ summed over all possible locations of the $p$ steps (see figure \ref{fig:G}), which reads
\begin{align}
 K_i(x_2,p|x_1)= \sum_{1\leq t_1<\ldots<t_p< i} \int_{a+\frac{\Delta}{2}}^\infty dy_{t_1}\ldots dy_{t_p} &G_{t_1}(y_{t_1}-x_1) G_{t_2-t_1}(y_{t_2}-y_{t_1})\ldots\nonumber\\
 & G_{t_p-t_{p-1}}(y_{t_p}-y_{q_{t-1}}) G_{i-t_p}(x_2-y_{q_{t}})\,,\label{eq:Gprop} 
\end{align}
where we recall that $G_j(x)=\int_{-\infty}^\infty \frac{dq}{2\pi} e^{-iqx} \hat [f(q)]^j$ is the free propagator of the original random walk starting from the origin. In Eq. (\ref{eq:SN2b}), $H_i(p|x_1)=\int_{-\infty}^{\infty} dx_2 K_i(x_2,p|x_1)$ is the propagator with a free end. Taking advantage of the convolution structure over the $i$'s and the $j$'s, and shifting all integration variables $x_{i_1},\ldots,x_{i_l}$ by $a+\frac{\Delta}{2}$, we rewrite (\ref{eq:SN2a}) as
\begin{align}
   \bar p_{r,z,s}(a,\Delta)&=  \sum_{l=0}^\infty (r-1)^l
  \int_{-\Delta}^{0}dx_{1}\ldots dx_{l}\bar K_s\left(x_{1},z\,\bigg|\,-a-\frac{\Delta}{2}\right) \bar K_s(x_{2},z|x_{1})\ldots \bar K_s(x_{l},z|x_{l-1})\bar H_{s}(z|x_{l}) \,,\label{eq:SN2b2}
\end{align}
where $\bar K_s(x_1,z|x_0)=\sum_{p=0}^\infty z^p K_s(x_1,p|x)$ and $\bar H_s(z,|x)=\sum_{p=0}^\infty z^pH_s(p|x)$ are the generating functions which, after shifting the integration variables $y_{t_1},\ldots,y_{t_p}$ in (\ref{eq:Gprop}) by $a+\frac{\Delta}{2}$, are given by
\begin{align}
   \bar K_s(x_2,z|x_1) &= \frac{s}{1-s}\sum_{p=0}^\infty u^p \int_{0}^\infty dy_{1}\ldots dy_{p} \mathrm{f}_{s}(y_{1}-x_1)\mathrm{f}_{s}(y_{2}-y_{1})\ldots\mathrm{f}_{s}(x_2-y_{p})\,,\label{eq:GEse}\\
   \bar H_{s}(z|x_1) &= \frac{1}{1-s}\sum_{p=0}^\infty u^p \int_{0}^\infty dy_{1}\ldots dy_{p} \mathrm{f}_{s}(y_{1}-x_1)\mathrm{f}_{s}(y_{2}-y_{1})\ldots \mathrm{f}_{s}(y_{p}-y_{p-1})\,,\label{eq:HEse}
\end{align}
where we again introduced the effective jump distribution $\mathrm{f}_s(\eta)$ (\ref{eq:Fse}) and defined $u=s(z-1)/(1-s)$ to ease notation.  We now interpret the multiple integrals (\ref{eq:GEse}) and (\ref{eq:HEse}). The former one can be expressed in terms of the excursion probability $E_s(p+1,x_2|x_1)$ of $p+1$ steps for a random walk with a jump distribution $\mathrm{f}_{s}(\eta)$ from $x_1$ to $x_2$, i.e., 
\begin{align}
  E_s(p+1,x_2|x_1) &= \int_{0}^\infty dy_{1}\ldots dy_{p} \mathrm{f}_{s}(y_{1}-x_1)\mathrm{f}_{s}(y_{2}-y_{1})\ldots\mathrm{f}_{s}(x_2-y_{p})\,.\label{eq:Esse}
\end{align}
The excursion probability $E_s(p,x_2|x_1)$ is the probability that a random walk, starting from $x_1$, reaches $x_2$ after $p$ steps while remaining above the origin during the intermediate steps
 (see the right panel in figure \ref{fig:surv}). Note that, contrary to the usual definition of the excursion where the initial and final positions are positive $x_1,x_2>0$, the current computation requires to extend it to a negative initial and final positions. 
It will turn out to be convenient later to introduce the generating function 
\begin{eqnarray}\label{GF_E}
\tilde E_s(z,x_2|x_1)=\sum_{p=1}^\infty z^p E_s(p,x_2,x_1) \;.
\end{eqnarray}
Using this generating function, the expressions in (\ref{eq:GEse}) and (\ref{eq:HEse}) can be re-written as
\begin{subequations}
\begin{align}
\bar K_s(x_2,z|x_1)&=\frac{s}{1-s}\sum_{p=0}^\infty u^p E(p+1,x_2|x_1) = \frac{s}{u(1-s)}\sum_{p=1}^\infty u^p E(p,x_2|x_1)=\frac{1}{z-1}\tilde E_s\left(u,x_2|x_1\right)\,,\\
\bar H_s(z|x_1)&=\frac{1}{1-s}\sum_{p=0}^\infty u^p S_s(p|x)=\frac{1}{1-s}\bar S_s\left(u|x_1\right)\,,
\end{align}
\label{eq:relGEHS}
\end{subequations}
where we used again the notation $u=s(z-1)/(1-s)$ and the probability $S_s(p|x)$ and its generating function $\bar S_s\left(u|x\right)$
have been defined respectively in Eqs. (\ref{eq:Ssse}) and (\ref{eq:genSszma}). 
Inserting the two expressions in (\ref{eq:relGEHS}) into (\ref{eq:SN2b2}) gives
\begin{align}
 \bar p_{r,z,s}(a,\Delta)&= \frac{1}{(1-s)}\bar S_s\left(u|-a-\frac{\Delta}{2}\right) \nonumber\\
&+\frac{1}{1-s}\sum_{l=1}^\infty \left(\frac{r-1}{z-1}\right)^l\int_{-\Delta}^0 dx_{1}\dots dx_{l}\tilde E_s\left(u,x_{1}|-a-\frac{\Delta}{2}\right)\tilde E_s\left(u,x_{2}|x_{1}\right)\ldots \nonumber\\
&\hspace{20em}\tilde E_s\left(u,x_{l}|x_{l-1}\right)\bar S_s\left(u|x_{l}\right)\,,\label{eq:barrhoe}
\end{align}
where we isolated the term $l=0$ in the sum for clarity.
We now compute the generating function of the gap distribution (\ref{rel_SPDF2})
\begin{align}
\tilde P_{z,s}(\Delta)=\sum_{n=0}^\infty \sum_{k=1}^{n-1} s^n z^k  P_{k,n}(\Delta) = \partial^2_{\Delta} \int_{-\infty}^\infty  \tilde p_{z,s}\left(a ,\Delta\right) \;da\,,\label{eq:tildeP}
\end{align}
where we recall that $\tilde p_{z,s}(a,\Delta)= \sum_{n=0}^\infty s^n   \sum_{k=1}^{n-1} z^k p_{k,n}(a,\Delta)$. Note that, as in the previous section, the generating function $\tilde p_{z,s}(a,\Delta)$ starts from $k=1$ and goes to $k=n-1$. In terms of $\bar p_{z,s}(a,\Delta)$, one must remove the terms $k=0$ and $k=n$, which reads
\begin{align}
   \tilde p_{z,s}(a,\Delta) &=  \bar p_{z,s}(a,\Delta)-\sum_{n=0}^\infty s^n p_{0,n}(a,\Delta)-\sum_{n=0}^\infty (zs)^n p_{n,n}(a,\Delta)+1\nonumber\\
   &= \bar p_{z,s}(a,\Delta)  - \bar S_0\left(s\,\bigg|\,a-\frac{\Delta}{2}\right) - \bar S_0\left(sz\,\bigg|\,-a-\frac{\Delta}{2}\right)+1\,,\label{eq:pkndag2}
\end{align}
where we used that $p_{0,n}(a,\Delta)=S_0(n,a-\Delta/2)$ and $p_{n,n}(a,\Delta)=S_0(n,-a-\Delta/2)$, where $\bar S_0(s|a)=\sum_{n=0}^\infty s^n S_0(n,a)$ is the generating function of the survival probability of a random walk with the original jump distribution $f(\eta)$ starting from $a$. Again, the $+1$ term arises from the double counting of $p_{0,0}(a,\Delta)=1$ in the two sums. Inserting (\ref{eq:pkndag2}) in (\ref{eq:tildeP}) and shifting the integration variable $a$ by $\Delta/2$, we get
\begin{align}
  \tilde P_{z,s}(\Delta)=\partial^2_{\Delta} \int_{-\infty}^\infty \left[ \bar p_{z,s}\left(a-\frac{\Delta}{2},\Delta\right)  - \bar S_0\left(s\,|\,a-\Delta\right) - \bar S_0\left(sz\,|\,-a\right)+1\right] \;da\,.\label{eq:tildePs}
\end{align}
Using $\bar p_{z,s}(a,\Delta)=\bar p_{z,s,r=0}(a,\Delta)$ where $\bar p_{z,s,r}(a,\Delta)$ is given in (\ref{eq:barrhoe}), we find that $\tilde P_{z,s}(\Delta)$ can be written as a sum of two contributions
\begin{align}
  \tilde P_{z,s}(\Delta)=  \partial^2_{\Delta}\left[ \tilde p^{(1)}_{z,s}(\Delta)+ \tilde p^{(2)}_{z,s}(\Delta)\right]\,,\label{eq:tildePd}
\end{align}
where
\begin{align}
  \tilde p^{(1)}_{z,s}(\Delta)&=\frac{u}{(1-s)}\sum_{l=1}^\infty \frac{1}{(1-z)^l}\int_{-\Delta}^0 dx_{1}\dots dx_{l}\bar S_s\left(u,x_{1}\right)\tilde E_s\left(u,x_{2}|x_{1}\right)\ldots \tilde E_s\left(u,x_{l}|x_{l-1}\right)\bar S_s\left(u,x_{l}\right)\,,
 \label{eq:barrhoei}\\
 \tilde p^{(2)}_{z,s}(\Delta)&=\int_{-\infty}^\infty \left[\frac{1}{(1-s)}\bar S_s\left(u,-a\right)- \bar S_0\left(s\,|\,a-\Delta\right) - \bar S_0\left(sz\,|\,-a\right)+1\right]da\,,\label{eq:barrhoei2}
\end{align}
with $\bar S_s\left(u,x_{1}\right)$ and $\tilde E_s\left(u,x_{2}|x_{1}\right)$ given respectively in Eq. (\ref{eq:genSszma}) and (\ref{GF_E}). 

To obtain (\ref{eq:barrhoei}), we performed the integration over $a$ and used that 
\begin{align}
\int_{-\infty}^\infty dx_1\tilde E_s(z,x_2|x_1)=\sum_{p=1}^\infty z^p \int_{-\infty}^\infty dx_1 E_s(p,x_2|x_1) = \sum_{p=1}^\infty z^p S_s(p-1|x_2)= z \bar S_s(z|x_2)\,.\label{eq:intEsSs}
\end{align} Let us now evaluate $\tilde p^{(2)}_{z,s}(\Delta)$ explicitly.
Upon taking a derivative with respect to $\Delta$ in (\ref{eq:barrhoei2}), we get
\begin{align}
  \partial_\Delta \tilde p^{(2)}_{z,s}(\Delta)=\int_{-\infty}^\infty \partial_a \bar S_0\left(s\,|\,a-\Delta\right) da &= \bar S_0\left(s\,|\,\infty\right)- \bar S_0\left(s\,|\,-\infty\right)= \frac{s}{1-s}\,,\label{eq:p2tD}
\end{align}
where we used the two identities
\begin{align}
\bar S_0\left(s\,|\,\infty\right) &= \sum_{n=0}^\infty S_0\left(n\,|\,\infty\right)s^n= \sum_{n=0}^\infty s^n =\frac{1}{1-s}\,,\label{eq:idS01}\\
\bar S_0\left(s\,|\,-\infty\right)&=\sum_{n=0}^\infty S_0\left(n\,|\,-\infty\right)s^n= S_0(0\,|\,-\infty)= 1\,.\label{eq:idS02}
\end{align}
Finally integrating (\ref{eq:p2tD}) with respect to $\Delta$ and recognizing from (\ref{eq:sym2}) that $\tilde p^{(2)}_{z,s}(0)= \mathbb{E}[\tilde \Delta_{z,s}] $, we find
\begin{align}
  \tilde p^{(2)}_{z,s}(\Delta)&=   \mathbb{E}[\tilde \Delta_{z,s}]  + \frac{s\Delta}{1-s}\,,\label{eq:p2e}
\end{align}
where $ \mathbb{E}[\tilde \Delta_{z,s}]$ is the generating function of the expected gap given in (\ref{eq:sym2}). As (\ref{eq:p2e}) is a polynomial function of order one in $\Delta$, we see that only $\tilde p^{(1)}_{z,s}(\Delta)$ will be non-zero after taking a double derivative with respect to $\Delta$ in (\ref{eq:tildePd}) which gives
\begin{align}
   \tilde P_{z,s}(\Delta)=  \sum_{n=0}^\infty \sum_{k=1}^{n-1} s^n z^k  P_{k,n}(\Delta) = \partial^2_{\Delta} \tilde p^{(1)}_{z,s}(\Delta)\,,\label{eq:tildePd2}
\end{align}
where $p^{(1)}_{z,s}(\Delta)$ is given in (\ref{eq:barrhoei}). This is formally our main result for the gap distribution $P_{k,n}(\Delta)$. 

\vspace*{0.5cm}
\noindent {\it Remark}: we end this subsection by noting that it is possible to write the infinite sum over $l$ entering the definition  $\tilde p^{(1)}_{z,s}(\Delta)$ in (\ref{eq:barrhoei}) in terms of the solution of an integral equation as follows. Let us define the function $\bar F_{s}(u,x|\Delta)$ as
\bea \label{def_F}
\bar F_{s}(u,x|\Delta) = \frac{1}{1-z} \bar S_s(u,x) + \sum_{l=2}^\infty \frac{1}{(1-z)^{l}} \int_{-\Delta}^0 dx_{2} \ldots \int_{-\Delta}^0 dx_{l} \tilde E_s(u,x|x_2) \ldots \tilde E_s(u,x_{l-1}|x_{l}) \bar S_s(u,x_l) \,,
\eea 
where $u  = s (z-1)/(1-s)$. It is straightforward to show that $\bar F_{s}(u,x|\Delta)$ satisfies the integral equation
\bea \label{int_Fs}
\int_{-\Delta}^0 \tilde E_s(u,x|x_1) \bar F_s(u,x_1|\Delta)\,dx_1 = (1-z) \bar{F}_s(u,x|\Delta) - \bar S_s(u,x) \;, \; x \in [-\Delta, 0] \;.
\eea
One can then write $\tilde p^{(1)}_{z,s}(\Delta)$ in terms of $\bar F_{s}(u,x|\Delta)$ as
\bea \label{ptilde_F}
 \tilde p^{(1)}_{z,s}(\Delta) = \frac{u}{1-s} \int_{-\Delta}^0 \bar S_s(u,x_1)\, \bar{F}_s(u,x_1|\Delta) \,dx_1 \;.
\eea

\subsection{Limiting stationary distribution $P_k(\Delta)$}
The expression for the generating function of the PDF of the gap in (\ref{eq:tildePd2}) is exact but in general, it is of course very difficult to invert it to obtain explicitly $P_{k,n}(\Delta)$ for any $k$ and $n$. However, as we show now, it is possible to study the large $n$ limit, by analysing $\tilde p_{z,s}^{(1)}$ in the limit $s \to 1$. From the Pollaczek-Spitzer formula and after a rather lengthy computation, we find that in the limit $s\to 1$, we have the following asymptotic behavior (see Appendix \ref{app:HIas})
\begin{align}
   \bar S_s\left(u|x\right) &\sim \frac{\sqrt{1-s}}{\sqrt{1-z}}\bar S_*(z|x) \,,\quad s\to 1\,,\nonumber\\
    \tilde E_s\left(u,y|x\right) &\sim \tilde E_*\left(z,y|x\right)\,,\quad s\to 1\,,\label{eq:SEbars1}
\end{align}
where $\bar S_*\left(z|x\right)$ and $\tilde E_*(z,y|x)$ are given by
\begin{align}
  \bar S_*\left(z|x\right) &=1-(1-z)\int_0^\infty dx'\left[g\left(x'-x\right)-g\left(x'\right)\right]\int_{\gamma_B} d\lambda \frac{e^{\lambda x'}}{2\pi i\lambda} \phi(\lambda,z)+\left\{ \begin{array}{ll}-x\sqrt{1-z}\,,& \mu=2\,,\\
 0 \,,& \mu<2\,,\end{array}\right.\label{eq:Ss}\\[1em]
  \tilde E_*(z,y|x) &= -\int_{0}^\infty \frac{dq}{\pi}\ln\left(1+ \frac{(1-z) \hat f(q)}{1-\hat f(q)}\right)-(1-z)\int_0^\infty \frac{dq}{\pi}\frac{\hat f(q)}{1-\hat f(q)}[\cos(q(y-x))-1]\nonumber\\
 & + (1-z)^2 \int_0^\infty dx' dy' \left[g(x'-x) g(y-y') - g(x') g(-y') \right]  \int_{\gamma_B} d\lambda d\lambda' \frac{e^{\lambda x'+\lambda' y'}}{(2\pi i)^2 (\lambda+\lambda')}\phi(\lambda,z)\phi(\lambda',z)\nonumber\\
 &+\left\{\begin{array}{ll}  (1-z)^{3/2} \int_0^\infty dy' \left[g(y-y')+g(x-y') - 2g(y') \right] \int_{\gamma_B} d\lambda' \frac{e^{\lambda' y'}}{(2\pi i)^2 \lambda'}\phi(\lambda',z)\,, &\mu=2\,,\\  0\,,& \mu<2\,.
  \end{array}\right.\label{eq:Est2}
\end{align}
In these expressions, the functions $g(x)$ and $\phi(\lambda,z)$ are given by
\begin{align}
 g(x) &= \int_{-\infty}^\infty \frac{dq}{2\pi} e^{-iqx} \frac{(1+|q|^\mu)\hat f(q)-1}{(1-\hat f(q))|q|^\mu}\,,\label{eq:defGm}\\
 \phi(\lambda,z) &= \exp\left( 
-\lambda\int_{0}^\infty \frac{dq}{\pi}\frac{\ln\left(1+\frac{(1-z) \hat f(q)}{1-\hat f(q)}\right)}{\lambda^2+q^2}\right) \;.\label{eq:defPhim}
\end{align}
These expressions are valid for sufficiently ``regular'' jump distributions where $\hat f(q)  = 1 - |q|^\mu + O(|q|^{2\mu})$ as $q \to 0$. If $\hat f(q)$ has a more singular behavior, the analysis may require a bit more care.  Inserting the asymptotic behaviors (\ref{eq:Ss}) and (\ref{eq:Est2}) into the generating function (\ref{eq:tildePd2}), we find that it behaves as 
\begin{align}
 \tilde P_{z,s} &=  \frac{1}{1-s}\partial^2_{\Delta}\tilde p^{(1)}_{z}(\Delta)\,,\quad s\to 1\,,\label{eq:p1zssl}
\end{align}
with 
 \begin{align}
  \tilde p^{(1)}_{z}(\Delta) &=-\sum_{l=1}^\infty \frac{1}{(1-z)^l}\int_{-\Delta}^0 dx_{1}\dots dx_{l} \bar S_*\left(z,x_{1}\right)\tilde E_*\left(z,x_{2}|x_{1}\right)\ldots \tilde E_*\left(z,x_{l}|x_{l-1}\right)\bar S_*\left(z,x_{l}\right) \;. \label{eq:barpst1}
\end{align}
Inverting the generating function with respect $s$ in (\ref{eq:p1zssl}), we find that the PDF of the gap becomes stationary $\lim_{n\to \infty}P_{k,n}(\Delta)=P_k(\Delta)$, and the generating function of the stationary PDF is given by  
\begin{align}
\tilde P_z(\Delta) = \sum_{k=0}^\infty z^k P_k(\Delta) &=  \partial^2_{\Delta}\tilde p^{(1)}_{z}(\Delta)\,.\label{eq:p1zss}
 \end{align}
We note that this result for the stationary gap distribution is actually exact for any jump distribution $f(\eta)$. Evidently, this is a rather formal result and one would first like to investigate if the known explicit result for the double exponential distribution (corresponding to $\mu=2$) can be recovered \cite{SM12}. Indeed, in Appendix \ref{App_exp}, we show how to recover this explicit result starting from Eq. (\ref{eq:p1zss}). However, our general result in Eq. (\ref{eq:p1zss}) allows us to go far beyond this specific distribution in deriving the exact asymptotic behaviors for large $k$ and large $\Delta$, and for any $1 \leq \mu \leq 2$, as detailed in the next section.

\subsection{Universal stationary distribution $P_k(\Delta)$ for $k\to \infty$ for $\mu\geq 1$}
The expression for the generating function of the stationary PDF in (\ref{eq:p1zss}) is exact and one can extract its asymptotic limit for $k\to\infty$, which corresponds to $z\to 1$. We find that $\tilde P_z(\Delta)$ takes a nontrivial scaling form in the limit $z \to 1$ and $\Delta \to 0$ keeping $\Delta(1-z)^{1/\mu-1}$ fixed. Indeed, for $z\to 1$ with $x,y=O(1)$, the asymptotic behaviors of $\bar S_*\left(z|x\right)$ and $\tilde E_*\left(z,y|x\right)$ given in (\ref{eq:Ss}) and (\ref{eq:Est2}), are strikingly simple
 \begin{subequations}
 \begin{align}
    \bar S_*\left(z|x\right) &\sim 1\,,\quad z\to 1\,,\label{eq:Ssam}\\
    \tilde E_*\left(z,y|x\right) &\sim -B_\mu (1-z)^{\frac{1}{\mu}}\,,\quad z\to 1\,,\quad \mu>1\,,
 \end{align}
 \label{eq:asSEsz}
 \end{subequations}
 where the constant $B_\mu$ is given by $B_\mu=[\sin(\frac{\pi}{\mu})]^{-1}$. These asymptotic behaviors can be easily checked upon noting that $\phi(z,\lambda)\to 1$ as $z\to 1$, $g(x)\propto x^{-1-\mu}$ for $x\to \infty$, and 
 \begin{align}
   -\int_{0}^\infty \frac{dq}{\pi}\ln\left(1+ \frac{(1-z) \hat f(q)}{1-\hat f(q)}\right)\sim  -B_\mu (1-z)^{\frac{1}{\mu}}\,,\quad z\to 1\,,\quad \mu >1\,.\label{eq:asintqz}
 \end{align}
To obtain the asymptotic behavior in (\ref{eq:asintqz}) we have rescaled $q$ by $(1-z)^{\frac{1}{\mu}}$ and used the small $q$ expansion of $\hat f(q)$ in (\ref{Fourier}).
Inserting the asymptotic expansions (\ref{eq:asSEsz}) into (\ref{eq:barpst1}) and (\ref{eq:p1zss}), we find
\begin{align}
  \tilde P_z(\Delta)&\sim
    \partial^2_{\Delta}\left(\frac{1}{B_\mu (1-z)^\frac{1}{\mu}\left(1+\Delta (1-z)^{\frac{1}{\mu}-1}B_\mu\right)}\right)\,,\quad z\to 1\,.\label{eq:rhoasz0}
\end{align}
Taking the double derivative with respect to $\Delta$ in (\ref{eq:rhoasz0}) and using the following inverse Laplace transform [see the formula (7.1) in \cite{Haubold11}]
\begin{align}
  \int_0^\infty e^{-s\zeta}\zeta^{m\alpha+\beta-1}E_{\alpha,\beta}^{(m)}\left(-a \zeta^{\alpha}\right)d\zeta = \frac{m!s^{\alpha-\beta}}{(s^{\alpha} + a)^{m+1}}\,\quad \mathrm{Re}(s)>0,\mathrm{Re}(\alpha)>0,\mathrm{Re}(\beta)>0\,,\label{eq:Hau}
\end{align}
where $E^{(m)}_{\alpha,\beta}(z)=\partial^{m}_z E_{\alpha,\beta}(z)$, together with the relation 
\begin{align}
E^{(2)}_{\alpha,\beta}(z)=\frac{E_{\alpha ,2 \alpha +\beta -2}(z)-(\alpha +2 \beta -3) E_{\alpha ,2 \alpha +\beta -1}(z)+(\beta -1) (\alpha +\beta -1)
   E_{\alpha ,2 \alpha +\beta }(z)}{\alpha ^2}\,,\label{eq:relE}
\end{align}
with $m=2$, $\alpha=\frac{1}{\mu}-1$ and $\beta=1$, we obtain the scaling function (\ref{eq:Pkas}) announced in the introduction.
For $\mu=2$, using the following relations for the Mittag-Leffler functions
\begin{align}
  E_{\frac{1}{2},-\frac{1}{2}}(-x) &=\frac{2 x^2-1}{2 \sqrt{\pi }}-e^{x^2} x^3
   \text{erfc}(x)\,, \quad
   E_{\frac{1}{2},\frac{1}{2}}(-x) = \frac{1}{\sqrt{\pi }}-e^{x^2} x\, \text{erfc}(x)\,,\label{eq:Mittag2}
\end{align}
the scaling function (\ref{eq:Px}) becomes the one in (\ref{exact_F}) obtained in \cite{SM12}.

\vspace*{0.5cm}
\noindent {\it Remark} 1: In the marginal case where $\mu=1$, the analysis has to be performed a bit differently since $\tilde E_*(z,y|x)$ exhibits an additional logarithmic correction as $z \to 1$, which is not present for $\mu >1$. Indeed one finds
\begin{align}
  \tilde E_*(z,y|x) \sim -\frac{(z-1)\ln(1-z)}{\pi}\,,\quad z\to 1\,,\quad \mu=1\,.\label{eq:Esamiu1}
\end{align}
This difference is due to the first term in (\ref{eq:Est2}) which becomes 
 \begin{align}
   -\int_{0}^\infty \frac{dq}{\pi}\ln\left(1+ \frac{(1-z) \hat f(q)}{1-\hat f(q)}\right)\sim -\frac{(z-1)\ln(1-z)}{\pi}\,,\quad z\to 1\,,\quad \mu =1\,.\label{eq:asintqz3}
 \end{align}
The result (\ref{eq:asintqz3}) can be shown by setting $x=(1-z)$ in the left-hand side and taking a double derivative with respect to $x$, which gives
\begin{align}
   -\partial^2_x\int_{0}^\infty \frac{dq}{\pi}\ln\left(1+ \frac{x \hat f(q)}{1-\hat f(q)}\right) &= \int_0^\infty \frac{dq}{\pi} \frac{\hat f(q)^2}{[1+(x-1)\hat f(q)]^2}\,, \label{eq:doubleder}
\end{align}
which upon rescaling $q$ by $x$ and letting $x\to 0$ gives
\begin{align}
  -\partial^2_x\int_{0}^\infty \frac{dq}{\pi}\ln\left(1+ \frac{x \hat f(q)}{1-\hat f(q)}\right) &\sim -\frac{1}{ x}\int_0^\infty \frac{dq}{\pi}\frac{1}{(q+1)^2}\sim -\frac{1}{\pi x}\,.\label{eq:doubleder1}
\end{align}
Integrating (\ref{eq:doubleder1}) twice with respect to $x$ gives the asymptotic result (\ref{eq:asintqz3}).
Inserting the asymptotic expansions (\ref{eq:Ssam})-(\ref{eq:Esamiu1}) into (\ref{eq:p1zss}), we find
\begin{align}
  \tilde P_z(\Delta)&\sim
    \partial^2_{\Delta}
    \frac{\pi^2}{(z-1)\ln(1-z)\left[\pi-\ln(1-z)\Delta\right]}\,,\quad z\to 1\,,\quad \mu=1\,.\label{eq:rhoasz0mu1}
\end{align}
Performing the double derivative and the inverse Laplace transform in the limit $k\to \infty$, we obtain 
\begin{align}
  P_k(\Delta) &\sim \int_{\gamma_B} \frac{dz e^{zk}}{2\pi i} \frac{-2\pi^2 \ln(z)}{z (\pi - \Delta \ln(z))^3}\,,\nonumber\\
  &\sim \int_{\gamma_B} \frac{du e^{u}}{2\pi i} \frac{-2\pi^2 [\ln(u)-\ln(k)]}{u (\pi - \Delta [\ln(u)-\ln(k)])^3}\,,\nonumber\\
  &\sim \int_{\gamma_B} \frac{du e^{u}}{2\pi i} \frac{2\pi^2 \ln(k)}{u (\pi + \Delta\ln(k))^3}\,,\quad k\to \infty\,,\quad \mu=1\,,
  \label{eq:pkmu1d}
\end{align}
where we changed variables $u=zk$ in the second line and took the limit $k\to \infty$ in the third one. Upon performing the inverse Laplace transform, we obtain the expression (\ref{eq:pkmu1}) announced in the introduction.

\vspace*{0.5cm}
\noindent {\it Remark} 2: It is natural to ask what happens in the case $\mu<1$. As discussed in the introduction, this corresponds to transient random walks, where we do not expect the gap distribution to become universal. Indeed, in this case, the asymptotic behavior of $\tilde E_*(z,y|x)$ in (\ref{eq:Est2}) is different and becomes
 \begin{align}
    \tilde E_*\left(z,y|x\right) &\sim -(1-z) f_*(y-x)\,,   
   \quad z\to 1\,,\label{eq:Esmul1}
 \end{align}
 where $f_*(\eta)$ is 
 \begin{align}
   f_*(\eta) = \int_{-\infty}^\infty\frac{dq}{2\pi} e^{-iq\eta} \frac{\hat f(\eta)}{1-\hat f(\eta)}\,.\label{eq:fs}
 \end{align}
 This difference is due to the first term in (\ref{eq:Est2}) which becomes
 \begin{align}
   -\int_{0}^\infty \frac{dq}{\pi}\ln\left(1+ \frac{(1-z) \hat f(q)}{1-\hat f(q)}\right)\sim  -(1-z)\int_{0}^\infty \frac{dq}{\pi}  \frac{ \hat f(q)}{1-\hat f(q)}\,,\quad z\to 1\,,\quad \mu <1\,,\label{eq:asintqz2}
 \end{align}
 where we used that $\ln(1+x)\sim x$ for $x\to 0$. Note that the integral over $q$ in (\ref{eq:asintqz2}) is convergent as the integrand diverges as $q^{-\mu}$  as $q\to 0$ with $\mu<1$. 
Inserting the asymptotic expansions (\ref{eq:Ssam})-(\ref{eq:Esmul1}) into (\ref{eq:p1zss}), we find
\begin{align}
  \tilde P_z(\Delta)&\sim
    \partial^2_{\Delta}
   \frac{1}{(1-z)} \sum_{l=1}^\infty (-1)^l \int_{-\Delta}^0 dx_1\ldots dx_l f_*(x_2-x_1)\ldots f_*(x_l-x_{l-1})\,, &\mu<1\,,
  \quad z\to 0\,.\label{eq:rhoasz0ml1}
\end{align}
Inverting the generating function with respect to $z$, we find
\begin{align}
  P_k(\Delta) &\sim \mathrm{P}(\Delta) \,,\quad k\to \infty\,.\label{eq:Pkasml1}
\end{align}
The function $\mathrm{P}(\Delta)$ is a non-universal function, independent of $k$, given by
\begin{align}
  \mathrm{P}(\Delta) = \partial^2_{\Delta}\sum_{l=1}^\infty (-1)^l \int_{-\Delta}^0 dx_1\ldots dx_l f_*(x_2-x_1)\ldots f_*(x_l-x_{l-1})\,.\label{eq:PD}
\end{align}
Upon using the definition of $f_*(\eta)$ in (\ref{eq:fs}), we find that it can be alternatively written as 
\begin{align}
  f_*(\eta) = \sum_{n=1}^\infty \int_{-\infty}^\infty \frac{dq}{2\pi}e^{-iq\eta} \hat f(\eta)^n  = \sum_{n=1}^\infty G_n(\eta)\,,\label{eq:fetasG}
\end{align}
where $G_n(\eta) =  \int_{-\infty}^\infty \frac{dq}{2\pi}e^{-iq\eta} \hat f(\eta)^n$ is the free propagator. Inserting (\ref{eq:fetasG}) into (\ref{eq:PD}), we get
\begin{align}
   \mathrm{P}(\Delta) &= \partial^2_{\Delta}\sum_{l=1}^\infty (-1)^l \sum_{n_1,\ldots,n_l=1}^\infty \int_{-\Delta}^0 dx_1\ldots dx_l G_{n_1}(x_2-x_1)\ldots G_{n_l}(x_l-x_{l-1})\nonumber\\
   &= \partial^2_{\Delta}\sum_{l=1}^\infty (-1)^l \sum_{1<i_1<\ldots<i_l<\infty} \int_{-\Delta}^0 dx_0\ldots dx_{l-1} G_{i_1}(x_1-x_0)\ldots G_{i_l-i_{l-1}}(x_{l-1}-x_{l-2})\,,\label{eq:PD2}
\end{align}
where we relabeled the integration variables $x_i=x_{i-1}$ and changed the summation indices $i_t=n_t-n_{t-1}$ in the second line.
The multiple integral summed over the $n_i$'s in (\ref{eq:PD2}) is the probability that a random walk of infinite duration spends $l$ steps in $[-\Delta,0]$ at the times $i_1,\ldots,i_l$, and integrated over its initial position $x_0\in [-\Delta,0]$. Equivalently, by translation along the $x$-axis, this is the probability that a random walk of infinite duration, starting from the origin, spends $l$ steps in the interval $[a+\Delta/2,a-\Delta/2]$ at the times $i_1,\ldots,i_l$, and integrated over the center of the interval $a\in [-\Delta/2,\Delta/2]$, i.e.,
\begin{align}
  \mathrm{P}(\Delta) &= \partial^2_{\Delta}\int_{-\Delta/2}^{\Delta/2} da\sum_{l=1}^\infty (-1)^l \sum_{1<i_1<\ldots<i_l<\infty} \text{Prob.}\left(\bigcap_{m=1}^l B_{i_m}(a,\Delta)\right)\,.\label{eq:PD3}
\end{align}
By the inclusion-exclusion formula, we can alternatively rewrite the sum over $l$ in (\ref{eq:PD3}) as the probability that a random walk of infinite duration spends none of its steps in the interval $[a+\Delta/2,a-\Delta/2]$:
\begin{align}
  \mathrm{P}(\Delta) = -\partial^2_{\Delta}\int_{\mathbb{R}\setminus]-\Delta/2,\Delta/2[} da\, \text{Prob.}\left(\bigcap_{m=1}^\infty \bar B_{i_m}(a,\Delta)\right)\,.\label{eq:altPD}
\end{align}
Note that $\text{Prob.}\left(\bigcap_{m=1}^\infty \bar B_{i_m}(a,\Delta)\right)$ is the solution of the integral equation
\begin{align}
  \text{Prob.}\left(\bigcap_{m=1}^\infty \bar B_{i_m}(a,\Delta)\right) = \int_{\mathbb{R}\setminus \left]a-\frac{\Delta}{2},a+\frac{\Delta}{2}\right[} d\eta f(\eta)\,\text{Prob.}\left(\bigcap_{m=1}^\infty \bar B_{i_m}(a-\eta,\Delta)\right)\,.\label{eq:inte}
\end{align} 
To our knowledge, this probability has not been computed explicitly in the literature.

\subsection{Condensation in the distribution of the $k^\text{th}$ gap for $\mu<2$}

As we have seen in the introduction, the average value of the universal distribution (\ref{eq:avgPmu}) does not match with the expected stationary gap in (\ref{eq:avgGass}) for $1<\mu<2$. This suggests the existence of ``condensation'' in the probability distribution, i.e. a part of the distribution which would not contribute to the normalization but would contribute to the first moment.
 We therefore need to look for a large $\Delta = O(k^{1/\mu})$ expansion of (\ref{eq:p1zss}), going beyond the typical scaling regime $\Delta = O(k^{1/\mu-1})$, describing the ``fluid'' part of the gap distribution, as discussed in the previous section. To derive the atypical scaling regime, i.e., the ``condensate'' part, stated in the second line of Eq. (\ref{eq:Pksum}), namely
 \begin{eqnarray} \label{scaling_M_text}
P_{k}(\Delta) \sim  \frac{1}{k^{1+\frac{1}{\mu}}} \mathcal{M}_\mu\left(\frac{\Delta}{k^{\frac{1}{\mu}}}\right) \quad, \quad \Delta = O(k^{1/\mu}) \;,
 \end{eqnarray}
it turns out that the approach used to compute the distribution of the typical fluctuations described in the previous sections is not very convenient. Instead, we need to use a slightly different technique that relies on a path decomposition of the ``true'' random walk and which is described in Appendix \ref{app:scalC}.    
  
%

\begin{figure}[t]
  \begin{center}
    \includegraphics[width=0.5\textwidth]{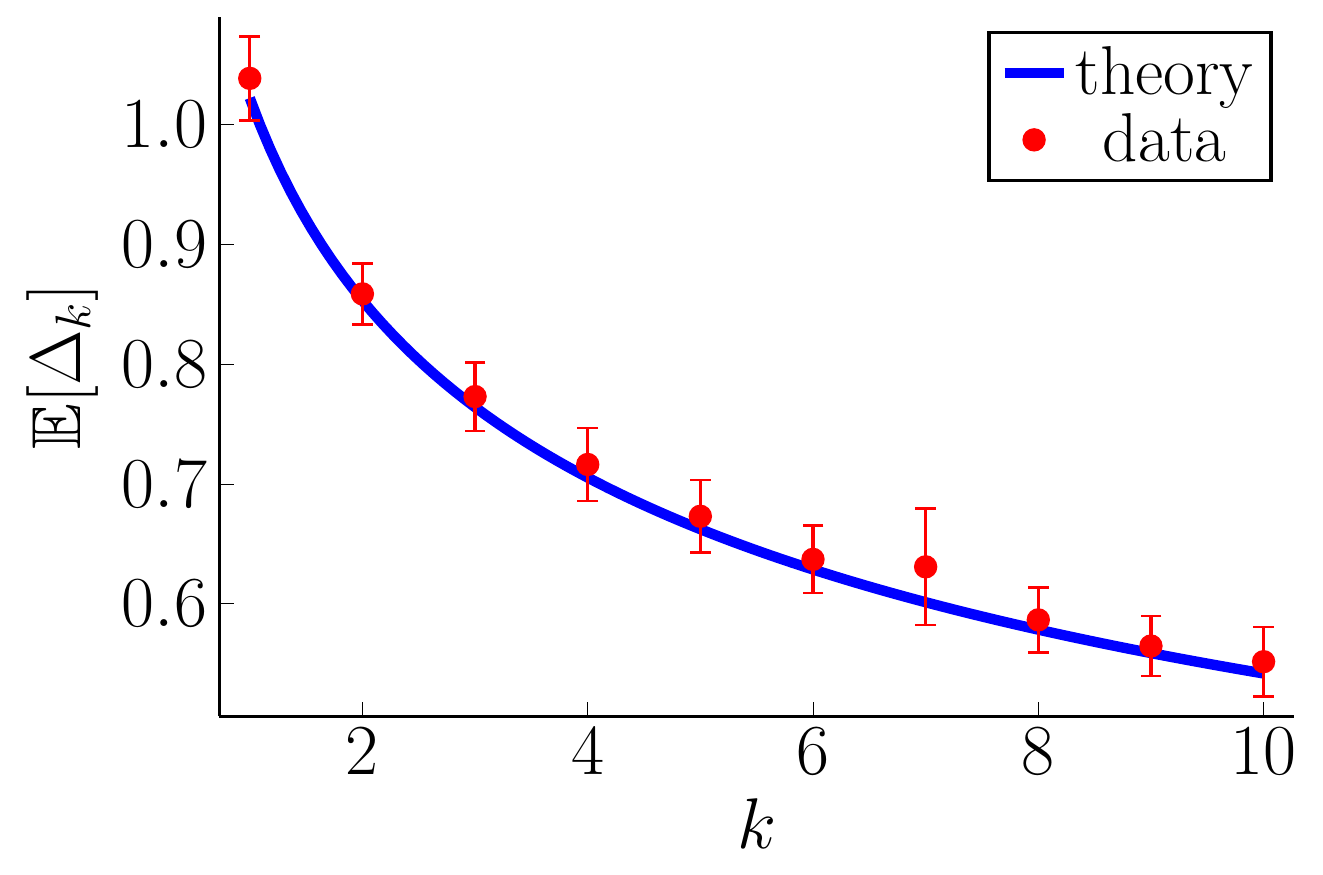}
    \caption{Expected stationary $k^\text{th}$ gap $\mathbb{E}[\Delta_k]$ as a function of $k$ for a Student's $t$ distribution with $\mu=1.5$, given in (\ref{eq:student}), and sampled over $10^8$ realisations each of $n=10^6$ steps. The numerical data (red dots) is compared to the theoretical prediction (blue line) given in (\ref{eq:statg}).}
    \label{fig:dk}
  \end{center}
\end{figure}
\section{Numerical results}
In this section, we compare our theoretical predictions with numerical results on gap statistics obtained by averaging random walk trajectories with various jump distributions with L\'evy index $1\leq \mu\leq 2$. To do so, we generate a random walk trajectory $\{x_1,\ldots,x_n\}$, which we order by decreasing order of magnitude to obtain $\{M_{1,n},\ldots,M_{n,n}\}$. We then collect the gaps $\Delta_{k,n}=M_{k,n}-M_{k+1,n}$, which serve as our data to compare with our theoretical results. 
\subsection{Stationary expected $k^\text{th}$ gap}
In figure \ref{fig:dk}, we compare our theoretical prediction for the expected stationary $k^\text{th}$ gap $\mathbb{E}[\Delta_k]$ in (\ref{eq:statg}) for a Student's $t$ jump distribution 
\begin{align}
  f(\eta) = \frac{\Gamma\left(\frac{\mu+1}{2}\right)}{a_1\sqrt{\mu \pi}\Gamma\left(\frac{\mu}{2}\right)}\left(1+\frac{\eta^2}{a_1^2\mu}\right)^{-\frac{\mu+1}{2}}\,,\label{eq:student}
\end{align}
and where the rescaling factor $a_1=2(-\mu ^{-\mu /2} \Gamma
   \left(\frac{\mu }{2}\right)/\Gamma \left(-\frac{\mu
   }{2}\right))^{\frac{1}{\mu }}$ has been chosen so that the Fourier transform of the distribution 
   \begin{align}
    \hat f(q) = \frac{2^{1-\frac{\mu }{2}} \mu ^{\mu /4} \left(a_1 |q|\right)^{\mu /2} K_{\frac{\mu }{2}}\left(a_1 \sqrt{\mu }
   | q| \right)}{\Gamma \left(\frac{\mu }{2}\right)}\,,\label{eq:studentF}
   \end{align}
where $K_\nu(x)$ is the modified Bessel function of the second kind,   behaves as $\hat f(q)\sim 1-|q|^\mu$ for $q\to 0$. The theoretical curve in figure \ref{fig:dk} has been obtained by inserting (\ref{eq:studentF}) into (\ref{eq:statg}) and performing the integration over $q$ numerically for various values of $k$. As one can see, the agreement is very good. 
\subsection{Universal stationary distribution for $k\to \infty$}
\begin{figure}[t]
  \begin{center}
    \includegraphics[width=0.4\textwidth]{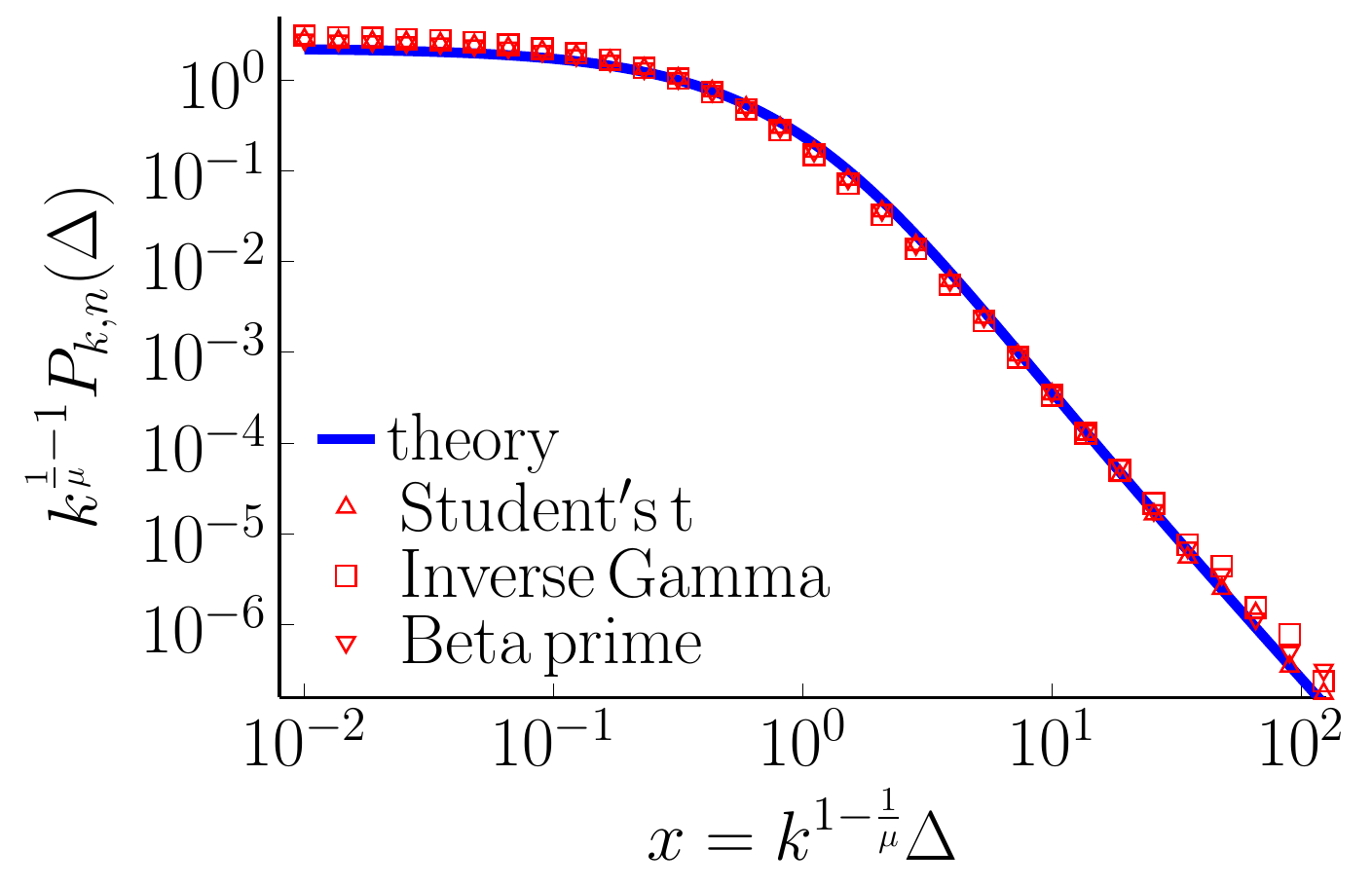}\includegraphics[width=0.4\textwidth]{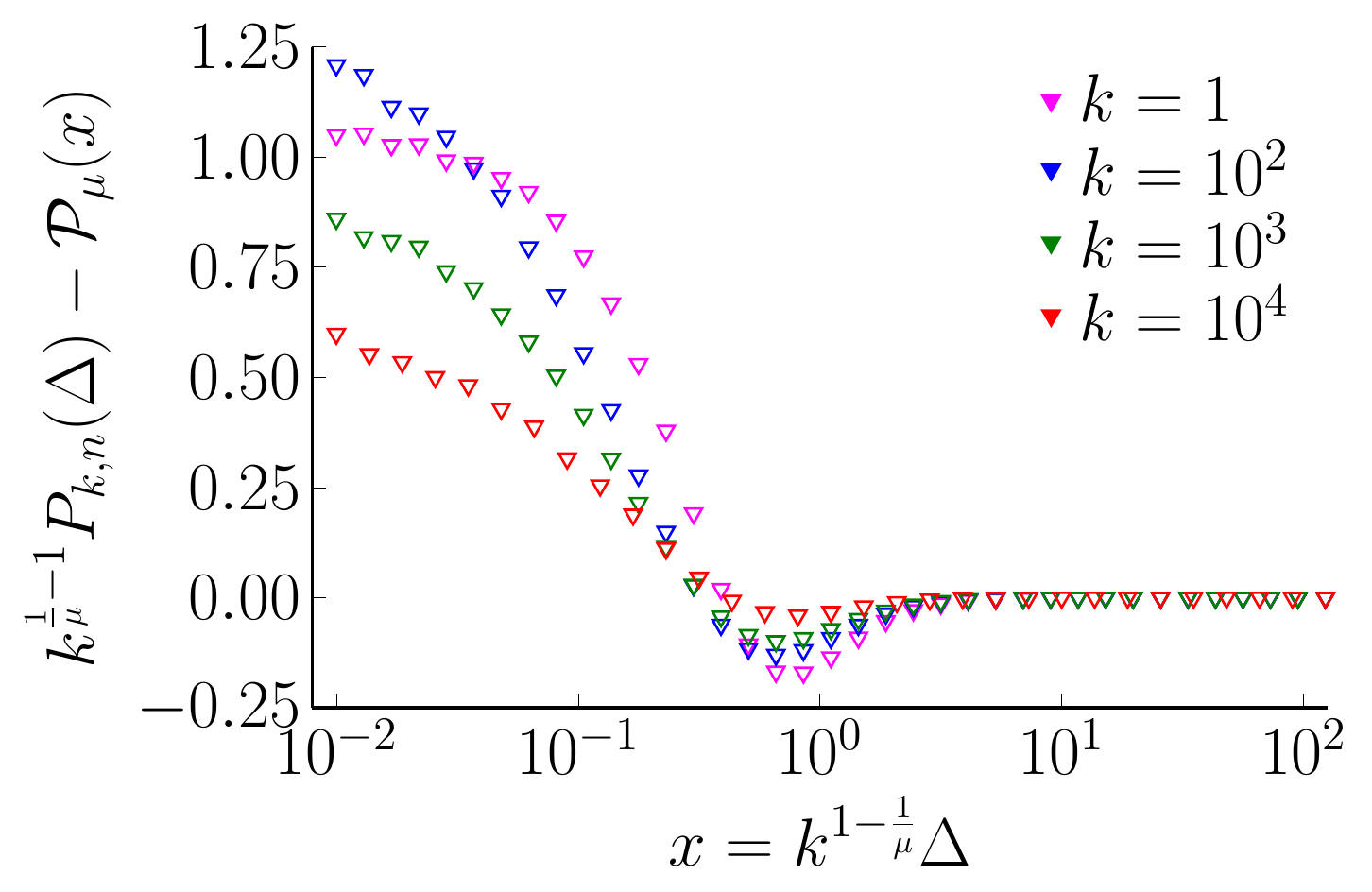}
    \caption{\textbf{Left panel:} Gap distribution $P_{k,n}(\Delta)$ in the stationary limit $n\to \infty$ and in the scaling regime $k\to \infty$ with $\Delta\to 0$ as a function of the scaling variable $x= k^{1-1/\mu}\Delta$. The theoretical prediction (blue line) in (\ref{eq:Px}) is compared to numerical distributions for $k=10^4$, $n=10^6$ obtained for various jump distributions: the Student's t distribution in (\ref{eq:student}), the inverse gamma distribution in (\ref{eq:invG}) and the beta prime distribution in (\ref{eq:betaP}). The L\'evy index has been set to $\mu=1.8$ and the histograms have been obtained by sampling over $10^7$ realisations. \textbf{Right panel:} Convergence to the gap distribution $P_{k,n}(\Delta)$ in the stationary limit $n\to \infty$ and in the scaling regime $k\to \infty$ with $\Delta\to 0$ as a function of the scaling variable $x= k^{1-1/\mu}\Delta$. The difference between the numerical distributions for the Student's t distribution in (\ref{eq:student} and the theoretical prediction in (\ref{eq:Px} is shown for increasing values of $k$ and for $n=10^6$. As $k$ increases, the difference goes to zero for all $x$ (albeit non-uniformly). The L\'evy index has been set to $\mu=1.8$ and the histograms have been obtained by sampling over $10^7$ realisations. }
    \label{fig:Pmu}
  \end{center}
\end{figure}
In the left panel in figure \ref{fig:Pmu}, we check the universal stationary distribution $\mathcal{P}_\mu(x)$ in (\ref{eq:Px}) for various jump distributions, namely the Student's $t$ distribution in (\ref{eq:student}), the symmetric inverse gamma distribution 
\begin{align}
  f(\eta) = \frac{a_2^\mu}{2\Gamma(\mu) |\eta|^{1+\mu}}\,e^{-\frac{a_2}{|\eta|}}\,,\label{eq:invG}
\end{align}
and the symmetric beta prime distribution 
\begin{align}
  f(\eta) = \frac{\mu}{2a_3\left(1+\frac{|x|}{a_3}\right)^{1+\mu}}\,,\label{eq:betaP}
\end{align}
where the rescaling factors $a_2= \left(\frac{2}{\pi }\sin
   \left(\frac{\pi  \mu }{2}\right) \Gamma (\mu )
   \Gamma (\mu +1)\right)^{\frac{1}{\mu }}$ and $a_3=  \left(\frac{2}{\pi }\sin \left(\frac{\pi  \mu }{2}\right) \Gamma (\mu )\right)^{\frac{1}{\mu }}$ have been added so that the Fourier transforms of the distributions behave as $\hat f(q)\sim 1-|q|^\mu$ for $q\to 0$. The theoretical curve in the left panel in figure \ref{fig:Pmu} is our prediction in (\ref{eq:Px}) and the numerical histograms have been obtained by sampling the gaps in the scaling limit $k\to \infty$ with $\Delta\to 0$ with $x=\Delta\, k^{1-\frac{1}{\mu}}$ fixed.  The good collapse of the different curves in the left panel in figure \ref{fig:Pmu} indicates that $\mathcal{P}_\mu(x)$ is universal. Note that it is numerically difficult to reach the scaling form of the stationary regime for $\mu<2$ as the convergence is quite slow with $k$ (see right panel in figure \ref{fig:Pmu}). It becomes even more difficult as the L\'evy index $\mu$ is lowered.

\subsection{Condensation in the distribution of the $k^\text{th}$ gap for $\mu<2$}
\begin{figure}[t]
  \begin{center}
    \includegraphics[width=0.5\textwidth]{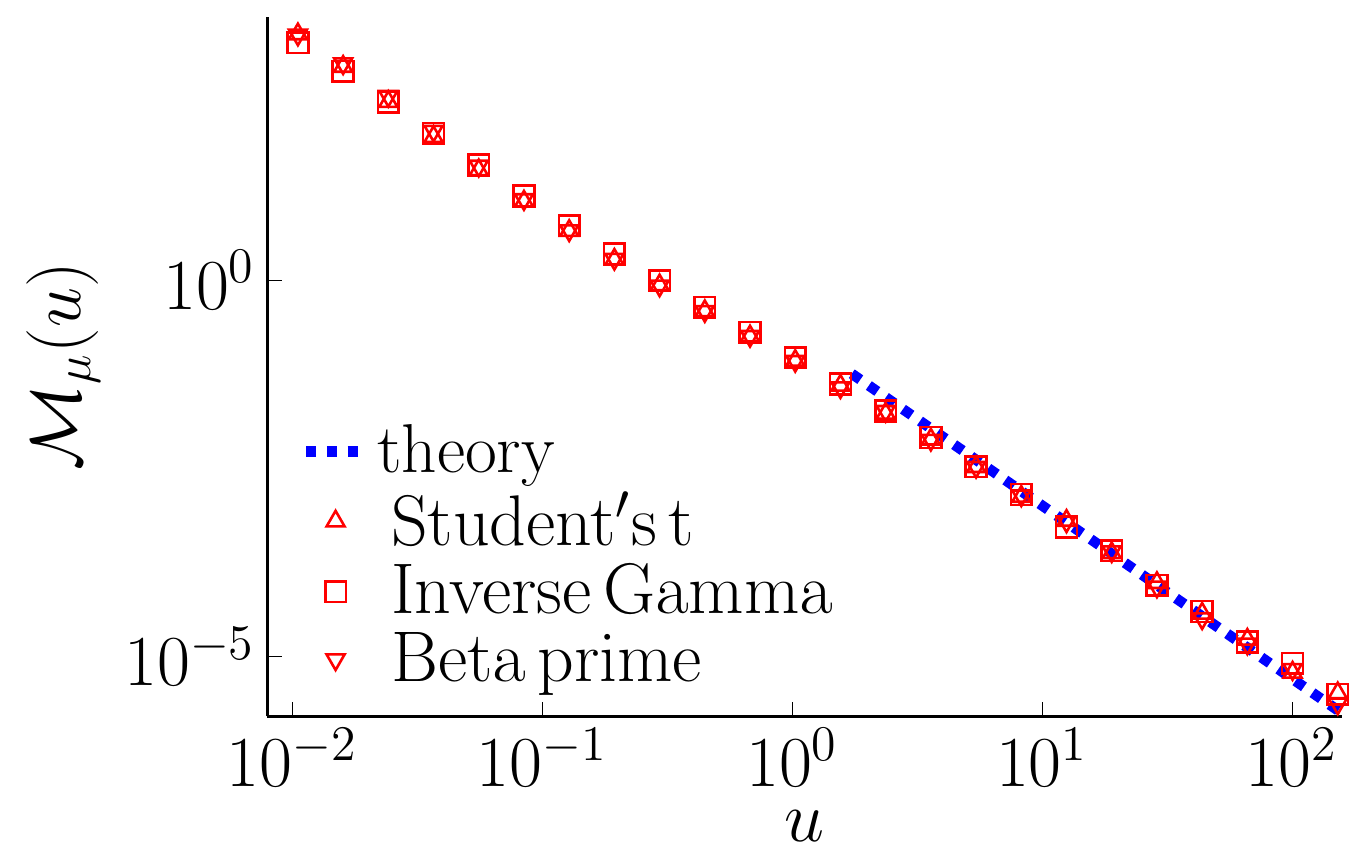}
    \caption{Scaling function $\mathcal{M}_\mu(u)$ as a function of $u$ for $\mu=1.3$. The theoretical prediction for the tail (dashed blue line) in (\ref{eq:expMmu}) is compared to numerical simulations for $k=10^2$ and $n=10^5$ obtained for various jump distributions: the Student's t distribution in (\ref{eq:student}), the inverse gamma distribution in (\ref{eq:invG}) and the beta prime distribution in (\ref{eq:betaP}). The histograms have been obtained by sampling over $10^7$ realisations.}
    \label{fig:Mmu}
  \end{center}
\end{figure}

In figure \ref{fig:Mmu}, we check the tail of the universal scaling function $\mathcal{M}_\mu(u)$ in (\ref{eq:expMmu}) for various jump distributions: the Student's t distribution in (\ref{eq:student}), the inverse gamma distribution in (\ref{eq:invG}) and the beta prime distribution in (\ref{eq:betaP}). The theoretical curve in figure \ref{fig:Mmu} is our prediction in (\ref{eq:expMmu}) and the numerical histograms have been obtained by sampling the gaps in the scaling limit $k\to \infty$ with $\Delta\to \infty$ with $u=\Delta\, k^{-\frac{1}{\mu}}$ fixed. As one can see, the agreement is quite good.
\section{Conclusion}
\label{sec:con}

We have studied a simple one-dimensional discrete-time random walk model where the walker starts at the origin and its position is incremented at each step independently by adding a noise drawn from a symmetric and continuous distribution $f(\eta)$. This includes L\'evy flights, where $f(\eta) \sim |\eta|^{-1-\mu}$ has a power law tail, with the L\'evy index $0<\mu \leq 2$. 
We observe the walk up to $n$ steps and order the positions at different times (discarding the initial position $x_0=0$) and denote the ordered positions by $M_{1,n} > M_{2,n} > \ldots > M_{n,n}$. Our main interest in this paper has been on the statistics of the gap $\Delta_{k,n} = M_{k,n} - M_{k+1,n}$. Our first interesting result was to show that in the limit $n \to \infty$, the gap distribution approaches an $n$ independent limit, which still depends on $k$. Moreover, we derived an exact formula, valid for arbitrary $k$, for the expected gap $\mathbb{E}[\Delta_{k,n \to \infty}]$ in this stationary limit, which is valid for arbitrary jump distribution $f(\eta)$. In addition, in the stationary limit, we could also compute the distribution of the typical gap, with size $\Delta_k = O(k^{1/\mu-1})$, in the scaling limit for large $k$, for all $1\leq \mu\leq 2$. Surprisingly, we find that the associated scaling function has a universal power law tail $\sim x^{-3}$ for large $x$, for all $1 \leq \mu < 2$, except in the limit $\mu \to 2$ where the tail changes abruptly to $\sim x^{-4}$. 

Another feature of the stationary distribution of the gap, for $1\leq \mu<2$, is the existence of an anomalous ``condensate part'' in the distribution that describes large atypical gaps of order $O(k^{1/\mu})$, which are much larger than the typical gaps of size $O(k^{1/\mu-1})$. This is somewhat similar to a ``condensate bump'' seen in many problems with real-space condensation, where a few large units (in our case the gaps) form a condensate that co-exists with the background fluid consisting of a thermodynamically large number of units of typical size.

In this paper, we were able to extract explicit results for the gap statistics only when the L\'evy index belongs to the range $1 \leq \mu \leq 2$. It is natural to wonder what happens to the gap statistics in the complementary range $0<\mu <1$. In this latter range, while our formalism presented here still holds, it is hard to extract explicit results. This is due to the fact that the typical gaps become of size $O(1)$ for all $0<\mu<1$, where one cannot use the asymptotic forms of the functions involved in the computations. It therefore remains an interesting challenge to compute the gap statistics for $0<\mu<1$. Another interesting question is to investigate the gap statistics when the gaps are deep inside the ``bulk'', i.e., when $k = \alpha n$, with $\alpha \in [0,1]$ fixed. Here, we have focused on the ``edge'' limit, where we took the limit where $n \to \infty$ but keeping $k$ fixed of order $O(1)$. The bulk limit would correspond to taking both $k$ and $n$ large simultaneously, with their ratio $\alpha = k/n$ fixed,  as it was done for the case of finite variance jump distributions in \cite{Lacroix}.

\begin{acknowledgements}
This work was partially supported by the Luxembourg National Research Fund (FNR) (App. ID 14548297).
\end{acknowledgements}

\appendix
\section{Proof of the Schuette-Nesbitt formula}\label{app:SN}

In this Appendix we provide two different derivations of the Schuette-Nesbitt formula. 

\subsection{Direct derivation}
In this section, we provide a proof of the Schuette-Nesbitt formula in (\ref{eq:SN}). Let $A_1,\ldots,A_n$ be a set of events. We introduce the indicator function $I(A)$ which takes the value $1$ if $A$ occurred and $0$ otherwise. We wish to compute the counting probability $\text{Prob.}(\nu=\sum_{m=1}^n I(A_m))$. To do so, let us first work with indicator functions and denote by $\mathcal{N}=\{1,\ldots,n\}$ the set of numbers from $1$ to $n$. For exactly $\nu$ events to happen, we sum over all possible combinations of sets of $\nu$ events and require that all the events in the set happened, while requiring that the remaining ones did not happen. This reads
\begin{align}
    I\left(\nu=\sum_{m=1}^n I(A_m)\right) = \sum_{\substack{J\subset \mathcal{N}\\|J|=\nu}} I\left(\bigcap_{j\in J} A_{j}\right)\times I\left(\bigcap_{l\in \mathcal{N}\setminus J}\bar A_l\right)\,,\label{eq:SN1} 
\end{align}
where $|J|$ denotes the cardinal number of the set $J$. By using the rule $I(B)=1-I(\bar B)$, where $\bar{B}$ is the complementary event of $B$, we can rewrite the last indicator function as 
\begin{align}
     I\left(\nu=\sum_{m=1}^n I(A_m)\right) = \sum_{\substack{J\subset \mathcal{N}\\|J|=\nu}} I\left(\bigcap_{j\in J} A_{j}\right)\times\left[ 1-I\left(\bigcup_{l\in \mathcal{N}\setminus J} A_l\right)\right]\,.\label{eq:SN2}
\end{align}
Making use of the inclusion-exclusion principle and that $|\mathcal{N}\setminus J|=n-\nu$, we rewrite the last indicator function as
\begin{align}
     I\left(\nu=\sum_{m=1}^n I(A_m)\right) = \sum_{\substack{J\subset \mathcal{N}\\|J|=\nu}} I\left(\bigcap_{j\in J} A_{j}\right)\times\left[ 1-\sum_{p=1}^{n-\nu}(-1)^{p-1}\sum_{\substack{K\subset \mathcal{N}\setminus J\\|K|=p}}I\left(\bigcap_{k\in K} A_k\right)\right]\,.\label{eq:SN3}
\end{align}
Distributing the product, we obtain
\begin{align}
     I\left(\nu=\sum_{m=1}^n I(A_m)\right) = \sum_{\substack{J\subset \mathcal{N}\\|J|=\nu}} I\left(\bigcap_{j\in J} A_{j}\right) - \sum_{p=1}^{n-\nu}(-1)^{p-1}\sum_{\substack{J\subset \mathcal{N}\\|J|=\nu}}\sum_{\substack{K\subset \mathcal{N}\setminus J\\|K|=p}}I\left(\bigcap_{q\in K\cup J} A_q\right)\,,\label{eq:SN4}
\end{align}
where we have used that $I\left(\bigcap_{j\in J} A_{j}\right)\times I\left(\bigcap_{k\in K} A_k\right)=I\left(\bigcap_{q\in K\cup J} A_q\right)$. The last double sum enumerates all the sets in $\mathcal{N}$ with $\nu+p$ elements. However, due to the double sum, we over-count them by a factor $\binom{\nu+p}{p}=\binom{\nu+p}{\nu}$, which is the number of ways one can separate a set of size $\nu+p$ into two sets of sizes $\nu$ and $p$. Therefore, it can be written as
\begin{align}
     I\left(\nu=\sum_{m=1}^n I(A_m)\right) = \sum_{\substack{J\subset \mathcal{N}\\|J|=\nu}} I\left(\bigcap_{j\in J} A_{j}\right) - \sum_{p=1}^{n-\nu}(-1)^{p-1}\binom{p+\nu}{\nu}\sum_{\substack{L\subset \mathcal{N}\\|L|=\nu+p}}I\left(\bigcap_{q\in L} A_q\right)\,.\label{eq:SN5}
\end{align}
Finally, noting that the first term on the rhs of (\ref{eq:SN5}) can be written as the $p=0$ term of the sum over $p$, we find
\begin{align}
     I\left(\nu=\sum_{m=1}^n I(A_m)\right) =  \sum_{p=0}^{n-\nu}(-1)^{p}\binom{p+\nu}{\nu}\sum_{\substack{L\subset \mathcal{N}\\|L|=\nu+p}}I\left(\bigcap_{q\in L} A_q\right)\,,\label{eq:SN6}
\end{align}
which, upon taking the expectation value and taking the generating function with respect to $\nu$, we get
\begin{align}
 \sum_{\nu=0}^n z^\nu \text{Prob.}\left(\nu=\sum_{m=1}^n I(A_m)\right) = \sum_{k=0}^n (z-1)^k \sum_{1\leq i_1<\ldots<i_k\leq n}\text{Prob.}\left(\bigcap_{s=1}^k A_{i_s}\right)\,,\label{eq:SNg}
\end{align}
which recovers the Schuette-Nesbitt formula in (\ref{eq:SN}).

\subsection{Alternative derivation}
An alternative proof of this last formula (\ref{eq:SNg}) is as follows. The generating function of $P_{\nu, n} =  \mathbb{E}[  I\left(\nu=\sum_{m=1}^n I(A_m)\right)]$ -- with respect to $\nu=0, \ldots, n$ -- can also be written as
\begin{eqnarray} \label{SN_gs1}
 \sum_{\nu=0}^n z^\nu P_{\nu,n}  =  \mathbb{E}\left[ z^{\sum_{m=1}^n I(A_m)}\right] =  \mathbb{E}\left[ \prod_{m=1}^n z^{I(A_m)}\right] \;.
\end{eqnarray}
Since $I(A_m)$, for $m=1, \ldots, n$, is a binary variable that takes only values $0$ or $1$, it is easy to check the identity
\begin{eqnarray}\label{SN_gs2}
z^{I(A_m)} = 1 + I(A_m)(z-1) \;.
\end{eqnarray}
By inserting this identity (\ref{SN_gs2}) in (\ref{SN_gs1}), one obtains
\begin{eqnarray}\label{SN_gs3}
\sum_{\nu=0}^n z^\nu P_{\nu,n} =   \mathbb{E}\left[ \prod_{m=1}^n \left( 1 + I(A_m)(z-1) \right) \right] \;.
\end{eqnarray}
We then expand the product in (\ref{SN_gs3}) to obtain
\begin{eqnarray} 
\sum_{\nu=0}^n z^\nu P_{\nu,n} &=& \sum_{k=0}^n (z-1)^k \sum_{1\leq i_1 < i_2< \ldots< i_k \leq n}  \mathbb{E}[ I(x_{i_1}) I(x_{i_2}) \ldots I(x_{i_k})]\nonumber\\
&=&  \sum_{k=0}^n (z-1)^k \sum_{1\leq i_1 < i_2< \ldots< i_k \leq n} \text{Prob.}\left(\bigcap_{s=1}^k A_{i_s}\right) \;,\label{SN_gs4} 
\end{eqnarray}
which is the result given in (\ref{eq:SNg}).

\section{Evaluation of the integral (\ref{eq:minf}) using the Pollaczek-Spitzer formula}
\label{app:pzsai}
In this appendix, we compute the integral in (\ref{eq:minf}), which we decompose in two parts
\begin{align}
  \int_{-\infty}^0 da \,\tilde p_{z,s}(a) = I_1(z,s) - I_2(s)\,,\label{eq:decint}
\end{align}
where
\begin{align}
 I_1(z,s) &=\int_0^\infty da  \left[\frac{1}{1-s} \bar S_s\left(u\Big|a\right) - \bar S_0(sz|a)\right]\label{eq:I1HI}\,,\\
 I_2(s) &=
 \int_0^\infty da  \left[\bar S_0(s|-a) -1\right]\,.\label{eq:I2HI}
\end{align}
Below, in subsections \ref{sub_B1} and \ref{sub_B2}, we show that $I_1(z,s)$ and $I_2(s)$ are given by
\begin{align}
 I_1(z,s) &=  \frac{1}{1-sz}\int_0^\infty \frac{dq}{\pi q^2}\ln\left(\frac{1-s \hat f(q)}{1-s}\right)\,,\label{eq:lambda00}\\
 I_2(s) &= \frac{s}{\sqrt{1-s}}\int_{-\infty}^0 da \int_0^\infty dy f(y-a)\int_{\gamma_B} \frac{d\lambda e^{\lambda a}}{2\pi i \lambda}\exp\left(-\lambda \int_0^\infty \frac{dq}{\pi} \frac{\ln(1-s \hat f(q))}{\lambda^2+q^2}\right)\,.\label{eq:I2sf}
\end{align}
Inserting these expressions into (\ref{eq:decint}), we recover the expression (\ref{eq:int2av}) from the main text. 
\subsection{Analysis of $I_1(z,s)$} \label{sub_B1}
To analyse $I_1(z,s)$ in (\ref{eq:I1HI}), we introduce an exponential cutoff in the integral which we let go to $0$ as
\begin{align}
  I_1(z,s)  = \lim_{\lambda\to 0}\int_0^\infty da \,e^{-\lambda a}\left[\frac{1}{1-s} \bar S_s\left(u\Big|a\right) - \bar S_0(sz|a)\right]\,.\label{eq:lambda0}
\end{align}
Using the Pollaczek-Spitzer formula in (\ref{eq:HI}) and the expression of the effective distribution (\ref{eq:effFb}), we find
\begin{align}
   I_1(z,s) = \lim_{\lambda\to 0}\frac{1}{\lambda}\Bigg[\frac{1}{(1-s)\sqrt{1-u}}\exp\left(-\lambda \int_0^\infty \frac{dq}{\pi}\frac{\ln\left[1-\frac{s(z-1)\hat f(q)}{1-s\hat f(q)}\right]}{\lambda^2+q^2}\right)
   - \frac{1}{\sqrt{1-sz}}\exp\left(-\lambda \int_0^\infty \frac{dq}{\pi}\frac{\ln[1-sz \hat f(q)]}{\lambda^2+q^2}\right)\Bigg]\,.\label{eq:lambda1}
\end{align}
We decompose the logarithms in (\ref{eq:lambda1}) as 
\begin{align}
  \ln\left[1-\frac{s(z-1)\hat f(q)}{1-s\hat f(q)}\right] &= \ln\left[1-\frac{s(z-1)}{1-s}\right] + \ln\left[\frac{1-\frac{s(z-1)\hat f(q)}{1-s\hat f(q)}}{1-\frac{s(z-1)}{1-s}}\right]\,,\label{eq:lnt}\\
  \ln[1-sz \hat f(q)] &= \ln[1-sz ] +\ln\left[\frac{1-sz \hat f(q)}{1-sz}\right]\,,\label{eq:lnt2}
\end{align}
to find that (\ref{eq:lambda1}) becomes
\begin{align}
   I_1(z,s) = \lim_{\lambda\to 0}\frac{1}{1-sz}\Bigg[\exp\left(-\lambda \int_0^\infty \frac{dq}{\pi}\frac{1}{\lambda^2+q^2}\ln\left[\frac{1-\frac{s(z-1)\hat f(q)}{1-s\hat f(q)}}{1-u}\right]\right)- \exp\left(-\lambda \int_0^\infty \frac{dq}{\pi}\frac{\ln\left[\frac{1-sz \hat f(q)}{1-sz}\right]}{\lambda^2+q^2}\right)\Bigg]\,.\label{eq:lambda2}
\end{align}
Taking the limit $\lambda\to 0$ and simplifying we get (\ref{eq:lambda00}).

\subsection{Analysis of $I_2(s)$} \label{sub_B2}
To analyse $I_2(s)$ in (\ref{eq:I2HI}), we cannot directly use Pollaczek-Spitzer formula in (\ref{eq:HI}) as the initial position in the generating function of the survival probability is negative. Therefore, we first decompose the generating function of the survival probability over the first step as
\begin{align}
  \bar S_0(s|-a) = 1 + s \int_0^\infty dy f(a+y)\bar S_0(s,y)\,,\label{eq:decS0}
\end{align}
where $f(\eta)$ is the jump distribution. We can then replace $\bar S_0(s,y)$ by Pollaczek-Spitzer formula (\ref{eq:HI}) in (\ref{eq:decS0}) as the initial position is now positive, which gives (\ref{eq:I2sf}). The limit $s\to 1$ of $I_2(s)$ can be obtained by rescaling $q$ by $\lambda$ and, $a$, $y$ and $\lambda$ by $(1-s)^{\frac{1}{\mu}}$, which gives
\begin{align}
   I_2(s)\sim \frac{1}{(1-s)^{\frac{1}{\mu}-\frac{1}{2}}}\int_{-\infty}^0 da \int_0^\infty dy \frac{c_\mu}{(y-a)^{1+\mu}}\int_{\gamma_B} \frac{d\lambda e^{\lambda a}}{2\pi i \lambda}\exp\left(- \int_0^\infty \frac{dq}{\pi} \frac{\ln\left[(1-s)(1+\lambda^\mu q^\mu)\right]}{1+q^2}\right)\,,\quad s\to 1\,,\label{eq:I2sfl}
\end{align}
where we used the expansion $f(\eta)\sim \frac{c_\mu}{\eta^{1+\mu}}$ for $\eta\to \infty$, where $c_\mu=\Gamma(1+\mu)\sin(\pi \mu/2)/\pi$, which can be obtained by inverting the small $q$ expression of the Fourier transform $\hat f(q)$ in (\ref{Fourier}).
Using that $\int_0^\infty \frac{dq}{\pi}\frac{1}{1+q^2}=\frac{1}{2}$, it gives
\begin{align}
   I_2(s)\sim \frac{1}{(1-s)^\frac{1}{\mu}}\int_{-\infty}^0 da \int_0^\infty dy \frac{c_\mu}{(y-a)^{1+\mu}}\int_{\gamma_B} \frac{d\lambda e^{\lambda a}}{2\pi i \lambda}\exp\left(- \int_0^\infty \frac{dq}{\pi} \frac{\ln\left(1+\lambda^\mu q^\mu\right)}{1+q^2}\right)\,,\quad s\to 1\,.\label{eq:I2sfl2}
\end{align}
One can easily check that the remaining integrals in (\ref{eq:I2sfl2}) are well-behaved for $\mu>1$.
\section{Alternative derivation of the expected gap}
\label{app:alt}
As in \cite{SM12}, we start from the celebrated Pollaczek-Wendel identity \cite{Pollackzek,Wendel} which states that the double generating function of the $k^\text{th}$ maximum is given by
\begin{align}
  \varphi(s,z,\rho) = \sum_{n=0}^\infty \sum_{k=0}^n s^n z^k  \mathbb{E}[ \exp\left(i \rho M_{k+1,n}\right)]= \exp\left(\sum_{n=1}^\infty \frac{s^n}{n} \mathbb{E}[ \exp\left(i\rho x_n^+\right) ]  + \frac{(sz)^n}{n} \mathbb{E}[ \exp\left(i\rho x_n^{-}\right)]\right)\,.\label{eq:phip}
\end{align}
Note that the $k^\text{th}$ maximum $M_{k,n}$ in the Pollaczek-Wendel identity is the $k^\text{th}$ maximum of the set $\{x_0,\ldots,x_n\}$, which includes the initial position of the random walk. This is different to the definition given in the main text above (\ref{def_order}), which does not include the initial position, but should not matter for the analysis of the expected stationary gap $ \mathbb{E}[\Delta_{k}]$.
As \cite{SM12} showed, this formula can be more conveniently written as 
\begin{align}
  \varphi(s,z,\rho) &= \frac{1}{\sqrt{1-s}\sqrt{1-sz}\sqrt{1-s\hat f(\rho)}\sqrt{1-sz\hat f(\rho)}} \exp\left(\frac{i}{2\pi}\text{PV}\int_{-\infty}^{\infty} \frac{dq}{q-\rho}\ln(1-s\hat f(q))\right)\nonumber\\
  &\qquad \times \exp\left(-\frac{i}{2\pi}\text{PV}\int_{-\infty}^\infty \frac{dq}{q-\rho}\ln(1-sz\hat f(q))\right)\,,\label{eq:phi2}
\end{align} 
where PV refers to the principal value of the integral. From this generating function, we can extract the generating function of the mean value $ \mathbb{E}[ M_{k,n}]$ of the $k^{\text{th}}$ maximum:
\begin{align}
  \sum_{n=0}^\infty \sum_{k=0}^n z^k  \mathbb{E}[ M_{n,k+1}] = \frac{1}{i}\partial_{\rho} \varphi(s,z,\rho)\bigg\rvert_{\rho=0}\,.\label{eq:phirho}
\end{align} 
Next, as in \cite{SM12}, we notice that (\ref{eq:phi2}) is not well suited to for an expansion close to $s=0$, and we use trick, similar to theirs but adapted to fat-tailed distributions (\ref{Fourier}), which consists in writing
\begin{align}
  \ln(1-s\hat f(q)) = \ln\left(1-s\left(1-|a q|^\mu\right)\right) + \ln\left(\frac{1-s\hat f(q)}{1-s\left(1-|q|^\mu\right)}\right)\,.\label{eq:trick}
\end{align}
 As in equation (45) in \cite{comtet_precise}, we denote
\begin{align}
 I_1(s,\rho) &=  \frac{\rho}{\pi}\int_{0}^{\infty} \frac{dq}{q^2+\rho^2}\ln\left(1-s\left(1-a^\mu q^\mu\right)\right)= \frac{1}{\pi}\int_{0}^{\infty} \frac{dq}{q^2+1}\ln\left(1-s\left(1-a^\mu \rho^\mu q^\mu\right)\right)\,,\label{eq:I1}
\end{align}
and \begin{align}
  I_2(s,\rho) &= \frac{\rho}{\pi} \int_0^\infty \frac{dq}{\rho^2+q^2} \ln\left[\frac{1-s\hat f(q)}{1-s\left(1- q^\mu\right)}\right]\,.\label{eq:I2}
\end{align}
Using the decomposition (\ref{eq:trick}) and the definitions (\ref{eq:I1}) and (\ref{eq:I2}), we find that $\varphi(s,z,\rho)$ writes
\begin{align}
  \varphi(s,z,\rho) = \frac{1}{\sqrt{1-s}\sqrt{1-sz}\sqrt{1-s\hat f(\rho)}\sqrt{1-sz\hat f(\rho)}}\exp\left(I_1(s,i\rho)+I_2(s,i\rho)-I_1(sz,i\rho)-I_2(sz,i\rho)\right)\,.\label{eq:phiI}
\end{align}
As in \cite{comtet_precise}, we take a derivative with respect to $\rho$ and evaluate at $\rho=0$, which gives
\begin{align}
 \sum_{n=0}^\infty \sum_{k=0}^n s^n z^k  \mathbb{E}[ M_{n,k+1}] &= \frac{1}{i}\partial_{\rho} \varphi(s,z,\rho)\bigg\rvert_{\rho=0} =\frac{1}{(1-s)(1-sz)}\left[\partial_\rho I_1(s,0)+\partial_\rho I_2(s,0)-\partial_\rho I_1(sz,0)-\partial_\rho I_2(sz,0) \right]\,,\label{eq:sII}
\end{align}
where we used that $I_1(s,0)=I_2(s,0)=0$. 
The derivative of $I_1(s,\rho)$ at $\rho=0$ is given by
\begin{align}
  \partial_\rho I_1(s,0) &=\lim_{\rho\to 0} \,\frac{\mu s \rho^{\mu-1}}{\pi}\int_{0}^{\infty} \frac{dq}{q^2+1}\frac{q^\mu}{\left[1-s\left(1- \rho^\mu q^\mu\right)\right]}= \frac{s^{1/\mu}}{(1-s)^{1/\mu}\sin(\frac{\pi}{\mu})}\,,\label{eq:dI1}
\end{align}
where we rescaled the integration variable by $\rho^\mu$. The derivative of $I_2(s,\rho)$ at $\rho=0$ is given by
\begin{align}
  \partial_\rho I_2(s,0)&= \frac{1}{\pi} \int_0^\infty \frac{dq}{q^2} \ln\left[\frac{1-s\hat f(q)}{1-s\left(1- q^\mu\right)}\right]\,.\label{eq:dI2}
\end{align}
Inserting the expressions (\ref{eq:dI1}) and (\ref{eq:dI2}) in (\ref{eq:sII}), we find
\begin{align}
  \sum_{n=0}^\infty \sum_{k=0}^n s^n z^k  \mathbb{E}[ M_{n,k+1}] &= \frac{1}{(1-s)(1-sz)}\Bigg( \frac{s^{1/\mu}}{(1-s)^{1/\mu}\sin(\frac{\pi}{\mu})} - \frac{(sz)^{1/\mu}}{(1-sz)^{1/\mu}\sin(\frac{\pi}{\mu})}\nonumber \\
  & + \frac{1}{\pi} \int_0^\infty \frac{dq}{q^2} \ln\left[\frac{1-s\hat f(q)}{1-s\left(1- q^\mu\right)}\right] - \frac{1}{\pi} \int_0^\infty \frac{dq}{q^2} \ln\left[\frac{1-sz\hat f(q)}{1-s\left(1- q^\mu\right)}\right]\Bigg)\,.\label{eq:dgfA}
\end{align}
Taking the limit $s\to 0$ gives
\begin{align}
  \sum_{n=0}^\infty \sum_{k=0}^n s^n z^k  \mathbb{E}[ M_{n,k+1}] &\sim \frac{1}{(1-s)^{1+1/\mu}(1-z)\sin(\frac{\pi}{\mu})}+\frac{1}{(1-s)(1-z)}\Bigg(\frac{1}{\pi} \int_0^\infty \frac{dq}{q^2} \ln\left[\frac{1-\hat f(q)}{ q^\mu}\right]\nonumber\\
   &\qquad-\frac{1}{\pi} \int_0^\infty \frac{dq}{q^2} \ln\left[\frac{1-z\hat f(q)}{1-z\left(1- q^\mu\right)}\right] - \frac{a z^{1/\mu}}{(1-z)^{1/\mu}\sin(\frac{\pi}{\mu})}\Bigg)\,,\quad s\to 1\,.\label{eq:s1avg}
\end{align}
Using Tauberian theorem, we find
\begin{align}
  \sum_{k=0}^n z^k  \mathbb{E}[ M_{n,k+1}] &\sim \frac{\mu \, \Gamma \left(1-\frac{1}{\mu }\right) n^{\frac{1}{\mu}}}{\pi(1-z) }+\frac{1}{(1-z)}\Bigg(\frac{1}{\pi} \int_0^\infty \frac{dq}{q^2} \ln\left[\frac{1-\hat f(q)}{ q^\mu}\right]-\frac{1}{\pi} \int_0^\infty \frac{dq}{q^2} \ln\left[\frac{1-z\hat f(q)}{1-z\left(1- q^\mu\right)}\right]\nonumber\\
  &\qquad- \frac{z^{1/\mu}}{(1-z)^{1/\mu}\sin(\frac{\pi}{\mu})}\Bigg)\,,\quad n\to \infty\,.\label{eq:inv1}
\end{align}
To invert the remaining generating function we use the identity
\begin{align}
  \frac{1}{\pi}\int_0^\infty \frac{dq}{q^2}\ln\left(\frac{1-\frac{z}{1+ q^\mu}}{1-z(1- q^\mu)}\right) = \frac{a}{\sin(\frac{\pi}{\mu})} \left(\frac{1}{(1-z)^{1/\mu}}-1-\frac{z^{1/\mu}}{(1-z)^{1/\mu}}\right)\,,\label{eq:id1}
\end{align}
in order to rewrite the term $\frac{z^{1/\mu}}{(1-z)^{1/\mu}\sin(\frac{\pi}{\mu})}$, which gives
\begin{align}
  \sum_{k=0}^n z^k  \mathbb{E}[ M_{n,k+1}] &\sim \frac{1}{1-z}\left( \frac{\mu \, \Gamma \left(1-\frac{1}{\mu }\right) n^{\frac{1}{\mu}}}{\pi }+\frac{1}{\pi} \int_0^\infty \frac{dq}{q^2} \ln\left[\frac{1-\hat f(q)}{  q^\mu}\right] +\frac{1}{\sin(\frac{\pi}{\mu})}\right)-\frac{1}{(1-z)^{1+1/\mu}\sin(\frac{\pi}{\mu})}\nonumber\\
   &+\frac{1}{(1-z)}\left(\frac{1}{\pi}\int_0^\infty \frac{dq}{q^2}\ln\left(1-\frac{z}{1+ q^\mu}\right)-\frac{1}{\pi} \int_0^\infty \frac{dq}{q^2} \ln\left[1-z\hat f(q)\right]\right)\,,\quad n\to \infty\,.
\end{align}
Using the following generating functions
\begin{align}
  \sum_{m=0}^\infty \frac{\Gamma\left(m+\frac{1}{\mu}\right)z^m}{\Gamma\left(\frac{1}{\mu}\right)\Gamma\left(m+1\right)} &= \frac{1}{(1-z)^{1/\mu}}\,,\quad
  \sum_{m=1}^\infty \frac{z^m}{m} = -\ln(1-z)\,,\label{eq:gen2}
\end{align}
and using the fact that a product of generating functions becomes a convolution, we find
\begin{align}
   \mathbb{E}[M_{n,k+1}] &\sim  \frac{\mu \, \Gamma \left(1-\frac{1}{\mu }\right) n^{\frac{1}{\mu}}}{\pi }+\frac{1}{\pi} \int_0^\infty \frac{dq}{q^2} \ln\left[\frac{1-\hat f(q)}{  q^\mu}\right] - \frac{1}{\pi}\,\Gamma\left(1-\frac{1}{\mu}\right)\sum_{m=1}^k \frac{\Gamma\left(m+\frac{1}{\mu}\right)}{\Gamma\left(m+1\right)}\nonumber\\
   &\quad +\frac{1}{\pi} \sum_{m=1}^k \frac{1}{m}\int_0^\infty \frac{dq}{q^2} \left[\hat f(q)^m-\frac{1}{(1+ q^\mu)^m}\right]\,.\label{eq:Mnksol}
\end{align}
The stationary gap is therefore given by
\begin{align}
   \mathbb{E}[\Delta_{k}] =  \lim_{n\to \infty} \mathbb{E}[M_{k,n}]- \mathbb{E}[M_{k+1,n}] =\frac{1}{\pi}\,\Gamma\left(1-\frac{1}{\mu}\right)\frac{\Gamma\left(k+\frac{1}{\mu}\right)}{\Gamma\left(k+1\right)}-\frac{1}{\pi} \frac{1}{k}\int_0^\infty \frac{dq}{q^2} \left[\hat f(q)^k-\frac{1}{(1+ q^\mu)^k}\right]\,.\label{eq:dksol}
\end{align}
Upon using the identity (\ref{eq:idcheck}), we recover the expression (\ref{eq:statg}) given in the introduction.

\section{Analysis of the general formula in Eq. (\ref{eq:p1zss}) in the case of a double sided exponential jump distribution}\label{App_exp}

To compute explicitly the generating function of the stationary gap distribution $\tilde P_z(\Delta) = \sum_{k=0} P_k(\Delta)$ given in Eqs. (\ref{eq:p1zss}) and (\ref{eq:barpst1}) for the double exponential jump distribution $f(\eta) = (1/2)\,e^{-|\eta|}$, it is useful to recall the remark done at the end of Section \ref{sec:3.1}. Indeed, as done there (see Eqs. (\ref{def_F})-(\ref{ptilde_F})), it is possible to express $\tilde p_z^{(1)}(\Delta)$ in (\ref{eq:barpst1}) in terms of the solution of an integral equation, similar to what was done in Eqs. (\ref{def_F})-(\ref{ptilde_F}). One finds indeed
\bea \label{p1_z}
\tilde p^{(1)}_{z}(\Delta) =  \int_{-\Delta}^0 dx_1 \bar S_*(z|x_1) \bar F_*(x_1|\Delta) \;,
\eea
where $\bar F_*(x_1|\Delta)$ satisfies the integral equation
\bea \label{int_eq}
\int_{-\Delta}^0 dx_1 \tilde E_*(z,x|x_1) \bar F_*(x_1|\Delta) = (1-z) \bar F_*(x|\Delta) + \bar S_*(z|x) \;.
\eea
In general, it is very difficult to solve this integral equation (\ref{int_eq}). However, it is possible to solve it explicitly for the double sided  
exponential distribution $f(\eta)=e^{-|\eta|}/2$, whose Fourier transform is 
\begin{align}
  \hat f(k) = \frac{1}{1+k^2}\label{eq:hatfdsj} \;.
\end{align}
In this case, the effective jump distribution (\ref{eq:effFb}) is given by
\begin{align}
  \mathrm{\hat f_s}(k) = \frac{1-s}{1-s+k^2}\,,\label{eq:effFbds}
\end{align}
which, in real space, reads $\mathrm{ f_s}(\eta)= \sqrt{1-s}\exp(-\sqrt{1-s}|\eta|)/2$.
Inserting this expression in (\ref{eq:Ss}) and (\ref{eq:Est2}), and using that in this case $g(\eta)=0$ -- see Eq. (\ref{eq:defGm}) -- we find
\begin{align}
  \bar S_*\left(z|x\right) &=1-x\sqrt{1-z} \,,\label{eq:Ssdj}\\
   \tilde E_*\left(z-1,y|x\right) 
  &=(1-z) \max(x,y)-\sqrt{1-z}\,,\label{eq:Esdsj3}
  \end{align}
where we used the identities
\begin{align}
  -\int_0^\infty \frac{dq}{\pi}\ln\left(1+\frac{1-z}{q^2}\right)&=-\sqrt{1-z}\,,\\
  -(1-z)\int_0^\infty \frac{dq}{\pi}\frac{1}{q^2}\left(\cos(q(y-x))-1\right) &=\frac{|y-x|(1-z)}{2}\,,
\end{align} 
together with the fact that the Laplace transform of the inverse Laplace transform of a function is the function itself, i.e., $\int_0^\infty dx e^{-x} \int_{\gamma_B} d\lambda\frac{e^{\lambda x}}{2\pi i} h(\lambda)=h(1)$ for any integrable function $h(\lambda)$. Inserting (\ref{eq:Ssdj}) and (\ref{eq:Esdsj3}) into the integral equation (\ref{int_eq}) gives
\begin{align}
  \int_{-\Delta}^0 dx_1 \left[(1-z) \max(x_1,x)-\sqrt{1-z}\right]\bar F_*(x_1|\Delta) = (1-z) \bar F_*(x|\Delta) + 1-x\sqrt{1-z}\,.\label{eq:inted}
\end{align}
One can now take two derivatives with respect to $x$, using $\partial^2_{x} \max(x_1,x) = \delta(x-x_1)$, to get the simple partial differential equation satisfied by $F(x)$, namely
\begin{align}
\bar F_*(x|\Delta) = \partial^2_{x}\bar F_*(x|\Delta)  \,, \quad x \in [- \Delta,0] \;.
 \label{int_eq3}
\end{align}
Therefore $\bar F_*(x|\Delta)$ is of the form
\begin{align}
\bar F_*(x|\Delta) =  A(\Delta)\,e^{x} +  B(\Delta)\,e^{-x} \;,
 \label{int_eq4}
\end{align}
where $A(\Delta)$ and $B(\Delta)$ are two unknown integration constants, i.e. independent of $x$. To determine them, we inject this solution (\ref{int_eq4}) in the original integral equation (\ref{eq:inted}), which yields a linear system of two independent equations relating $A(\Delta)$ and $B(\Delta)$. By solving this linear system, one finds
\begin{align}
A(\Delta) =\frac{\sqrt{1-z}-1}{2 \left(\cosh (\Delta )+\sqrt{1-z}
   \sinh (\Delta )-z \cosh (\Delta )\right)} \quad, \quad B(\Delta) =\frac{\sqrt{1-z}+1}{2 (z-1) \cosh (\Delta )-2 \sqrt{1-z} \sinh (\Delta )}\;.
    \label{sol_AB}
\end{align}
Substituting these expressions in the one for $\bar F_*(x|\Delta)$, we find
\begin{align}
 \bar F_*(x|\Delta)= \frac{\sqrt{1-z} \sinh (x)-\cosh (x)}{\cosh (\Delta )+\sqrt{1-z} \sinh (\Delta )-z \cosh (\Delta )}\,.\label{eq:Fss}
\end{align}
Inserting it in (\ref{p1_z}) and performing the integration, we find a fairly simple expression, namely
\begin{align}
  \tilde p_z^{(1)}(\Delta) &= \int_{-\Delta}^0 dx_1 (1-x_1 \sqrt{1-z})\,\frac{\sqrt{1-z} \sinh (x_1)-\cosh (x_1)}{\cosh (\Delta )+\sqrt{1-z} \sinh (\Delta )-z \cosh (\Delta )}\nonumber\\
  &=  - \frac{1}{\sqrt{1-z}}- \Delta + \frac{\sqrt{1-z} \sinh \Delta+\cosh \Delta}{\sqrt{1-z} \cosh \Delta +\sinh \Delta}\,.\label{eq:pz1sol}
\end{align}
On the other hand, the term $\tilde p_{z,s}^{(2)}(\Delta)$ in (\ref{eq:p2e}) in the limit $s \to 1$ reads
\begin{align} 
\tilde p_{z,s}^{(2)}(\Delta) = \mathbb{E}(\tilde \Delta_{z,s}) +  \frac{s\Delta}{1-s} &\sim \frac{1}{1-s} \left(\int_0^\infty \frac{dq}{\pi} \frac{1}{q^2} \ln\left(\frac{1-z\hat f(q)}{1-z} \right) + \Delta \right)  \\
&\sim \frac{1}{1-s}  \left( \frac{1}{\sqrt{1-z}} - 1 + \Delta \right)\,,\quad s\to 1\,,
\label{tildep2_exp}
\end{align}
where we used (\ref{eq:genstat}) for the expected gap in the limit $s\to 1$.
Finally, combining the results in (\ref{eq:pz1sol}) and (\ref{tildep2_exp}) one finds
\bea \label{ptilde_exp}
\tilde p_{z,s}(\Delta) = \tilde p_{z,s}^{(1)}(\Delta) +   \tilde p_{z,s}^{(2)}(\Delta) \underset{s \to 1}{\sim} \frac{1}{1-s}  \left(\frac{\sqrt{1-z} \sinh \Delta+\cosh \Delta}{\sqrt{1-z} \cosh \Delta +\sinh \Delta} - 1 \right) \;.
\eea 
Hence we recover the expression obtained in \cite{SM12} and \cite{BM17} [see equation (70) in that reference]. Note that there the generating function of $P_{k,n}(\Delta)$ includes the term $k=0$ (with by convention $P_{0,n} = 1$), which is not the case here: this explains the $-1$ in (\ref{ptilde_exp}), which is not present in \cite{SM12, BM17}. Note that it is also possible to compute explicitly the expressions of $\bar S_*(z|x)$ and $\tilde E_*(z,y|x)$ for the case of the jump distribution $f(\eta) \propto |\eta|e^{-|\eta|}$ (see Appendix \ref{app:ds2j}).

\section{Small $s$ limit of $\bar S_s\left(u|x\right)$ and $\bar E_s\left(u,y|x\right)$ for $x<0$ and $y<0$}\label{app:HIas}

In this Appendix, we analyse the small $s$ limit of $\bar S_s\left(u|x\right)$ and $\bar E_s\left(u,y|x\right)$, with $u=s(z-1)/(1-s)$ for $x<0$ and $y<0$.
\subsection{$\bar S_s\left(u|x\right)$}
From the Pollaczek-Spitzer formula \cite{Pollackzek}, we know that for $x>0$, the generating function of the survival probability is given by (\ref{eq:HI}). In particular, we have the Sparre Andersen result
\begin{align}
  \bar S_s(z,0) &= \frac{1}{\sqrt{1-z}}\,.
  \label{eq:SEz}
\end{align}
To obtain an expression for $x<0$, we expand the survival probability over the first jump, which reads
\begin{align}
  S_s(n,x) = \int_0^\infty dy f_s(y-x) S_s(n-1,y)\,,\label{eq:fstep}
\end{align}
where $\mathrm{f}_s(\eta)=\int_{-\infty}^\infty \frac{dq}{2\pi}e^{-iq\eta}\,\mathrm{\hat f_s}(q)$ is the effective jump distribution (\ref{eq:effFb}). The expression (\ref{eq:fstep}) states that for the random walk to survive from $x$ during $n$ steps, it must jump to $y>0$ at the first step and survive $n-1$ steps from $y$. Upon taking a generating function of (\ref{eq:fstep}) and using that by definition in (\ref{eq:Ssse}) one has $S_s(0,x)=1$ for $x\in \mathbb{R}$, the generating function of the survival probability reads
\begin{align}
  \bar S_s(z,x) &= 1 + z \int_0^\infty dx' \mathrm{f_s}(x'-x) \bar S_s(z,x')\,,
  \label{eq:SEgen}
\end{align}
which is valid for all $x\in \mathbb{R}$. We now analyse (\ref{eq:SEgen}) in the limit $s\to 1$. To do so, we found it convenient to add and subtract the following terms
\begin{align}
  \bar S_s(z,x) &= 1 +z \int_0^\infty dx' \mathrm{f_s}(x') \bar S_s(z,x')+ z \int_0^\infty dx' [\mathrm{f_s}(x'-x)-\mathrm{f_s}(x')] \bar S_s(z,x')   \,.
  \label{eq:SEgens}
\end{align}
Recognizing that the first two terms correspond to the generating function of the survival probability evaluated at $x=0$, we get
\begin{align}
  \bar S_s(z,x) &= \bar S_s(z,0) + z \int_0^\infty dx' [\mathrm{f_s}(x'-x)-\mathrm{f_s}(x')] \bar S_s(z,x')\,,\nonumber\\
   &= \frac{1}{\sqrt{1-z}} + z \int_0^\infty dx' [\mathrm{f_s}(x'-x)-\mathrm{f_s}(x')] \bar S_s(z,x')\,,
\label{eq:SEgen3}
\end{align}
where we used the Sparre Andersen result (\ref{eq:SEz}).
As $\bar S_s(z,x')$ in the integral is only evaluated for positive argument $x'>0$, we can replace it by the Pollaczek-Spitzer formula given in (\ref{eq:HI}). We find
\begin{align}
   \bar S_s(z,x) &= \frac{1}{\sqrt{1-z}} + \frac{z}{\sqrt{1-z}} \int_0^\infty dx' [\mathrm{f_s}(x'-x)-\mathrm{f_s}(x')]\int_{\gamma_B} d\lambda \frac{e^{\lambda x'}}{2\pi i\lambda} \exp\left(-\lambda \int_0^\infty \frac{dq}{\pi}\frac{\ln(1-z\, \mathrm{\hat f_s}(q))}{\lambda^2+q^2}\right) \,.
\label{eq:Sgen4b}
\end{align}
We now evaluate this expression at $u=s(z-1)/(1-s)$ in the limit $s \to 1$. It reads explicitly
\bea \label{attempt1}
  \bar S_s( u,x) &\sim \sqrt{\frac{1-s}{{1-z}}} -   \frac{\sqrt{1-z}}{(1-s)^{3/2}}  \int_0^\infty dx'[\mathrm{f_s}(x'-x)-\mathrm{f_s}(x')]\int_{\gamma_B} d\lambda \frac{e^{\lambda x'}}{2\pi i\lambda} \exp\left(-\lambda \int_0^\infty \frac{dq}{\pi}\frac{\ln(1+ \frac{(1-z)\hat f(q)}{1-s \hat f(q)})}{\lambda^2+q^2}\right) \,.
\eea 
Naively, it seems that one can take directly the limit $s \to 1$ in (\ref{attempt1}) by setting $s=1$ in the integral. This is however too naive, as it can be shown by an explicit calculation for the double exponential case. In fact, this can be seen from the fact that the expansion of the effective distribution as $s \to 1$ leads, to leading order
\bea \label{gs_small_s}
\frac{\mathrm{\hat f_s}(q)}{(1-s)} = \frac{\hat f(q)}{1-s \hat f(q)}  = \frac{\hat f(q)}{1 - \hat{f}(q)} - (1-s) \frac{\hat f(q)}{(1 - \hat f(q))^2} + O((s-1)^2) \;.
\eea
Once inserted in (\ref{attempt1}), we see that the integral of the term of order $O(1-s)$ is diverging for $q \to 0$ since $1/(1-\hat f(q))^2 \propto 1/|q|^{2\mu}$. This signals the fact that one must be careful to take this limit $s \to 1$. To analyse this limit, it is convenient to use the following trick which amounts to add and subtract $ \mathrm{\hat{f}^\mu_s}(q) = 1/(1+|q|^\mu)$ to $\mathrm{\hat{f}_s}(q)$
\bea \label{trick}
\mathrm{\hat{f}_s}(q) = \mathrm{\hat{f}^\mu_s}(q) + \mathrm{\hat{F}_s}(q) \quad {\rm where} \quad \mathrm{\hat{F}_s}(q)  = (1-s)\frac{\hat f(q)}{1-s\, \hat f(q)} - (1-s) \frac{1}{(1-s)+|q|^\mu} \;.
\eea
An interesting property of $\mathrm{\hat{F}_s}(q)$ is that its expansion for $s \to 1$, i.e. the analogue of (\ref{gs_small_s}) reads, to leading order
\bea \label{asympt_Fs}
\frac{\mathrm{\hat F_s}(q)}{1-s} =  \frac{(1+|q|^\mu)\hat f(q)-1}{(1-\hat f(q))|q|^\mu} - (1-s) \left(\frac{\hat f(q)^2}{(1-\hat f(q))^2} - \frac{1}{|q|^{2\mu}}\right)  + O((1-s)^2)\;.
\eea
Since $\hat f(q) \sim 1 - |q|^\mu$ for small $q$ we see that the divergence $\propto 1/q^{2\mu}$ is thus suppressed in the term of order $O(1-s)$ in (\ref{asympt_Fs}), while the first term goes to a constant as $q \to 0$ -- assuming that $\hat f(q)  = 1 - q^\mu + O(q^{2\mu})$ as $q \to 0$. In the limit $s\to 1$, we find that the Fourier transform of the effective distribution (\ref{trick}) takes the limiting form
\begin{subequations}
\begin{align}
 \mathrm{\hat f_s}(q) &\sim \mathcal{\hat F}_\mu\left(\frac{q}{(1-s)^\frac{1}{\mu}}\right)+(1-s)\hat g(q)\,,\quad s\to 1\,,
\end{align}
which in real space reads
\begin{align}
   \mathrm{f_s}(x) = \int_{-\infty}^\infty \frac{dq e^{-iqx}}{2\pi} \mathrm{\hat f_s}(q) &\sim (1-s)^{\frac{1}{\mu}} \mathcal{F}_\mu\left((1-s)^\frac{1}{\mu}x\right)+(1-s)g\left(x\right)\,,\quad s\to 1\,,
\end{align}
  \label{eq:scalfs}
\end{subequations}
where 
\begin{align}
  \mathcal{\hat F}_\mu(u)&= \frac{1}{1+u^\mu}\,,\label{eq:mathf}\\
  \hat g(q)&=\frac{(1+|q|^\mu)\hat f(q)-1}{(1-\hat f(q))|q|^\mu}\,,\label{eq:math2}
\end{align}
and $\mathcal{ F}_\mu(x)=\int_{-\infty}^\infty \frac{dq }{2\pi}e^{-ixu}\mathcal{\hat F}(u)$, and $g(x)=\int_{-\infty}^\infty \frac{dq }{2\pi}e^{-ixq}\hat g(q)$. Upon inserting the scaling (\ref{eq:scalfs}) in (\ref{attempt1}), and rescaling all integration variables by $(1-s)^\frac{1}{\mu}$, we get upon letting $s\to 1$,
\begin{align}
   \bar S_s\left(u|x\right) &\sim \frac{\sqrt{1-s}}{\sqrt{1-z}} - \frac{\sqrt{1-z}}{\sqrt{1-s}}\int_0^\infty dx'\left[\mathcal{F}_\mu\left(x'-(1-s)^\frac{1}{\mu}x\right)-\mathcal{F}_\mu\left(x'\right)\right] \int_{\gamma_B} d\lambda \frac{e^{\lambda x'}}{2\pi i\lambda} \exp\left(-\lambda \int_0^\infty \frac{dq}{\pi}\frac{\ln(1+\frac{1-z}{1-s}\, \frac{1}{1+q^\mu})}{\lambda^2+q^2}\right) \nonumber\\
   &-\sqrt{(1-z)(1-s)}\int_0^\infty dx'\left[g\left(x'-x\right)-g\left(x'\right)\right]\int_{\gamma_B} d\lambda \frac{e^{\lambda x'}}{2\pi i\lambda} \exp\left(-\lambda \int_0^\infty \frac{dq}{\pi}\frac{\ln(1+ \frac{(1-z)\hat f(q)}{1- \hat f(q)})}{\lambda^2+q^2}\right)\,.
\label{eq:Sgen41}
\end{align}
Further letting $s\to 1$ in the logarithm, and using that $\lambda \int_0^\infty \frac{dq}{\pi} \frac{1}{\lambda^2+q^2}=\frac{1}{2}$, we find
\begin{align}
   \bar S_s\left(u|x\right) &\sim \frac{\sqrt{1-s}}{\sqrt{1-z}} +(1-s)^\frac{1}{\mu}x\int_0^\infty dx'\mathcal{F}'_\mu\left(x'\right) \int_{\gamma_B} d\lambda \frac{e^{\lambda x'}}{2\pi i\lambda} \exp\left(\lambda \int_0^\infty \frac{dq}{\pi}\frac{\ln( 1+q^\mu)}{\lambda^2+q^2}\right) \nonumber\\
   &-\sqrt{1-s}\sqrt{1-z}\int_0^\infty dx'\left[g\left(x'-x\right)-g\left(x'\right)\right]\int_{\gamma_B} d\lambda \frac{e^{\lambda x'}}{2\pi i\lambda} \exp\left(-\lambda \int_0^\infty \frac{dq}{\pi}\frac{\ln(1+ \frac{(1-z)\hat f(q)}{1- \hat f(q)})}{\lambda^2+q^2}\right)\,,
\label{eq:Sgen42}
\end{align}
where $\mathcal{F}'_\mu$ is the derivative of $\mathcal{F}_\mu$. Keeping only the highest order terms in (\ref{eq:Sgen42}), which means the first and the last terms for $\mu<2$ and all of them for $\mu=2$, we find the expression (\ref{eq:SEbars1}) given in the main text. For $\mu=2$, we used that 
  \begin{align}
  \int_0^\infty dx'\mathcal{F}_2'\left(x'\right) \int_{\gamma_B} d\lambda \frac{e^{\lambda x'}}{2\pi i\lambda} \exp\left(\lambda \int_0^\infty \frac{dq}{\pi}\frac{\ln( 1+q^2)}{\lambda^2+q^2}\right) &= -1\,,\label{eq:phiN}
  \end{align}
  which can be easily shown by noting that $\mathcal{F}_2(x) = e^{-|x|}/2$ and using the fact that the Laplace transform of the inverse Laplace transform of a function is the function itself, i.e. that $\int_0^\infty dx e^{-x} \int_{\gamma_B} d\lambda\frac{e^{\lambda x}}{2\pi i} h(\lambda)=h(1)$ for a function $h$.

\subsection{$\tilde E_s\left(u,y|x\right)$}
From the Pollaczek-Spitzer formula we know that for $x>0$ and $y>0$, the generating function $\bar E_s(z,y|x)$ is given by
\begin{align}
 \bar E_s(z,y|x) &= \sum_{n=0}^\infty z^n E_s(n,y|x) = \int_{\gamma_B} d\lambda d\lambda' \frac{e^{\lambda x}}{2\pi i} \frac{e^{\lambda' y}}{2\pi i }\frac{1}{\lambda+\lambda'} \exp\left(-\lambda \int_0^\infty \frac{dq}{\pi}\frac{\ln(1-z\, \mathrm{\hat f_s}(q))}{\lambda^2+q^2}-\lambda' \int_0^\infty \frac{dq}{\pi}\frac{\ln(1-z\, \mathrm{\hat f_s}(q))}{\lambda'^2+q^2}\right)\,.\label{eq:Ebarxyp}
 \end{align}
Note that in our case, we are interested in $\tilde E_s(z,y|x)=\sum_{n=1}^\infty z^n E_s(n,y|x)$, where the index starts from $1$ (hence the different notations $\tilde E_s(z,y|x)$ and $\bar E_s(z,y|x)$). We therefore need to remove the term $n=0$ in the sum in (\ref{eq:Ebarxyp}), which gives
 \begin{align}
   \tilde E_s(z,y|x) = \bar E_s(z,y|x) -\delta(y-x)\,,\label{eq:relEbEt}
 \end{align}
 where we used that $E_s(0,y|x)=\delta(y-x)$.
 In particular for $x=y=0$, we have (see Appendix \ref{app:E00})
 \begin{align}
  \tilde E_s(z,0|0)= -\int_{0}^\infty \frac{dq}{\pi}\ln\left(1-z \,\mathrm{\hat f}_s(q)\right)\,.\label{eq:barE00s}
\end{align}
 To obtain an expression for $x<0$ and $y<0$, similarly to the previous section, we expand the excursion over the first jump and last jump
 \begin{align}
   E_s(n,y|x) =  \int_0^\infty dx'dy' \mathrm{f_s}(x'-x)  E_s(n-2,y'|x')  \mathrm{f_s}(y-y')\,,\label{eq:Esjump}
 \end{align}
 where $\mathrm{f}_s(\eta)=\int_{-\infty}^\infty \frac{dq}{2\pi}e^{-iq\eta}\,\mathrm{\hat f_s}(q)$ is the effective jump distribution (\ref{eq:effFb}). The expression (\ref{eq:Esjump}) states that for the random walk to make an excursion from $x$ to $y$ during $n$ steps, it must perform a first jump to $x'>0$ at the first step, make an excursion of $n-2$ steps to $y'$, then jump from $y'$ to $y$. Upon taking a generating function of (\ref{eq:Esjump}) and using that by definition in (\ref{eq:Esse}) $E_s(0,y|x)=\delta(y-x)$ and $E_s(1,y|x)=f_s(y-x)$ for $x,y\in \mathbb{R}$, we get that the generating function of the survival probability reads
 \begin{align}
  \bar E_s(z,y|x) &= \sum_{n=0}^\infty z^n E_s(n,y|x)= \delta(y-x) + z f_s(y-x) + z^2 \int_0^\infty dx'dy' \mathrm{f_s}(x'-x) \bar E_s(z,y'|x')  \mathrm{f_s}(y-y')\,.\label{eq:Ebarxyn0}
  \end{align}
  Inserting this in \eqref{eq:relEbEt}, we get
  \begin{align}
    \tilde E_s(z,y|x) =z f_s(y-x) + z^2 \int_0^\infty dx'dy' \mathrm{f_s}(x'-x) \bar E_s(z,y'|x')  \mathrm{f_s}(y-y')\,.\label{eq:Ebarxyn}
  \end{align}
  We now analyse the small $s$ limit of (\ref{eq:Ebarxyn}).
As in the previous section, we find it convenient to add and subtract the following terms
   \begin{align}
  \tilde E_s(z,y|x) &= z [f_s(y-x)-f_s(0)] + z^2 \int_0^\infty dx'dy' [\mathrm{f_s}(x'-x)\mathrm{f_s}(y-y')-\mathrm{f_s}(x')\mathrm{f_s}(-y')] \bar E_s(z,y'|x') \nonumber\\
  &+z f_s(0)-z^2 \int_0^\infty dx'dy' \mathrm{f_s}(x') \bar E_s(z,y'|x')  \mathrm{f_s}(-y')\,.\label{eq:Ebarxyn2}
  \end{align}
  Using (\ref{eq:Ebarxyn}), we recognize that the last line is the generating function evaluated at $x=y=0$, namely
  \begin{align}
      \tilde E_s(z,y|x) &= z [f_s(y-x)-f_s(0)] + z^2 \int_0^\infty dx'dy'[\mathrm{f_s}(x'-x)\mathrm{f_s}(y-y')-\mathrm{f_s}(x')\mathrm{f_s}(-y')] \bar E_s(z,y'|x') +  \tilde E_s(z,0|0)\nonumber\\
      &= z [f_s(y-x)-f_s(0)] + z^2 \int_0^\infty dx'dy'[\mathrm{f_s}(x'-x)\mathrm{f_s}(y-y')-\mathrm{f_s}(x')\mathrm{f_s}(-y')] \bar E_s(z,y'|x') -\int_{0}^\infty \frac{dq}{\pi}\ln\left(1-z \,\mathrm{\hat f}_s(q)\right) \,,\label{eq:barxyn3}
  \end{align}
  where we used the expression of $\tilde E_s(z,0|0)$ in (\ref{eq:barE00s}).
  Upon replacing $z$ by $u=s(z-1)/(1-s)$ and using the same trick as explained before in (\ref{trick}) and using the expansion (\ref{eq:scalfs}), one finds in the limit $s\to 1$ that
   \begin{align}
 \tilde E_s\left(u,0|0\right) &\sim   -\frac{1-z}{1-s} [f_s(y-x)-f_s(0)]\nonumber\\
&-(1-z)(1-s)^{\frac{2}{\mu}-1}(x+y)\int_0^\infty dx'dy' \mathcal{F}'_\mu\left(x'\right)\mathcal{F}_\mu\left(y'\right)\nonumber \\
& \times \int_{\gamma_B} d\lambda d\lambda' \frac{e^{\lambda x'+\lambda' y'}}{(2\pi i)^2 (\lambda+\lambda')}\exp\left(\lambda \int_0^\infty \frac{dq}{\pi}\frac{\ln( 1+q^\mu)}{\lambda^2+q^2}\right)\exp\left(\lambda' \int_0^\infty \frac{dq}{\pi}\frac{\ln( 1+q^\mu)}{\lambda'^2+q^2}\right)\nonumber\\
& + (1-z)^{3/2}(1-s)^{\frac{1}{\mu}-\frac{1}{2}} \int_0^\infty dx' dy' \left[{\cal F}_\mu(x') g(y-y') - {\cal F}_{\mu}(x') g(-y') \right]  \nonumber\\
& \times \int_{\gamma_B} d\lambda d\lambda' \frac{e^{\lambda x'+\lambda' y'}}{(2\pi i)^2 \lambda'}\exp\left(\lambda \int_0^\infty \frac{dq}{\pi}\frac{\ln( 1+q^\mu)}{\lambda^2+q^2}\right)\phi(\lambda',z)\nonumber\\
& + (1-z)^{3/2}(1-s)^{\frac{1}{\mu}-\frac{1}{2}} \int_0^\infty dx' dy' \left[g(x'-x){\cal F}_{\mu}(-y') - g(x'){\cal F}_{\mu}(-y') \right] \nonumber\\
& \times \int_{\gamma_B} d\lambda d\lambda' \frac{e^{\lambda x'+\lambda' y'}}{(2\pi i)^2 \lambda}\phi(\lambda,z)\exp\left(\lambda' \int_0^\infty \frac{dq}{\pi}\frac{\ln( 1+q^\mu)}{\lambda'^2+q^2}\right)\nonumber\\
 & + (1-z)^2 \int_0^\infty dx' dy' \left[g(x'-x) g(y-y') - g(x') g(-y') \right]  \int_{\gamma_B} d\lambda d\lambda' \frac{e^{\lambda x'+\lambda' y'}}{(2\pi i)^2 (\lambda+\lambda')}\phi(\lambda,z)\phi(\lambda',z) \nonumber \\
       & -\int_{0}^\infty \frac{dq}{\pi}\ln\left(1+ \frac{(1-z) \hat f(q)}{1-\hat f(q)}\right)\,. \label{eq:Es22}
\end{align}
Next, we note that in the limit $s\to 1$ with $x=O(1)$ and $y=O(1)$, the first line becomes
\begin{align}
  -\frac{1-z}{1-s} [f_s(y-x)-f_s(0)] \sim -(1-z)\int_0^\infty \frac{dq}{\pi}\frac{\hat f(q)}{1-\hat f(q)}[\cos(q(y-x))-1]\,,\quad s\to 1\,.\label{eq:fslimi}
\end{align}
Finally, keeping only the highest order terms in the limit $s\to 1$, which corresponds to all the lines in (\ref{eq:Es22}) for $\mu=2$ and all the lines except the terms of order $O[(1-s)^{\frac{1}{\mu}-\frac{1}{2}}]$,
we recover the expression (\ref{eq:SEbars1}) given in the main text. In doing so, we used that, for $\mu=2$, we have
\begin{align}
    \int_0^\infty dx'dy' \left[\mathcal{F'}_2\left(x'\right)\mathcal{F}_2\left(y'\right)\right]\int_{\gamma_B} d\lambda d\lambda' \frac{e^{\lambda x'+\lambda' y'}}{(2\pi i)^2 (\lambda+\lambda')}\exp\left(\lambda \int_0^\infty \frac{dq}{\pi}\frac{\ln( 1+q^2)}{\lambda^2+q^2}\right)\exp\left(\lambda' \int_0^\infty \frac{dq}{\pi}\frac{\ln( 1+q^2)}{\lambda'^2+q^2}\right)&=-\frac{1}{2}\,,\\
   \int_0^\infty dx' \mathcal{F}_2\left(x'\right)\int_{\gamma_B} d\lambda \frac{e^{\lambda x'}}{(2\pi i)^2}\exp\left(\lambda \int_0^\infty \frac{dq}{\pi}\frac{\ln( 1+q^2)}{\lambda^2+q^2}\right)&=1\,,\label{eq:Est2b}
\end{align}
 which can be easily shown by noting that $\mathcal{F}_2(x) = e^{-|x|}/2$ and using the fact that the Laplace transform of the inverse Laplace transform of a function is the function itself, i.e. that $\int_0^\infty dx e^{-x} \int_{\gamma_B} d\lambda\frac{e^{\lambda x}}{2\pi i} h(\lambda)=h(1)$ for a function $h$.

  \subsection{$\tilde E_s(z,0|0)$}
\label{app:E00}
We take the limit $\lambda\to\infty$ and $\lambda'\to \infty$ in (\ref{eq:Ebarxyp}), which gives
\begin{align}
   \bar E_s(z,0|0) = \int_{\gamma_B} d\lambda d\lambda' \frac{e^{\lambda x}}{2\pi i} \frac{e^{\lambda' y}}{2\pi i }\frac{1}{\lambda+\lambda'}\left(1 -\left(\frac{1}{\lambda}+\frac{1}{\lambda'}\right)\int_0^\infty \frac{dq}{\pi} \ln(1-z\,\mathrm{\hat f}_s(q))\right)\,.\label{eq:llp0}
\end{align}
Inverting the Laplace transforms with respect to $\lambda$ and $\lambda'$ and inserting it in (\ref{eq:relEbEt}), we recover (\ref{eq:barE00s}).

\section{The special case of the distribution $f(\eta) = \frac{3}{2} |\eta|\,e^{-\sqrt{3}|\eta|}$}
\label{app:ds2j}
In this appendix, we focus on the special case of 
\begin{align}
f(\eta) = \frac{3}{2} |\eta|\,e^{-\sqrt{3}|\eta|}\,,\label{eq:feds2}
\end{align}
for which the Fourier transform is given by
\begin{align}
\hat f(q)=\frac{3 \left(3-q^2\right)}{\left(q^2+3\right)^2}\,.\label{eq:fedsd2}
\end{align} In this case, $\hat g(q)$, which is defined in (\ref{eq:math2}), reads
\begin{align}
  \hat g(q)=-\frac{4}{q^2+9}\,,\label{eq:getadef}
\end{align}
so that 
\begin{align}
  g(\eta) = -\frac{2}{3} e^{-3|\eta|}\,.\label{eq:getadfe}
\end{align} 
Inserting this expression (\ref{eq:getadfe}) in (\ref{eq:Ss}) and (\ref{eq:Est2}), we find
\begin{align}
S_*(z|x) &=  1 - x \sqrt{1-z}  + \frac{2}{3}(z-1) (1-e^{3x}) \frac{\phi(\lambda=3,z)}{3}   \quad \;,\\
E_*(z,y|x) &= -\int_{0}^\infty \frac{dq}{\pi}\ln\left(1+(1-z)\left(\frac{1}{q^2}-\frac{4}{q^2+9}\right)\right) - \frac{(1-z)}{6} \left(4 - 4e^{-3|x-y|} - 3 |x-y| \right)+ \frac{(1-z)}{2}(x+y)  \\
&+ \frac{2(1-z)^2}{27} \phi^2(3,z) (e^{3x+3y}-1) + \frac{2}{9}(1-z)^{3/2} \phi(3,z) \left(2- e^{3x}- e^{3y}\right) \,,
\label{Star_modx}
\end{align}
where the function $\phi(\lambda, z)$ is given in (\ref{eq:defPhim}) with $\hat f(k) = 3(3-k^2)/(k^2+3)^2$ and where we used the fact that the Laplace transform of the inverse Laplace transform of a function is the function itself, i.e. that $\int_0^\infty dx e^{-x} \int_{\gamma_B} d\lambda\frac{e^{\lambda x}}{2\pi i} h(\lambda)=h(1)$ for a function $h$, as well as 
\begin{align}
 \int_0^\infty \frac{dq}{\pi}\left(\frac{1}{q^2}-\frac{4}{q^2+9}\right)[\cos(q(y-x))-1] =  \frac{1}{6} \left(4 - 4e^{-3|x-y|} - 3 |x-y| \right)\,.\label{eq:intdsf2}
\end{align}In particular, the function $\phi(\lambda,z)$ admits the small $z$ expansion
\bea \label{phi_smallz}
\frac{1}{3}\phi(\lambda=3,z) = (2 - \sqrt{3}) + \frac{z}{4}(2 - \sqrt{3})^2 + O(z^2) \;,
\eea
from which one obtains the small $z$ expansion of $S_*(z|x)$ as
\begin{align}
S_*(z|x) &= \frac{2\sqrt{3}-1}{3} - x + \frac{4-2 \sqrt{3}}{3}\,e^{3x}+ \frac{z}{6} \left( 3x+1-e^{3x} \right)  + O(z^2) \,.
\label{Star_modx_smallz}
\end{align}
Quite remarkably, one observes from (\ref{Star_modx_smallz}), that $S_*(z=0,x)$ coincides exactly with the function $\sqrt{\pi}\, U_0(-x)$ defined
in \cite{MMS17} -- see (77). Note that this also holds for the double exponential case in (\ref{eq:Ssdj}). Inserting the small $z$ expansion (\ref{phi_smallz}) in (\ref{Star_modx}) one can similarly obtain the small $z$ expansion of $E_*(z,y|x)$. These expansions, up to order $O(z)$, can eventually be used in Eq. \eqref{eq:barpst1} to obtain the small $\Delta$ expansion of the distribution of the first gap for this jump distribution $f(\eta) = \frac{3}{2} |\eta|\,e^{-\sqrt{3}|\eta|}$. We have carefully checked that this small $\Delta$ expansion exactly coincides (up to order $O(\Delta^2)$) with the small $\Delta$ expansion of the exact result for the distribution of the first gap, obtained from a completely different method in Refs. \cite{MMS13, MMS14}.

\section{Scaling function of the condensed phase}\label{app:scalC}
\begin{figure}[t]
  \begin{center}
    \includegraphics[width=0.5\textwidth]{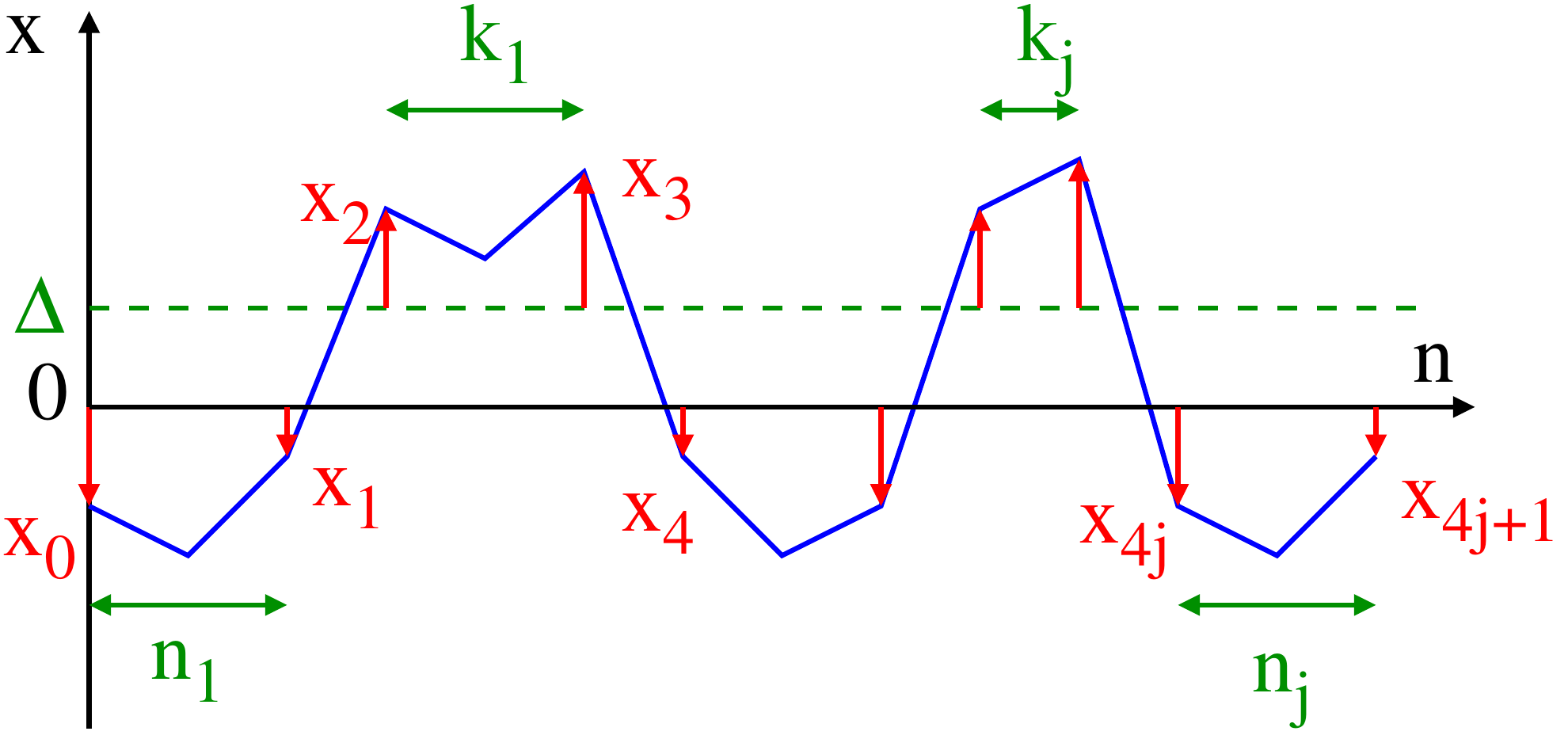}
    \caption{In the limit of $n\to \infty$ with $k$ fixed, the trajectories that contribute to the gap probability $\int_{-\infty}^\infty da\,p_{k,n}(a,\Delta)$ are the trajectories that jump an even number of times $2j$ over the gap. We parametrise these trajectories by letting the random walk start from a distance $x_0$ from the gap, then propagates to a point located a distance $x_1$ during $n_1$ steps while remaining below the gap, after which it jumps over the gap to a point located a distance $x_2$ from the gap, then it propagates to $x_3$ during $k_1$ steps while remaining above the gap, and so on, so that $k=k_1+\ldots+k_j+j$ and that $n-k=n_1+\ldots+n_j+j$ (where the last term $j$ accounts for the first jump in each sequences). We then sum over all possible values of these variables, which corresponds to summing over all possible locations and configurations of the gap.}
    \label{fig:brute}
  \end{center}
\end{figure}
In this Appendix, we are interested in the stationary probability distribution of the $k^\text{th}$ gap $P_k(\Delta)$ in the limit of large gaps $\Delta$. To study this limit, we have found convenient to leave aside ``Spitzer's idea'' for a moment, and to write out explicitly the random walk path contributions. We argue that all the contributions can be written as (see figure \ref{fig:brute}):
\begin{align}
 \int_{-\infty}^\infty da p_{k,n}(a,\Delta) &= \sum_{j=1}^\infty\sum_{k_1,\ldots,k_j=0}^\infty \sum_{n_1,\ldots,n_j=0}^\infty  \delta(k_1+\ldots +k_{j}+j-k)\delta(n_1+\ldots +n_{j}+j-[n-k])\nonumber\\
  &\int_0^\infty dx_0 \ldots dx_{4j+1}  E_0(n_1,x_1|x_0)f(x_2+x_1+\Delta)E_0(k_1,x_3|x_2)f(x_3+x_4+\Delta)\nonumber\\
  &\quad E_0(n_2,x_5|x_4)f(x_5+x_6+\Delta)E_0(k_2,x_7|x_6)f(x_7+x_8+\Delta)
 \nonumber \\&\quad \ldots  \nonumber\\
  &\quad E_0(n_{j-1},x_{4j-3}|x_{4j-4})f(x_{4j-2}+x_{4j-3}+\Delta)E_0(x_{4j-1}|x_{4j-2},k_j)f(x_{4j-1}+x_{4j}+\Delta)\nonumber\\
  &\quad E_0(n_j,x_{4j+1}|x_{4j})\,, \label{eq:pnub}
\end{align}
where $E_0(n,y|x)$ is the propagator of the random walk starting from $x>0$ and reaching $y>0$ after $n$ steps while remaining above the origin. It is given by the Pollaczek-Spitzer formula in (\ref{eq:Ebarxyp}) with $s=0$. 
Using the following expansions
   \begin{align}
   f(\eta) &\sim c_\mu\,\eta^{-1-\mu}\,,\quad \eta \to \infty\,,\label{eq:assf}\\
   \int_0^\infty dy E_0\left(n,y|x\right) &\sim \frac{1}{\sqrt{n}}  U_\mu(x)\,,\quad x=O(1)\label{eq:scalpp1}\,,\\
  \sum_{n=0}^\infty E_0\left(n,y|x\right) &\sim V_\mu(x,y)\,,\quad x=O(1)\,, y=O(1)\,,\label{eq:assPs}\\
 E_0\left(k,y|x\right) &\sim \frac{1}{k^\frac{1}{\mu}} \mathcal{J}_\mu \left(u=\frac{y}{k^\frac{1}{\mu}},v=\frac{x}{k^\frac{1}{\mu}}\right)\,,\quad x=O(k^\frac{1}{\mu})\,, y=O(k^\frac{1}{\mu})\,,\label{eq:scalpp2}
\end{align}
where 
\begin{align}
U_\mu(x) &= \int_{\gamma_B} \frac{d\lambda e^{\lambda x}}{2\pi i} \frac{1}{\lambda \sqrt{\pi}}\exp\left[-\frac{\lambda}{\pi}\int_0^\infty \frac{dq}{\lambda^2+q^2}\ln\left(1-\hat f(q)\right)\right]\,,\\
V_\mu(x,y)  &= \int_{\gamma_B} \frac{d\lambda e^{\lambda x}}{2\pi i}\int_{\gamma_B} \frac{d\lambda' e^{\lambda' y}}{2\pi i} \frac{1}{\lambda +\lambda'}\exp\left[-\frac{\lambda}{\pi}\int_0^\infty \frac{dq}{\lambda^2+q^2}\ln\left(1-\hat f(q)\right)-\frac{\lambda'}{\pi}\int_0^\infty \frac{dq}{\lambda'^2+q^2}\ln\left(1-\hat f(q)\right)\right]\,,\\
  \mathcal{J}_\mu(u,v) &= \int_{\gamma_B} \frac{ds e^s}{2\pi i}\int_{\gamma_B} \frac{d\lambda e^{\lambda u}}{2\pi i} \int_{\gamma_B} \frac{d\lambda'e^{\lambda' v}}{2\pi i} \frac{1}{s(\lambda+\lambda')} \exp\left[-\frac{1}{\pi}\int_0^\infty dq\left( \frac{\ln\left(1+\frac{\lambda^\mu q^\mu}{s}\right)}{1+q^2}+\frac{\ln\left(1+\frac{\lambda'^\mu q^\mu}{s}\right)}{1+q^2}\right)\right]\,,\label{eq:scal}
\end{align}
and $c_\mu=\Gamma(1+\mu)\sin(\pi\mu/2)/\pi$.
we find that (\ref{eq:pnub}) becomes $\lim_{n\to\infty}\int_{-\infty}^\infty da p_{k,n}(a,\Delta)= \int_{-\infty}^\infty da p_{k}(a,\Delta)$ where
\begin{align}
 \int_{-\infty}^\infty da p_{k}(a,\Delta) &\sim \pi\sum_{j=1}^\infty\sum_{k_1,\ldots,k_j=0}^\infty  \delta(k_1+\ldots +k_{j}+j-k)\nonumber\\
  &\int_0^\infty  dx_1 \ldots dx_{4j} U_\mu(x_1)\frac{c_\mu}{(x_1+x_2+\Delta)^{1+\mu}}\frac{1}{k_1^\frac{1}{\mu}}\mathcal{J}_\mu\left(\frac{x_2}{k_1^\frac{1}{\mu}},\frac{x_3}{k_1^\frac{1}{\mu}}\right)\frac{c_\mu}{(x_3+x_4+\Delta)^{1+\mu}}\nonumber\\
  &\quad V_\mu(x_{5},x_{4})\frac{c_\mu}{(x_{5}+x_{6}+\Delta)^{1+\mu}}\frac{1}{k_2^\frac{1}{\mu}}\mathcal{J}_\mu\left(\frac{x_{6}}{k_2^\frac{1}{\mu}},\frac{x_{7}}{k_2^\frac{1}{\mu}}\right)\frac{c_\mu}{(x_{7}+x_{8}+\Delta)^{1+\mu}}\nonumber\\
  &\quad\ldots \nonumber\\
  &\quad V_\mu(x_{4j-4},x_{4j-3})\frac{c_\mu}{(x_{4j-2}+x_{4j-3}+\Delta)^{1+\mu}}\frac{1}{k_j^\frac{1}{\mu}}\mathcal{J}_\mu\left(\frac{x_{4j-1}}{k_j^\frac{1}{\mu}},\frac{x_{4j-2}}{k_j^\frac{1}{\mu}}\right)\frac{c_\mu}{(x_{4j-1}+x_{4j}+\Delta)^{1+\mu}}\nonumber\\
  &\quad  U_\mu(x_{4j})\,,\quad k\to \infty\,, \label{eq:pnub2}
\end{align}
where we used that $\sum_{j=1}^{n-1} j^{-1/2}(n-j)^{-1/2}\sim \pi$ for $n\to \infty$. Rescaling all the integration variables by $\Delta$ and using the following expansions
\begin{align}
  U_\mu(x) &\sim A_\mu x^{\frac{\mu}{2}}\,,\quad x\to\infty\,, \label{eq:expUm}\\
  V_\mu(\alpha\, x,\alpha\, y)&\sim  C_\mu\,\alpha^{\mu-1} v_\mu(x,y)\,,\quad \alpha \to \infty\,, \label{eq:expVm}
\end{align}
where $A_\mu=1/[\sqrt{\pi}\,\Gamma(1+\mu/2)]$, $C_\mu=1/\Gamma(\mu/2)^2$ and $v_\mu(x,y)= \int_0^{\min(x,y)}dz (y-z)^{\frac{\mu}{2}-1}(x-z)^{\frac{\mu}{2}-1}$,
gives
\begin{align}
 \int_{-\infty}^\infty da p_{k}(a,\Delta) &\sim \pi\sum_{j=1}^\infty\sum_{k_1,\ldots,k_j=0}^\infty   \delta(k_1+\ldots +k_{j}+j-k)\,\frac{\Delta^{j+1-j\mu}}{k_1^{\frac{1}{\mu}}\ldots k_j^{\frac{1}{\mu}}}\,A_\mu^2 c_\mu^{2j} C_\mu^{j-1}\nonumber\\
  &\int_0^\infty  dx_1 \ldots dx_{4j} x_1^\frac{\mu}{2}\frac{1}{(x_1+x_2+1)^{1+\mu}}\mathcal{J}_\mu\left(\frac{\Delta x_2}{k_1^\frac{1}{\mu}},\frac{\Delta x_3}{k_1^\frac{1}{\mu}}\right)\frac{1}{(x_3+x_4+1)^{1+\mu}}\nonumber\\
  &\quad v_\mu(x_{5},x_{4})\frac{1}{(x_{5}+x_{6}+1)^{1+\mu}}\mathcal{J}_\mu\left(\frac{ \Delta x_{6}}{k_2^\frac{1}{\mu}},\frac{\Delta x_{7}}{k_2^\frac{1}{\mu}}\right)\frac{1}{(x_{7}+x_{8}+1)^{1+\mu}}\nonumber\\
  &\quad\ldots \nonumber\\
  &\quad v_\mu(x_{4j-4},x_{4j-3})\frac{1}{(x_{4j-2}+x_{4j-3}+1)^{1+\mu}}\mathcal{J}_\mu\left(\frac{\Delta x_{4j-1}}{k_j^\frac{1}{\mu}},\frac{\Delta x_{4j-2}}{k_j^\frac{1}{\mu}}\right)\frac{1}{(x_{4j-1}+x_{4j}+1)^{1+\mu}}\nonumber\\
  &  x_{4j}^\frac{\mu}{2}\,,\quad k\to \infty\,. \label{eq:pnub3}
\end{align}
 Integrating over $x_1$ and $x_{4j}$ gives
\begin{align}
 \int_{-\infty}^\infty da p_{k}(a,\Delta) &\sim \pi\sum_{j=1}^\infty\sum_{k_1,\ldots,k_j=0}^\infty   \delta(k_1+\ldots +k_{j}+j-k)\,\frac{\Delta^{j+1-j\mu}}{k_1^{\frac{1}{\mu}}\ldots k_j^{\frac{1}{\mu}}}\,A_\mu^2 c_\mu^{2j} C_\mu^{j-1}D_\mu^2\nonumber\\
  &\int_0^\infty  dx_2 \ldots dx_{4j-1} \frac{1}{(x_2+1)^{\frac{\mu}{2}}}\mathcal{J}_\mu\left(\frac{\Delta x_2}{k_1^\frac{1}{\mu}},\frac{\Delta x_3}{k_1^\frac{1}{\mu}}\right)\frac{1}{(x_3+x_4+1)^{1+\mu}}\nonumber\\
  &\quad v_\mu(x_{5},x_{4})\frac{1}{(x_{5}+x_{6}+1)^{1+\mu}}\mathcal{J}_\mu\left(\frac{\Delta x_{6}}{k_2^\frac{1}{\mu}},\frac{\Delta x_{7}}{k_2^\frac{1}{\mu}}\right)\frac{1}{(x_{7}+x_{8}+1)^{1+\mu}}\nonumber\\
  &\quad\ldots \nonumber\\
  &\quad v_\mu(x_{4j-4},x_{4j-3})\frac{1}{(x_{4j-2}+x_{4j-3}+1)^{1+\mu}}\mathcal{J}_\mu\left(\frac{\Delta x_{4j-2}}{k_j^\frac{1}{\mu}},\frac{\Delta x_{4j-1}}{k_j^\frac{1}{\mu}}\right)\frac{1}{(x_{4j-1}+1)^{\frac{\mu}{2}}}\,,\quad k\to \infty\,, \label{eq:pnub4}
\end{align}
where $D_\mu=\sqrt{\pi } 2^{-\mu } \Gamma \left(\frac{\mu }{2}\right)/\Gamma
   \left(\frac{\mu +1}{2}\right)$.
Formally taking a double derivative with respect to $\Delta$, we recover the scaling form (\ref{eq:Pksum}) announced in the introduction.
In the limit of large $\Delta$, we only take the term in $j=1$ in (\ref{eq:pnub4}) which gives
\begin{align}
   \int_{-\infty}^\infty da p_{k}(a,\Delta) &\sim \pi\,\frac{\Delta^{2-\mu}}{k^{\frac{1}{\mu}}}\,A_\mu^2 c_\mu^{2} D_\mu^2 \int_0^\infty  dx_2 dx_{3} \frac{1}{(x_2+1)^{\frac{\mu}{2}}}\mathcal{J}_\mu\left(\frac{\Delta x_2}{k^\frac{1}{\mu}},\frac{\Delta x_3}{k^\frac{1}{\mu}}\right)\frac{1}{(x_3+1)^{\frac{\mu}{2}}}\,,\quad \Delta \to \infty\,.\label{eq:pnub5}
\end{align}
Using that $\mathcal{J}_\mu(x_1,x_2)\to \mathcal{L}_\mu(x_2-x_1)$ for large $x_1$ and $x_2$ gives
\begin{align}
   \int_{-\infty}^\infty da p_{k}(a,\Delta) &\sim \pi\,\frac{\Delta^{2-\mu}}{k^{\frac{1}{\mu}}}\,A_\mu^2 c_\mu^{2} D_\mu^2 \int_0^\infty  dx_2 dx_{3} \frac{1}{(x_2+2)^{\frac{\mu}{2}}}\mathcal{L}_\mu\left(\frac{\Delta (x_3-x_2)}{k^\frac{1}{\mu}}\right)\frac{1}{(x_3+2)^{\frac{\mu}{2}}}\,,\quad \Delta \to \infty\,.\label{eq:pnub6}
\end{align}
Changing coordinates $u=(x_3-x_2)/\sqrt{2}$ and $v=(x_3+x_2)/\sqrt{2}$, and rescaling $u$ by $\Delta k^\frac{1}{\mu}$ it gives
\begin{align}
   \int_{-\infty}^\infty da p_{k}(a,\Delta) &\sim \pi\,\Delta^{1-\mu}\,A_\mu^2 c_\mu^{2} D_\mu^2 \int_0^\infty dv\int_{-\infty}^\infty du \frac{1}{(\frac{v}{\sqrt{2}}+1)^{\mu}}\mathcal{L}_\mu\left(u\sqrt{2}\right)\,,\nonumber\\
   &\sim \pi\,\Delta^{1-\mu}\,A_\mu^2 c_\mu^{2} D_\mu^2 E_\mu \,,\quad \Delta \to \infty\,, \label{eq:pnub62}
\end{align}
where $E_\mu=1/(\mu -1)$. Inserting this result in (\ref{rel_SPDF2}), we find the large argument of $\mathcal{M}_\mu(u)$ in (\ref{eq:expMmu}).


\begin{thebibliography}{}


\bibitem{gumbel}
E.~J. Gumbel, {\it Statistics of Extremes}, Dover, (1958).

\bibitem{katz}
R.~W. Katz, M.~P.~Parlange and P.~Naveau, Adv. Water Resour. {\bf 25},
1287 (2002). 

\bibitem{embrecht}
P. Embrecht, C. Kl\"uppelberg, T. Mikosh, {\it Modelling Extremal
  Events for Insurance and Finance} (Springer), Berlin (1997).  

\bibitem{bouchaud_satya}
S.~N. Majumdar, J.-P. Bouchaud, Quantitative Finance {\bf 8}, 753 (2008).


\bibitem{bm97}
J.-P. Bouchaud, M. M\'ezard, J. Phys. A {\bf 30}, 7997 (1997).

\bibitem{sg}
M. M\'ezard, G. Parisi, and M. A. Virasoro, {\it Spin glass theory and beyond: An Introduction to the Replica Method and Its Applications}, World Scientific Publishing Company (1987).

\bibitem{PLDCecile}
P. Le Doussal and C. Monthus,  Physica A {\bf 317}, 140 (2003).

\bibitem{Dahmen}
M. Leblanc, L. Angheluta, K. Dahmen, and N. Goldenfeld, Phys. Rev. E {\bf 87}, 022126 (2013). 


\bibitem{Shapir}
S. Raychaudhuri, M. Cranston, C. Przybla and Y. Shapir, Phys. Rev. Lett. {\bf 87}, 136101 (2001).

\bibitem{GHPZ}
G. Gyorgyi, P. C. Holdsworth, B. Portelli and Z. Racz, Phys. Rev. E {\bf 68}, 056116 (2003).

\bibitem{Satya_Airy1}
S. N. Majumdar and A. Comtet, Phys. Rev. Lett. {\bf 92}, 225501 (2004). 

\bibitem{Satya_Airy2}
S. N. Majumdar and A. Comtet J. Stat. Phys. {\bf 119}, 777 (2005).


\bibitem{SOS_Airy}
G. Schehr, S. N. Majumdar, Phys. Rev. E {\bf 73}, 056103 (2006).


\bibitem{TW}
C. A. Tracy and H. Widom, Commun. Math. Phys. {\bf 159}, 151 (1994).

\bibitem{SMS14}
S. N. Majumdar and G. Schehr, J. Stat. Mech., 01012 (2014).


\bibitem{reviewMPS}
S. N. Majumdar, A. Pal, G. Schehr, Physics Reports \textbf{840}, 1 (2020).

\bibitem{SMS14r}
G. Schehr and S. N. Majumdar, \textit{Exact record and order statistics of random walks via first-passage ideas}, in First-passage phenomena and their applications, 226 (2014).


\bibitem{Vivo15}
P. Vivo, Eur. J. Phys. \textbf{36}, 055037 (2015).

\bibitem{SSM07}
S. Sabhapandhit and S. N. Majumdar, Phys. Rev. Lett. \textbf{98}, 140201 (2007).

\bibitem{SSM08}
S. Sabhapandit, S. N. Majumdar and S. Redner, J. Stat. Mech., 03001 (2008).

\bibitem{Arnold}
B. C. Arnold, N. Balakrishnan and H. N. Nagaraja, \textit{A first course in order statistics}, Wiley, New York (1992).

\bibitem{Nagaraja}
H. N. Nagaraja, H. A. David, \textit{Order statistics (third ed.)}, Wiley, New Jersey (2003).


\bibitem{Feller}
W. Feller, \textit{An introduction to probability theory and its applications}, Vol. I. and II, Third edition, John Wiley \& Sons, Inc., New York-London-Sydney, (1968).

\bibitem{dean_majumdar}
D.~S. Dean, S.~N. Majumdar, Phys. Rev. E {\bf 64}, 046121 (2001).



\bibitem{pld_carpentier}
D. Carpentier, P. Le Doussal, Phys.Rev. E {\bf 63}, 026110 (2001);
Erratum-ibid. {\bf 73}, 019910 (2006); Y.V. Fyodorov and
J.-P. Bouchaud {\it J. Phys. A: Math. Theor.} {\bf 
41}, 372001 (2008); Y.~V. Fyodorov, P. Le Doussal and A. Rosso,
J. Stat. Mech. 10005 (2009).  


\bibitem{satya_airy}
S.~N.~Majumdar, A.~Comtet, Phys. Rev. Lett. {\bf 92}, 225501 (2004);
J. Stat. Phys. {\bf 119}, 777 (2005).

\bibitem{schehr_airy}
G. Schehr, S. N. Majumdar, Phys. Rev. E {\bf 73}, 056103 (2006).


\bibitem{gyorgyi}
G. Gy\"orgyi, N. Moloney, G. Ozog{\'a}ny, Z. R{\'a}cz, Phys. Rev. E {\bf 75}, 021123 (2007). 


\bibitem{satya_yor}
S.~N.~Majumdar, J.~Randon-Furling, M.~J.~Kearney, M.~Yor, J. Phys. A
Math. Theor. {\bf 41}, 365005 (2008).  



\bibitem{comtet_precise}
A. Comtet, S.~N. Majumdar,  J. Stat. Mech. Theor. Exp. {\bf 06}, 06013 (2005).

\bibitem{schehr_rsrg}
G. Schehr, P. Le Doussal, J. Stat. Mech., 01009 (2010). 


\bibitem{SM12}
G. Schehr, S. N. Majumdar, Phys. Rev. Lett. {\bf 108}, 040601 (2012).

\bibitem{MMS13}
S. N. Majumdar, Ph. Mounaix, and G. Schehr, Phys. Rev. Lett. \textbf{111}, 070601 (2013).

\bibitem{MMS14}
S. N. Majumdar, Ph. Mounaix, and G. Schehr, J. Stat. Mech. \textbf{2014}, 09013 (2014).

\bibitem{Lacroix}
B. Lacroix-A-Chez-Toine, S. N. Majumdar and G. Schehr  J. Phys. A: Math. Theor. \textbf{52}, 315003 (2019).

\bibitem{Mori1}
F. Mori, S. N. Majumdar, G. Schehr, Phys. Rev. Lett. \textbf{123}, 200201 (2019).

\bibitem{Mori2}
F. Mori, S. N. Majumdar, G. Schehr, Phys. Rev. E \textbf{101}, 052111 (2020).


\bibitem{BM17}
M. Battilana, S. N. Majumdar, and G. Schehr, Markov Processes Relat. Fields {\bf 26}, 57 (2020).

\bibitem{PT20}
 J. Pitman, W. Tang, arXiv:2007.13991 (2020).

\bibitem{Mori3}
F. Mori, S. N. Majumdar, G. Schehr, EPL \textbf{135}, 30003 (2021).


\bibitem{PT21}
 J. Pitman, W. Tang, arXiv:2107.05095 (2021).

\bibitem{BSG21a}
B. De Bruyne, S. N. Majumdar and G. Schehr 
J. Stat. Mech., 083215 (2021).

\bibitem{BSG21b}
B. De Bruyne, O. B\'enichou, S. N. Majumdar and G. Schehr 
J. Phys. A: Math. Theor. \textbf{55}, 144002 (2021).




\bibitem{Erdos46}
P. Erd\"os, M. Kac, Bull. Am. Math. Soc. \textbf{52}, 292 (1946).

\bibitem{Darling56}
D. A. Darling, Trans. Am. Math. Soc. \textbf{83}, 164 (1956).

\bibitem{VVI94}
 V. V. Ivanov, Astron. Astrophys. \textbf{286}, 328 (1994).


\bibitem{Pollackzek}
F. Pollaczek C. R. Acad. Sci. Paris, \textbf{234}, 2334 (1952); J. Appl. Probab. \textbf{12}, 390 (1975).

\bibitem{Wendel}
J. G. Wendel, Ann. Math. Statist. \textbf{31}, 1034 (1960). 


\bibitem{Port}
S. C. Port, J. Math. Anal. Appl. \textbf{6}, 109 (1963).

\bibitem{Dassios}
A. Dassios, Ann. Appl. Probab. \textbf{6}, 1041 (1996).

\bibitem{Chaumont}
L. Chaumont, J. London Math. Soc. \textbf{59}, 729 (1999).

\bibitem{Embrechts}
P. Embrechts, L.C.G. Rogers and M. Yor, Ann. Appl. Prob. \textbf{5}, 757 (1995).


\bibitem{Dassiosa}
A. Dassios, Ann. Appl. Prob. \textbf{5}, 389 (1995).



 \bibitem{Spi56}
F. Spitzer, Proc. Am. Math. Soc. {\bf 7}, 164 (1956).

\bibitem{EH05}
M. R. Evans, T. Hanney, J. Physics A: Math. Gen. {\bf 38}, 195 (2005).

\bibitem{MEZ05}
S. N. Majumdar, M. R. Evans, R. K. P. Zia, Phys. Rev. Lett. {\bf 94}, 180601 (2005).

\bibitem{EMZ06}
M. R. Evans, S. N. Majumdar, R. K. P. Zia, J. Stat. Phys. \textbf{123}, 357 (2006).

\bibitem{Satya_Houches}
S. N. Majumdar, in {\it Exact Methods in Low-dimensional Statistical Physics and Quantum Computing}: Lecture Notes of the Les Houches Summer School: Vol. 89, (2010).

\bibitem{Gradenigo}
G. Gradenigo, S. N. Majumdar, J. Stat. Mech., 053206 (2019).



\bibitem{MGM21}
F. Mori, G. Gradenigo, S. N. Majumdar, J. Stat. Mech., 103208 (2021).

\bibitem{MoriPRE}
F. Mori, P. Le Doussal, S. N. Majumdar, G. Schehr, Phys. Rev. E {\bf 103}, 062134 (2021).

\bibitem{Smith}
N. R. Smith, S. N. Majumdar, J. Stat. Mech., 053212 (2022).

\bibitem{Haubold11}
H. J. Haubold, A. M. Mathai, R. K. Saxena, J. Appl. Math. \textbf{2011}, 1 (2011).

\bibitem{MMS17}
S. N. Majumdar, Ph. Mounaix, G. Schehr, J. Phys. A: Math. Theor. \textbf{50}, 465002 (2017).







\end{thebibliography}
\end{document}